\documentclass[aps,prd,twocolumn,amsmath,superscriptaddress,preprintnumbers,amssymb,showpacs,floatfix,nofootinbib,longbibliography]{revtex4-1}

\usepackage{graphicx,dcolumn,hyperref,xcolor,bm,natbib}
\usepackage{calligra}
\usepackage{egothic}
\usepackage[T1]{fontenc}
\newfont{\rsfsten}{rsfs10 scaled 1200}
\newfont{\rsfsseven}{rsfs10 scaled 1200}
\newfont{\rsfsfive}{rsfs10 scaled 1200}
\usepackage{epsfig}
\usepackage{units}
\usepackage[utf8]{inputenc}
\usepackage{enumerate}
\usepackage{fontawesome}

\newcommand{\aap}{{Astron.~Astrophys.}}

\newcommand{\mnras}{{Mon.~Not.~R.~Astron.~Soc.}}

\newcommand{\be}{\begin{equation}}
\newcommand{\ee}{\end{equation}}
\newcommand{\bea}{\begin{eqnarray}}
\newcommand{\eea}{\end{eqnarray}}

\def\lsim{\mathrel{\raise.3ex\hbox{$<$\kern-.75em\lower1ex\hbox{$\sim$}}}}
\def\gsim{\mathrel{\raise.3ex\hbox{$>$\kern-.75em\lower1ex\hbox{$\sim$}}}}

\newcommand{\beq}{\begin{equation}} \newcommand{\eeq}{\end{equation}}

\newcommand{\Sec}[1]{Sec.~\ref{#1}}

\begin{document}

\title{
Robustness of the Galactic Center Excess Morphology Against Masking
}

\author{Yi-Ming Zhong}
\affiliation{Department of Physics, City University of Hong Kong, Kowloon, Hong Kong SAR, China}
\email{yimzhong@cityu.edu.hk}
\affiliation{Kavli Institute for Cosmological Physics, University of Chicago, Chicago, IL 60637, USA}

\author{Ilias Cholis}
\affiliation{Department of Physics, Oakland University, Rochester, Michigan, 48309, USA}
\email{cholis@oakland.edu}

\date{\today}

\begin{abstract}
The Galactic Center Excess (GCE) remains an enduring mystery, with leading explanations being annihilating dark matter or an unresolved population of millisecond pulsars. Analyzing the morphology of the GCE provides critical clues to identify its exact origin. We investigate the robustness of the inferred GCE morphology against the effects of masking, an important step in the analysis where the gamma-ray emission from point sources and the galactic disk are excluded. Using different masks constructed from \textit{Fermi} point source catalogs and a wavelet method, we find that the GCE morphology, particularly its ellipticity and cuspiness, is relatively independent of the choice of mask for energies above $2-3$ GeV. The GCE morphology systematically favors an approximately spherical shape, as expected for dark matter annihilation. Compared to various stellar bulge profiles, a spherical dark matter annihilation profile better fits the data across different masks and galactic diffuse emission backgrounds, except for the stellar bulge profile from Coleman et al. (2020), which provides a similar fit to the data. Modeling the GCE with two components, one from dark matter annihilation and one tracing the Coleman Bulge, we find this two-component model outperforms any single component or combinations of dark matter annihilation and other stellar bulge profiles. Uncertainty remains about the exact fraction contributed by each component across different background models and masks. However, when the Coleman Bulge dominates, its corresponding spectrum lacks characteristics typically associated with millisecond pulsars, suggesting that it mostly models the emission from other sources instead of the GCE that is still present and spherically symmetric.
\href{https://github.com/ymzhong/gce_mask}{\faGithub}  
\end{abstract}

\maketitle

\section{Introduction}
\label{sec:introduction}

The Galactic Center Excess (GCE) has remained 
an unexplained component 
of gamma-ray emission for a decade and a half since it was 
first identified in the 
\textit{Fermi}-LAT 
data by Ref.~\cite{Goodenough:2009gk}. The GCE peaks in its energy output around a GeV. It appears to be a diffuse and extended component of gamma rays that cannot be explained 
by steady-state galactic diffuse emission, originating from the interaction with the interstellar medium of cosmic rays, typically produced 
by astrophysical sources such as pulsars or supernova remnants \cite{Hooper:2010mq, 2009arXiv0912.3828V, Abazajian:2010zy, Hooper:2011ti, Hooper:2013rwa, Gordon:2013vta, Abazajian:2014fta, Daylan:2014rsa, Calore:2014xka, Zhou:2014lva, TheFermi-LAT:2015kwa, Linden:2016rcf, Cholis:2021rpp}. 
The exact origin of the GCE is still a topic of ongoing research debate, with the two most intriguing explanations being dark matter annihilation \cite{Gordon:2013vta, Daylan:2014rsa, Calore:2014nla, Agrawal:2014oha, Berlin:2015wwa, Karwin:2016tsw, TheFermi-LAT:2017vmf, Leane:2019xiy} and an unresolved population of millisecond pulsars (MSPs) \cite{Abazajian:2012pn, Hooper:2013nhl, Lee:2015fea, Petrovic:2014xra, Cholis:2014lta, Bartels:2017vsx, Buschmann:2020adf, Gautam:2021wqn}; while also a sequence of cosmic-ray burst events from the center of the galaxy provides a third alternative \cite{Petrovic:2014uda, Carlson:2014cwa, Cholis:2015dea}. 
If the dark matter explanation is true, the GCE will not only serve as the first evidence that dark matter interacts with ordinary matter beyond its gravitational pull, but it will also shed light on its production mechanism. If the MSPs explanation is true, it could reveal a new population of MSPs with luminosity function that is different from the known pulsars in globular clusters or the galactic disk \cite{Cholis:2014lta, Zhong:2019ycb}. Finally, if the GCE comes from a sequence of burst events, it would help us understand the conditions and recent history of the inner region of the Milky Way. 

Various efforts have been pursued to identify the origin of the GCE, which include analyzing the details of its energy spectrum, its photon statistics, and its morphology \cite{Daylan:2014rsa, Calore:2014xka, Bartels:2015aea, TheFermi-LAT:2015kwa, Macias:2016nev, Macias:2019omb, Zhong:2019ycb, Buschmann:2020adf, DiMauro:2021raz, Cholis:2021rpp, Pohl:2022nnd, McDermott:2022zmq} (see also~\cite{Miller:2023qph}). A common step adopted in many GCE analyses is masking certain areas of the galactic center (GC). Masks are applied to exclude 
pixels of the sky where conventional sources of gamma-ray emission are expected to be so bright that 
even the uncertainty on their expected flux dominates the possible signal. 
Those include the galactic disk and the known point (or extended) sources that could interfere with the GCE signal. Various methods exist for creating such masks. One standard approach involves using the \textit{Fermi}-LAT point source catalog to exclude regions around the cataloged sources, combined with band-shaped masks along $b=0^\circ$ to exclude the disk.

The choice of the mask and method to apply them to the data can potentially affect the results of the GCE analysis. For example, earlier works have demonstrated such masking effects on the Non-Poissonian Template Fitting~\cite{Lee:2015fea}, where switching from masks created using the \textit{Fermi}-LAT 3FGL catalog to the 4FGL catalog diminishes the statistical power of the method on the GCE photon statistics~\cite{Buschmann:2020adf}. Refs.~\cite{Zhong:2019ycb, Cholis:2021rpp} investigated how the GCE characteristics are affected using alternative masks for the known \textit{Fermi}-LAT point sources. They found its energy spectrum and morphology robust to the different updates of the \textit{Fermi}-LAT catalog. Given the significance of the GCE morphology in its interpretation, we perform 
an extensive analysis of the robustness of the GCE morphology to alternative masking procedures in this work.
We do that by combining with the vast publicly available library of galactic diffuse emission models produced in Ref.~\cite{Cholis:2021rpp} to account for the astrophysical modeling uncertainties of the Milky Way conditions. 
Such an analysis is especially timely given the increasing number of point sources released in the latest \textit{Fermi}-LAT catalogs~\cite{Fermi-LAT:2019yla, Ballet:2020hze, Fermi-LAT:2022byn,Ballet:2023qzs}.

We begin in Sec.~\ref{sec:setup} by introducing the \textit{Fermi} data, the templates, the masks tested, and the statistical procedure used in our analysis. Sec.~\ref{sec:injection} presents our injection tests for the template fitting method in a controlled environment. Sec.~\ref{sec:morphism} shows the results of the GCE morphology under different masks. 
We find that the GCE morphology remains largely spherical and resembles that of a dark matter annihilation profile rather than that of a population of dim gamma-ray point sources tracing known stellar populations. 
An additional population of faint gamma-ray point sources tracing the stellar distribution known as the ``Coleman Bulge''~\cite{Coleman:2019kax} may exist. However, the spectrum associated with this component does not possess the characteristics expected from MSPs. We conclude in Sec.~\ref{sec:conclusions}.

\section{Setup for the template fitting}
\label{sec:setup}

Here, we describe the \textit{Fermi} data and the templates used for the diffuse emission and the GCE. We also describe the masks applied and the statistical procedure used in our template fitting analysis. Except for the mask construction, our analysis framework largely follows that of Refs.~\cite{Cholis:2021rpp, McDermott:2022zmq}.

\subsection{Data}
\label{sec:data}

We use \textit{Fermi} Pass 8 data, version P8R3, recorded from August 4, 2008, to April 14, 2021, corresponding to weeks $9-670$ of \textit{Fermi}-LAT observations.\footnote{\url{https://fermi.gsfc.nasa.gov/ssc/data/access/}} We use \textit{Fermi} {\tt ScienceTools P8v27h5b5c8} for selection cuts and to calculate the relevant exposure-cube files and exposure maps,\footnote{\url{https://fermi.gsfc.nasa.gov/ssc/data/analysis/}} which allow us to pass from fluxes to expected counts. The exposure is calculated for each pixel using the \textit{Fermi} {\tt ScienceTools P8v27h5b5c8}. 

We keep only FRONT-converted {\tt CLEAN} data. In addition, we set the following filters: {\tt zmax = $100^{\circ}$}, {\tt DATA$\_$QUAL==1}, {\tt LAT$\_$CONFIG==1}, and {\tt ABS(ROCK$\_$ANGLE) < 52}. Our data maps are centered at the galactic center and cover a square window of $40^{\circ}$ per side in galactic coordinates in Cartesian pixels of size $0.1^{\circ} \times 0.1^{\circ}$. Unlike a {\tt HEALPix} pixelization, our pixels do not have equal area, but we account for this in our fits.

\setlength{\tabcolsep}{6pt}
\begin{table}[t]
    \begin{tabular}{ccrrrrr}
    \hline 
             & $E_{\textrm{min}}-E_{\textrm{max}} {\rm\,[GeV]}$ &  $\theta_{s} [^{\circ}]$ &  $\theta_{l} [^{\circ}]$\\
            \hline 
            0 & $0.275-0.357$ & 1.125 & 3.75 \\
            1 & $0.357-0.464$ &  0.975 & 3.25 \\
            2 & $0.464-0.603$ & 0.788 & 2.63 \\
            3 & $0.603-0.784$ & 0.600 & 2.00 \\
            4&  $0.784-1.02$ &  0.450 & 1.50 \\
            5 & $1.02-1.32$ & 0.375 & 1.25 \\
            6 & $1.32-1.72$ & 0.300 & 1.00 \\
            7 & $1.72-2.24$ & 0.225 & 0.750 \\
            8 & $2.24-2.91$ & 0.188 & 0.625 \\
            9 & $2.91-3.78$ & 0.162 & 0.540 \\
            10 & $3.78-4.91$ & 0.125 & 0.417 \\
            11 & $4.91-10.8$ & 0.100 & 0.333 \\
            12 & $10.8-23.7$ & 0.060 & 0.200 \\
            13 & $23.7-51.9$ & 0.053 & 0.175 \\
             \hline 
        \end{tabular}
       \caption{The energy bins and the energy-dependence of the fiducial radii $\theta_{s}$ ($\theta_{l}$)  used to mask known point sources with $\rm{TS}<49$ ($\rm{TS}\geq 49$).}
    \label{tab:PSF_vsE}
\end{table}

We bin the gamma-ray data in 14 energy bins spanning energies from 0.275 GeV to 51.9 GeV, given in the first column of Tab.~\ref{tab:PSF_vsE}. The first eleven energy bins have a constant log width; because the gamma-ray flux drops at higher energy, the final three energy bins are wider, so each bin has a roughly similar statistical impact in our fits. 

\subsection{Galactic Diffuse Emission and GCE Templates}
\label{sec:template}

We use the 80 astrophysical models that predict an equal number of sets of high-resolution energy-dependent templates developed in Ref.~\cite{Cholis:2021rpp} for the galactic diffuse emission background.\footnote{We use the term ``background'' to describe the combination of the galactic diffuse emission, the isotropic and mostly extragalactic emission, and the emission from the \textit{Fermi} bubbles. If the GCE is the ``signal,'' the term ``background'' is used in this context as the non-signal emission. 
For our region of interest (ROI), depending on the latitude and energy, about 50--80$\%$ of the total emission 
and 60--90$\%$ of the galactic diffuse emission is truly foreground emission, i.e., emission generated between our location and where the inner galaxy region begins (inner 3 kpc from the center of the Milky Way). 
See Ref.~\cite{Calore:2014xka} for a more detailed analysis of this.} Each set includes templates for $\pi^0$, Inverse Compton Scattering (ICS), and Bremsstrahlung, calculated for each energy bin used in this analysis. In addition, we include an energy-independent template for the isotropic background and a template for the \textit{Fermi} bubbles. In all cases, we projected the templates 
onto a $0.1^{\circ} \times 0.1^{\circ}$ Cartesian pixel grid near the galactic center that covers a square window of $40^{\circ} \times 40^{\circ}$. All templates are subsequently smoothed by the energy-dependent \textit{Fermi}-LAT point spread function~\cite{psf-vals}.

\begin{figure*}[t!]
\centering
\includegraphics[width=\textwidth]{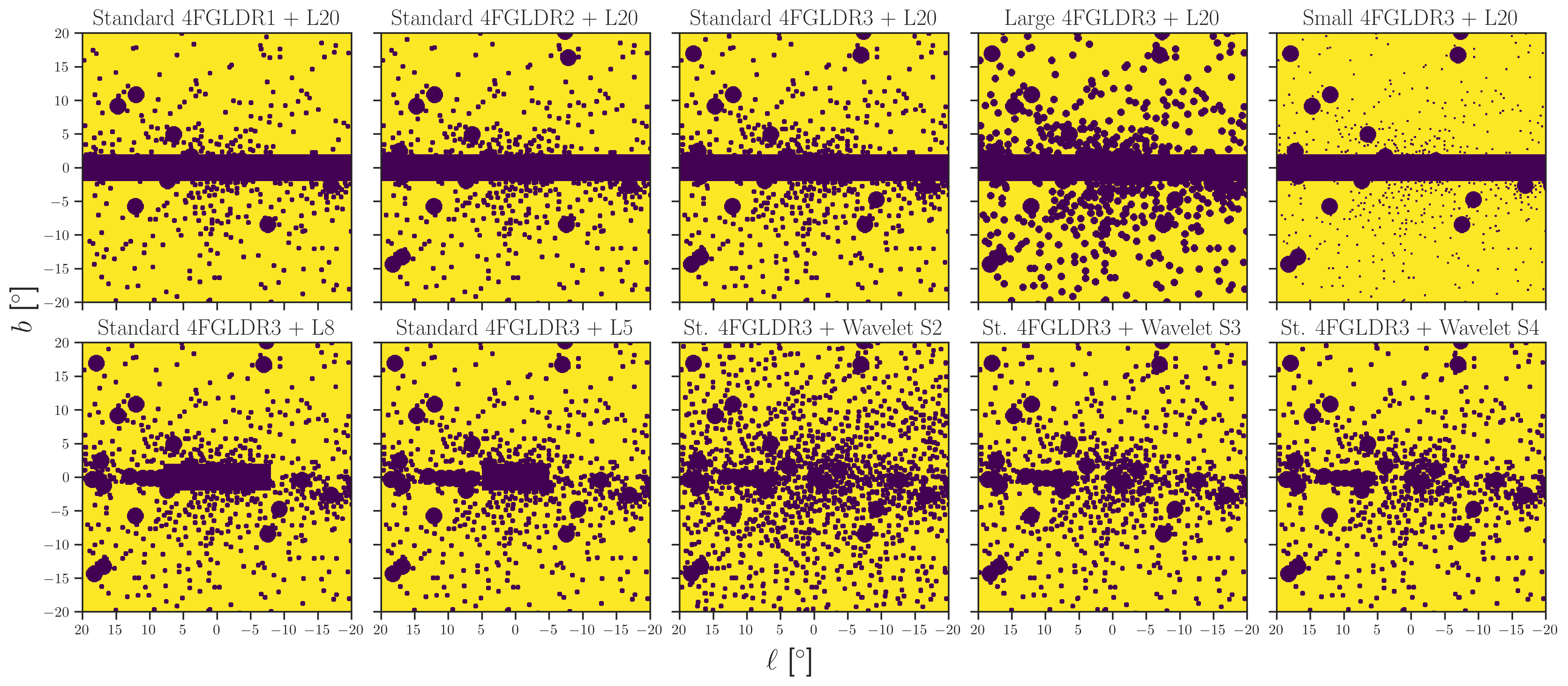}
\caption{The masks used in this work. Here, we show the masked regions (dark purple) for the energy bin 5 ($1.02\,\text{GeV}< E<1.32\,\text{GeV}$) only.}
\label{fig:mask_only}
\end{figure*}

For the GCE, we consider contracted, squared, and integrated along the line-of-sight Navarro-Freak-White (NFW) profiles of the form,
\beq
\rho(r) = \frac{\rho_0}{(r/r_c)^\gamma (1+r/r_c)^{3-\gamma}}, 
\label{eq:DM}
\eeq
where the cuspiness parameter $\gamma$ represents the inner slope of the density profile, $r_c = 20\,\text{kpc}$ is the scale radius and $\rho_0$ is the scale density that is fixed by $\rho(r=8.5\,\text{kpc}) = 0.4\,\text{GeV}/\text{cm}^3$. We further deform the NFW profile by introducing an ellipticity (elongation) parameter $\epsilon$, such that the opening angle from the GC, $\psi$, is related to the galactic coordinates ($b$, $\ell$) by
\beq
\cos(\psi) = \cos(b) \cos(\ell/\epsilon).
\label{eq:ellipticity}
\eeq
A spherically symmetric NFW template has $\epsilon =1$ while an oblate (prolate) NFW template has $\epsilon >1 (<1)$. We construct the GCE template with a combination of $\gamma$ and $\epsilon$ values. All the GCE templates are projected to the same Cartesian grid as the galactic diffuse emission and isotropic templates and smoothed by the same point spread functions.

\subsection{Masks}
\label{sec:mask}

We construct a variety of masks for the template fitting analysis.\footnote{The masks are available at \url{https://github.com/ymzhong/gce_mask}.} Most of the masks are a combination of the point source mask and the disk mask.

\emph{Point source masks}: We take the location information of the point sources from the 4FGL-DR1~\cite{Fermi-LAT:2019yla}, DR2~\cite{Ballet:2020hze}, and DR3~\cite{Fermi-LAT:2022byn} catalogs. The point sources are classified into those with a test statistic (TS) < 49 and those with TS$\geq 49$. We masked the two classes of point sources with disk-shaped masks that have energy-dependent radii $\theta_s$ and $\theta_l$ respectively (the same radii used in~\cite{Cholis:2021rpp}), as listed in Tab.~\ref{tab:PSF_vsE}. We call those point sources masks the ``Standard 4FGLDRX'' with X$=1,2,3$.

On top of the ``Standard 4FGLDRX'' masks, we introduce two variations on the mask radii. For the ``Large 4FGLDRX'' masks, we increase the fiducial radii by a factor of 1.5 for the point sources with TS$<49$ while keeping radii for the sources with TS$\geq 49$ unchanged. For the ``Small 4FGLDRX'' masks, we decrease the fiducial radii by a factor of 0.5 for the point sources with TS$<49$. The radii for the sources with TS$\geq 49$ are again unchanged.

\emph{Disk masks}: We set up three band-shaped masks for the galactic disk. For the standard disk mask (``L20''), we mask the region $|\ell| < 20^\circ$ and $|b| < 2^\circ$, i.e., the entire disk region of ROI. For the ``L8'' (``L5'') disk mask, we mask the region $|\ell| < 8^\circ$ ($|\ell| < 5^\circ$) and $|b| < 2^\circ$. 

\emph{Combined regular masks}: Combing the point source masks and the disk masks described above, we set up the following ``regular'' masks for our analysis: ``Standard 4FGLDRX + L20'', ``Large 4FGLDRX + L20'', ``Small 4FGLDRX + L20'', ``Standard 4FGLDRX + L8'', and ``Standard 4FGLDRX + L5''. For ``Standard 4FGLDRX + L20'', we create masks for X=$1,2,3$. For other variations, the masks are only built for X=$2,3$. We show some of those combined masks in the first seven panels of Fig.~\ref{fig:mask_only}.

\emph{Wavelet-based masks}: The wavelet method~\cite{Bartels:2015aea,Zhong:2019ycb} is a powerful way to identify potential point sources. Here, we utilize the wavelet peaks found in~\cite{Zhong:2019ycb}. We select peaks with the statistic $\mathcal S >2$, $3$, and $4$ (as defined in Refs.~\cite{Bartels:2015aea,Zhong:2019ycb}), respectively, masking each peak with a disk-shaped mask that has energy-dependent radii identical to those used for the point sources with TS$<49$. We then combine these wavelet peak masks with the ``Standard 4FGLDRX'' point source masks and leave out the disk masks. The resulting masks are ``Standard 4FGLDRX + Wavelet SY'' with X=$2,3$ and Y$=2,3,4$, where the Y values represent the thresholds of $\mathcal S$. We show the masks in the last three panels of Fig.~\ref{fig:mask_only}.

\begin{figure}[t!]
    \centering
    \includegraphics[width=0.48\textwidth]{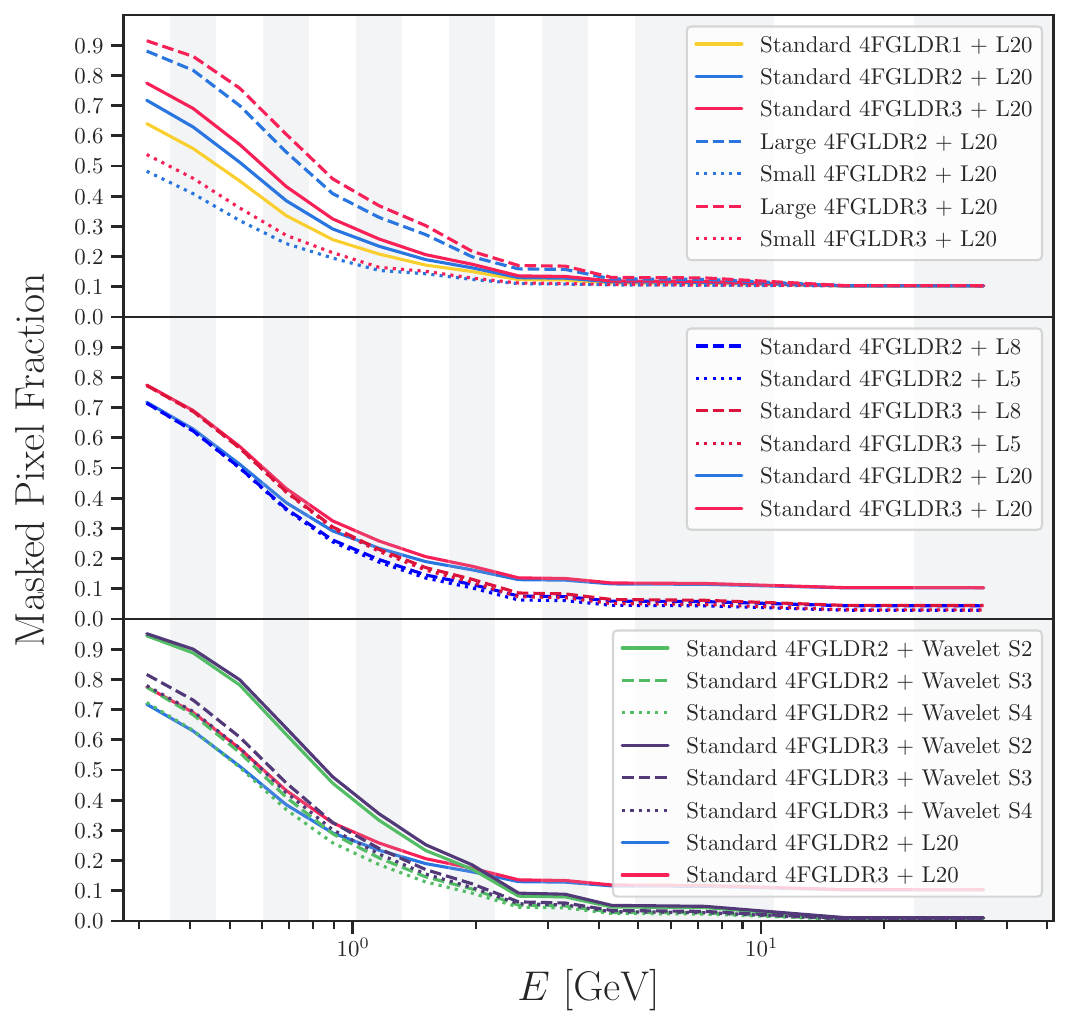}
    \caption{Masked pixel fraction for various masks used in the analysis. The vertical bands indicate the setup of 14 energy bins.}
    \label{fig:mpf}
    \end{figure}

For each mask, we compute the fraction of the masked pixels in each energy bin to the total number of pixels in the inner $40^\circ \times 40^\circ$ galactic center region ($1.6\times 10^5$ pixels). The results are shown in Fig.~\ref{fig:mpf}. Overall, the masked area decreases as the energy bin's energy increases. This reflects that the mask radii of the point sources decrease, following the \textit{Fermi} point spread function as the energy increases. Looking into more detail, the top panel of Fig.~\ref{fig:mpf} highlights the differences between masks created by different point source catalogs as well as by different radii criteria. As the number of point sources in the catalog increases, the masked pixel fraction significantly increases. The difference is significant for $E< 2.24\,\text{GeV}$ and gets diminished for higher energy bins as the number of masked pixels by the disk dominates over those by the point sources. Increasing or decreasing the radii criteria introduces a significant variation in the masked pixel fraction, which is the most significant for the lower energy bins. 
The middle panel of Fig.~\ref{fig:mpf} highlights the difference in how we mask the galactic disk. 
Here, the difference between masking the entire galactic disk or parts of it is most significant at the higher energy bins ($E>0.603\,\text{GeV}$). For lower energy bins, the two sets of masks yield a similar masked pixel fraction, given 
that the point sources mask dominates the lower latitudes.
Combining the information from the two panels, we can conclude that energy bins 0--3 are 
mostly affected 
by point sources masks, energy bins 8--13 are 
mostly affected by the disk mask, while energy bins 4--7 are influenced by both types. 
The lower panel of Fig.~\ref{fig:mpf} highlights the difference between different wavelet-based masks and those with a band mask for the disk. The wavelet-based masks, in general, mask a greater fraction of the GC region than the ``Standard 4FGLDRX + L20'' masks at lower energy bins. However, as energy increases and the mask radii shrink, the wavelet-based masks are not blocked by the disk mask and have a sub 10\% masked pixel fractions for $E>3\,\text{GeV}$. Also, at energies around 1 GeV, typically we only get 30\% of the inner $40^{\circ} \times 40^{\circ}$ region masked.

\subsection{Statistical Procedure}
\label{sec:stat}

We construct the log-likelihood $\ln \mathcal{L}_j$ for a given energy bin $j$ as,
\begin{align}
- 2 \ln \mathcal{L}_j ={}& \Big\{
2 \sum_{p} \left[\mathcal M_{j,p} \mathcal C_{j,p} + \ln\left[(\mathcal M_{j,p}\mathcal D_{j,p})!\right] \right. \nonumber\\
& \left.- (\mathcal M_{j,p}\mathcal D_{j,p}) \ln (\mathcal M_{j,p}\mathcal C_{j,p})  \right]\Big\} 
\nonumber  \\
{}& + \chi_{\textrm{Bubbles},j}^{2} + \chi_{\textrm{Iso},j}^{2}.
\label{eq:Likelihood}
\end{align}
The index $p$ runs over the pixels of the ROI. The data $\mathcal D_{j,p}$ and the mask $\mathcal M_{j,p}$ are described in detail in \Sec{sec:data} and \Sec{sec:mask}, respectively. The expected counts $\mathcal C_{j,p} = \mathcal E_{j,p} \sum_i c^i_j \Phi^i_{j,p} $ are obtained from summing over all the diffuse emission component templates $\Phi^i_{j,p}$ ($i$ runs over different components) with independent normalizations $c_j^i$. We run fits without the GCE and with one or two templates to describe it. In all cases, our fits assume independent normalizations for each template. After linearly adding the templates, we multiply their sum by the exposure $\mathcal E_{j,p}$. The ``external $\chi^2$'' functions $\chi^2_{\textrm{Bubbles},j}$ and $\chi^2_{\textrm{Iso},j}$ are constraints that act as penalties when the Bubbles and isotropic normalizations deviate too much from their spectra measured at high latitudes, see Ref.~\cite{Fermi-LAT:2014sfa, Ackermann:2014usa}.

We use \texttt{emcee}~\cite{ForemanMackey:2012ig}, a Markov chain Monte Carlo (MCMC) program, to sample the coefficient parameters $c^i_j$ and to assess their posterior probability distribution. This process allows us to conveniently obtain the coefficient parameters that minimize the negative log-likelihood, Eq.~\ref{eq:Likelihood}, and identify the parameters' uncertainties. We use \texttt{EnsembleSampler} with 100 walkers and 1000 steps to assess convergence. We then discard the first 300 steps of the resulting chain from \texttt{emcee} and draw posterior probabilities for the parameters with \texttt{ChainConsumer}~\cite{Hinton2016}. The total negative log-likelihood is a sum over the value of all the energy bins,
\beq
-2 \ln \mathcal L \equiv -2 \sum_j \ln \mathcal L_j.
\eeq

\begin{figure*}[t]
\centering
\includegraphics[width=0.5\textwidth]{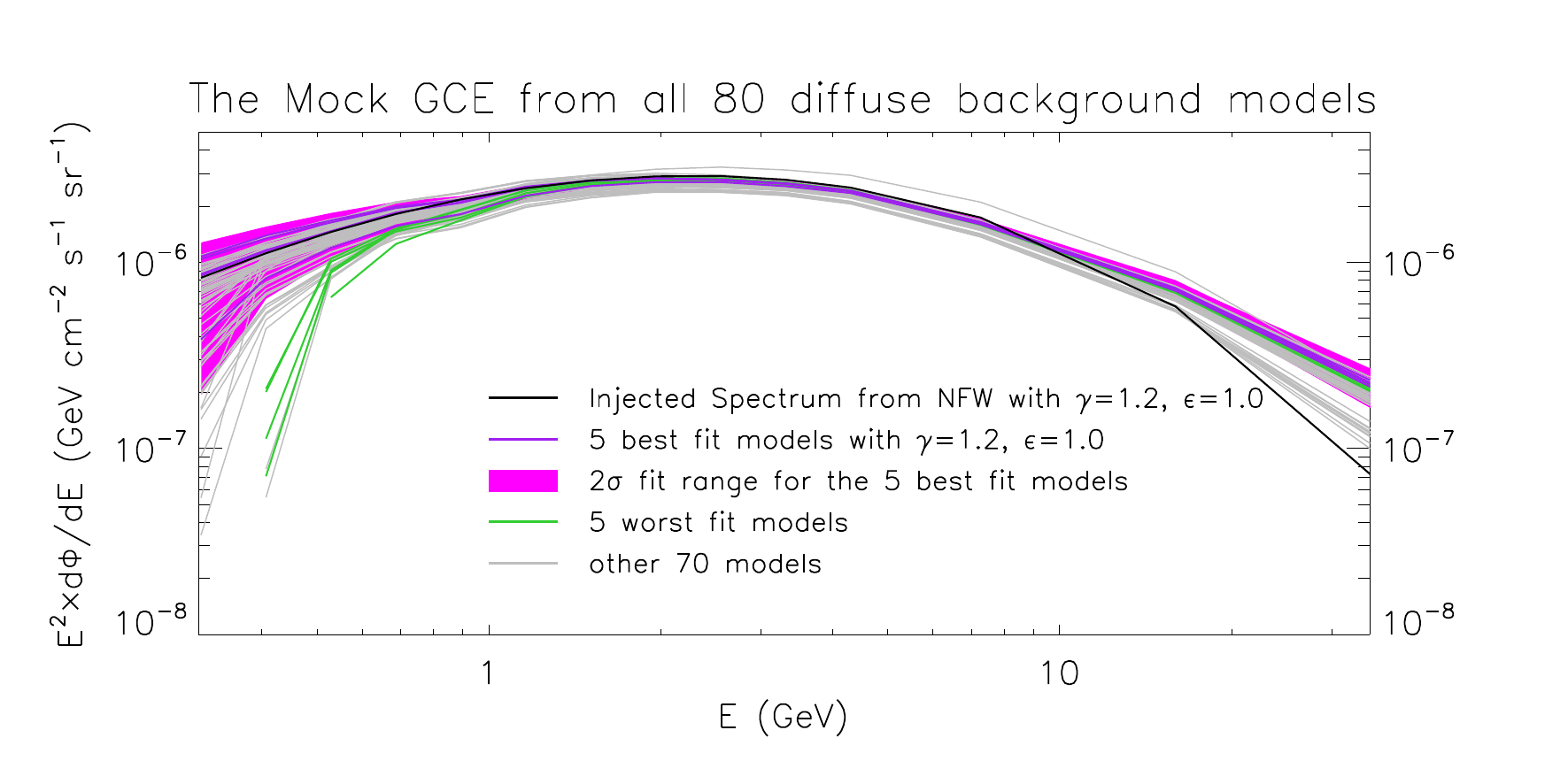}~\includegraphics[width=0.5\textwidth]{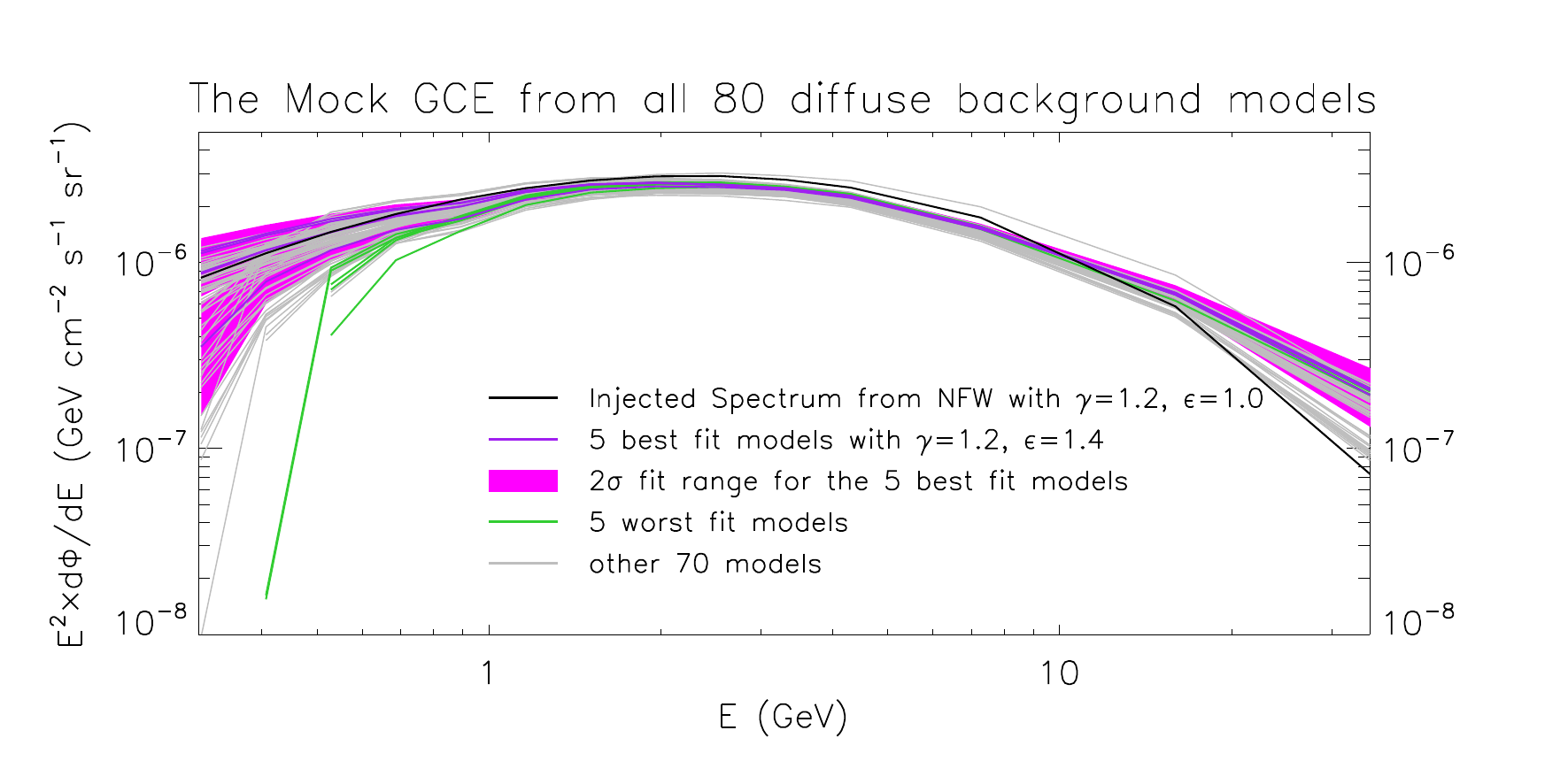}\\ 
\includegraphics[width=0.5\textwidth]{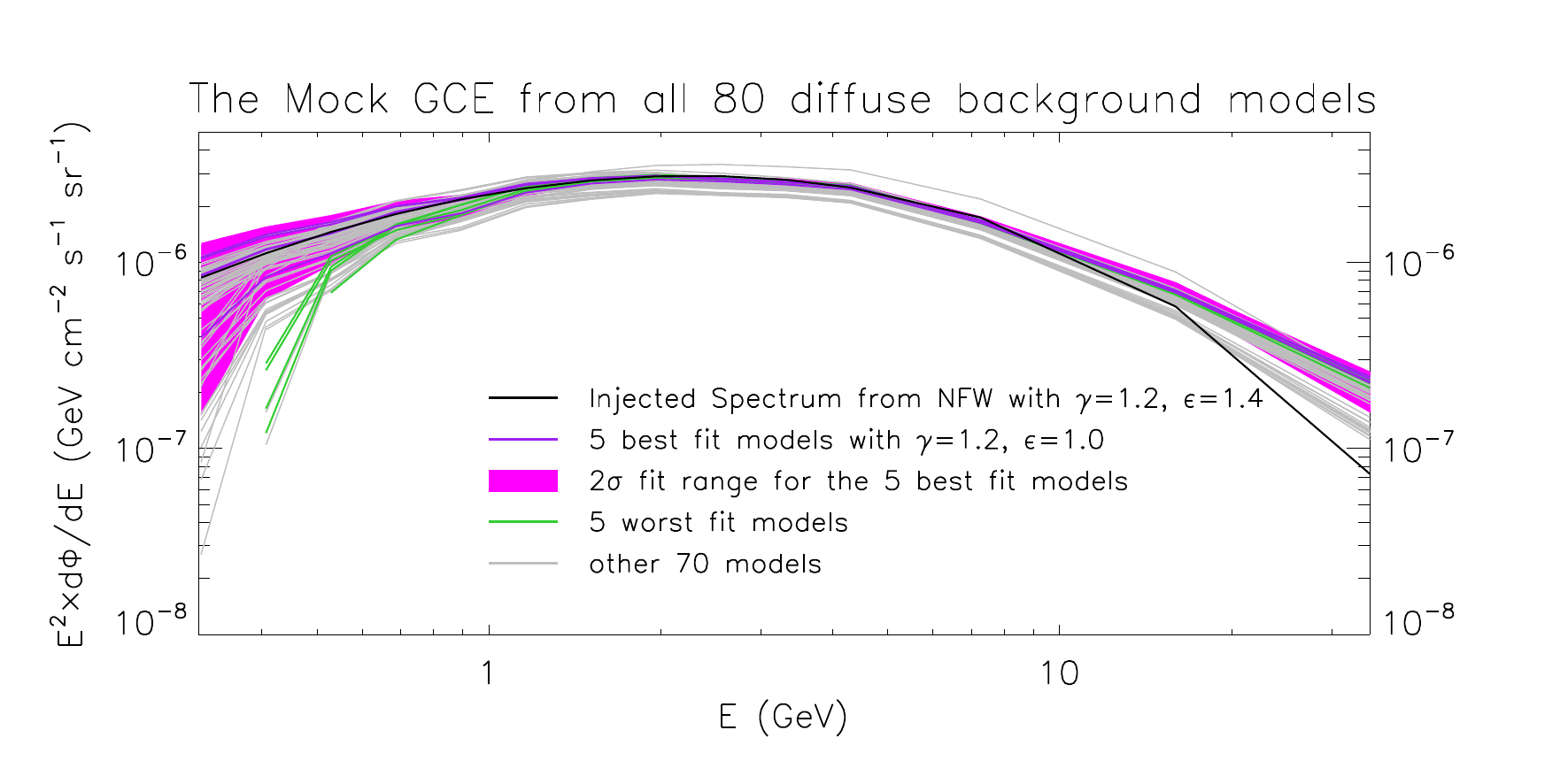}~
\includegraphics[width=0.5\textwidth]{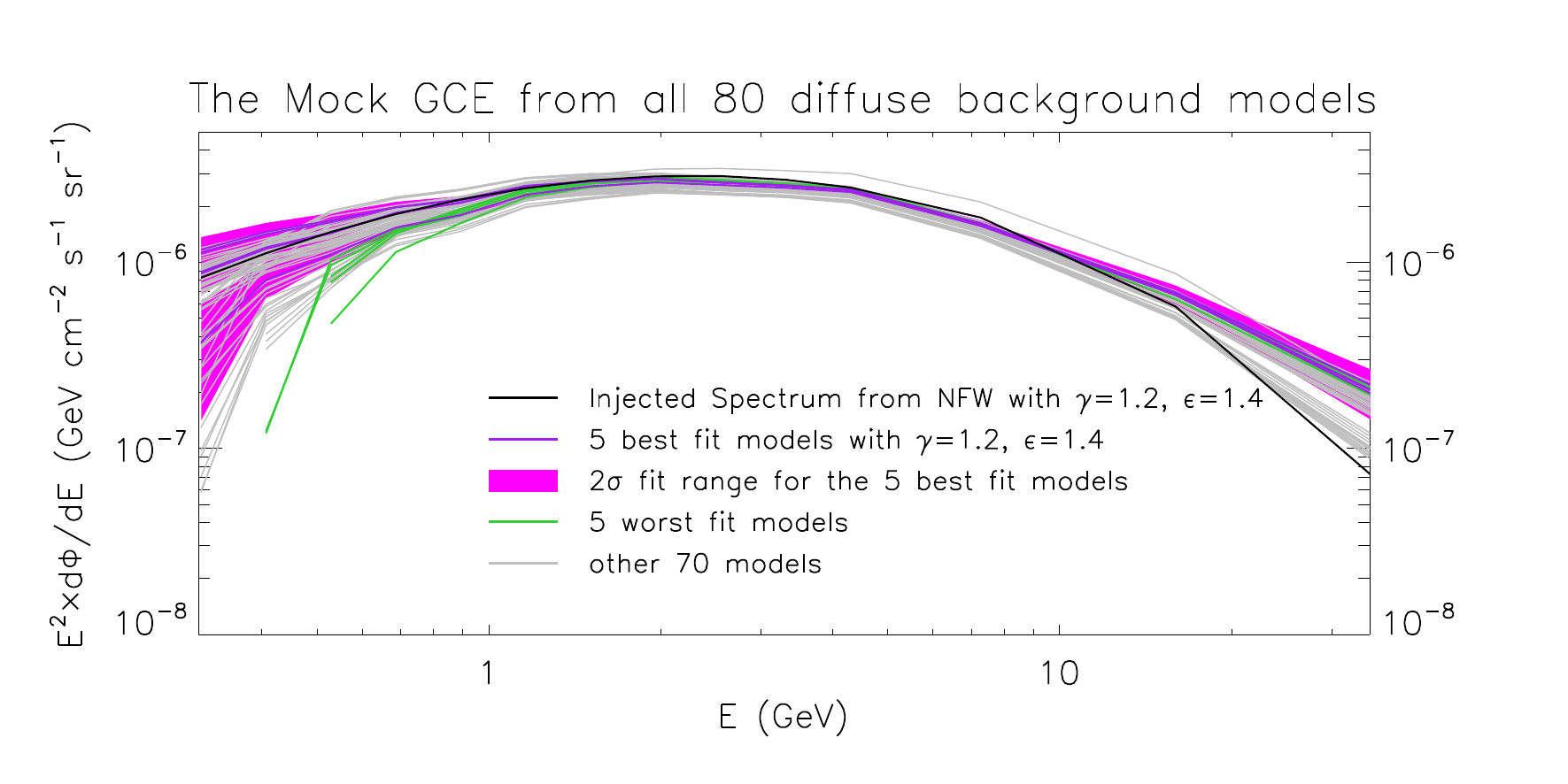} \\
\includegraphics[width=0.5\textwidth]{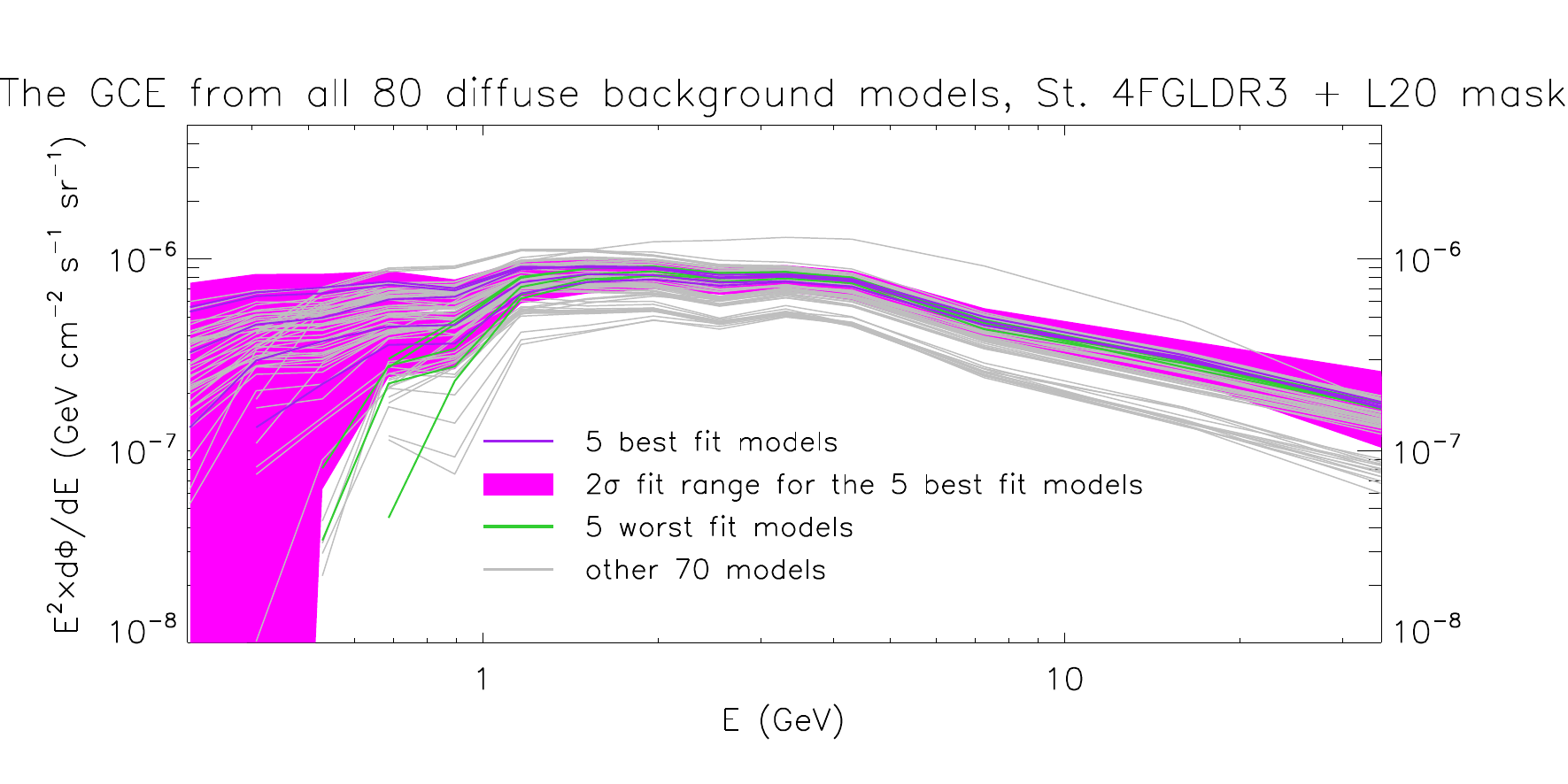}
\caption{Top two rows:  The inferred energy spectra of the GCE from the mock \textit{Fermi} data for the four injection tests, utilizing all 80 background galactic diffuse emission models of Ref~\cite{Cholis:2021rpp} and the ``Standard 4FGLDR3 + L20'' mask. Further details are provided in the text. Bottom panel: the inferred GCE spectrum from the actual \textit{Fermi} data for comparison. In the fit, the GCE flux is permitted to be negative, although this occurs in only a small subset of models and exclusively at energies below 0.7 GeV (indicated by the termination of lines). For the five background models out of the 80 that yield the best overall fit to the gamma-ray data, the best-fit GCE normalization and its corresponding $2\sigma$ uncertainty range are represented by purple lines and magenta bands, respectively. The green lines give the GCE normalization from the five galactic diffuse emission models that provide the worst fit to the gamma-ray data out of the 80 considered.}
\label{fig:injection_test}
\end{figure*}

\section{Injection test}
\label{sec:injection}

Injecting a GCE component into the \textit{Fermi} data has been utilized to assess the sensitivity and robustness of a method in determining the characteristics of the GCE~\cite{Leane:2019xiy}. In this study, we conduct an injection test to evaluate how well the template fitting can identify the ellipticity parameter $\epsilon$ of the GCE. We investigate two GCE signals: a spherical NFW profile with parameters $\gamma = 1.2, \epsilon=1.0$  and an oblate NFW profile with  $\gamma = 1.2, \epsilon=1.4$. The GCE signals are injected into the \textit{Fermi} data with a flux twice the fiducial value. Note that the template fit of the actual \textit{Fermi} data indicates that the observed flux of the GCE is around the fiducial values for both signals. Utilizing these mock \textit{Fermi} datasets, we conduct four separate template-fitting analyses as outlined in Tab.~\ref{tab:injection}. In these tests,  scenarios (1) and (4) match the injected data with the correct GCE template, whereas scenarios (2) and (3) involve a mismatch between the injected data and the template.

\begin{table}[htbp]
   \centering
   \begin{tabular}{@{} ccc @{}} 
     \hline 
           & Mock data & Fitting GCE Template\\
           \hline
      (1)   & \textit{Fermi}+2$\times$NFW($\epsilon = 1.0$) & 80 GDEs, NFW($\epsilon = 1.0$) \\
      (2)  & \textit{Fermi}+2$\times$NFW($\epsilon = 1.0$) & 80 GDEs, NFW($\epsilon = 1.4$) \\
      (3)   & \textit{Fermi}+2$\times$NFW($\epsilon = 1.4$) & 80 GDEs, NFW($\epsilon = 1.0$) \\
      (4)   & \textit{Fermi}+2$\times$NFW($\epsilon = 1.4$) & 80 GDEs, NFW($\epsilon = 1.4$) \\
        \hline 
   \end{tabular}
   \caption{The mock \textit{Fermi} data and templates used for the injection tests. For the GCE templates, we use a spherical ($\epsilon=1.0$) or oblate  ($\epsilon=1.4$) NFW profile with the cuspiness parameter of $\gamma=1.2$. ``GDE'' stands for galactic diffuse emission model.}
   \label{tab:injection}
\end{table}

\begin{figure*}
\centering
\includegraphics[width=0.48\textwidth]{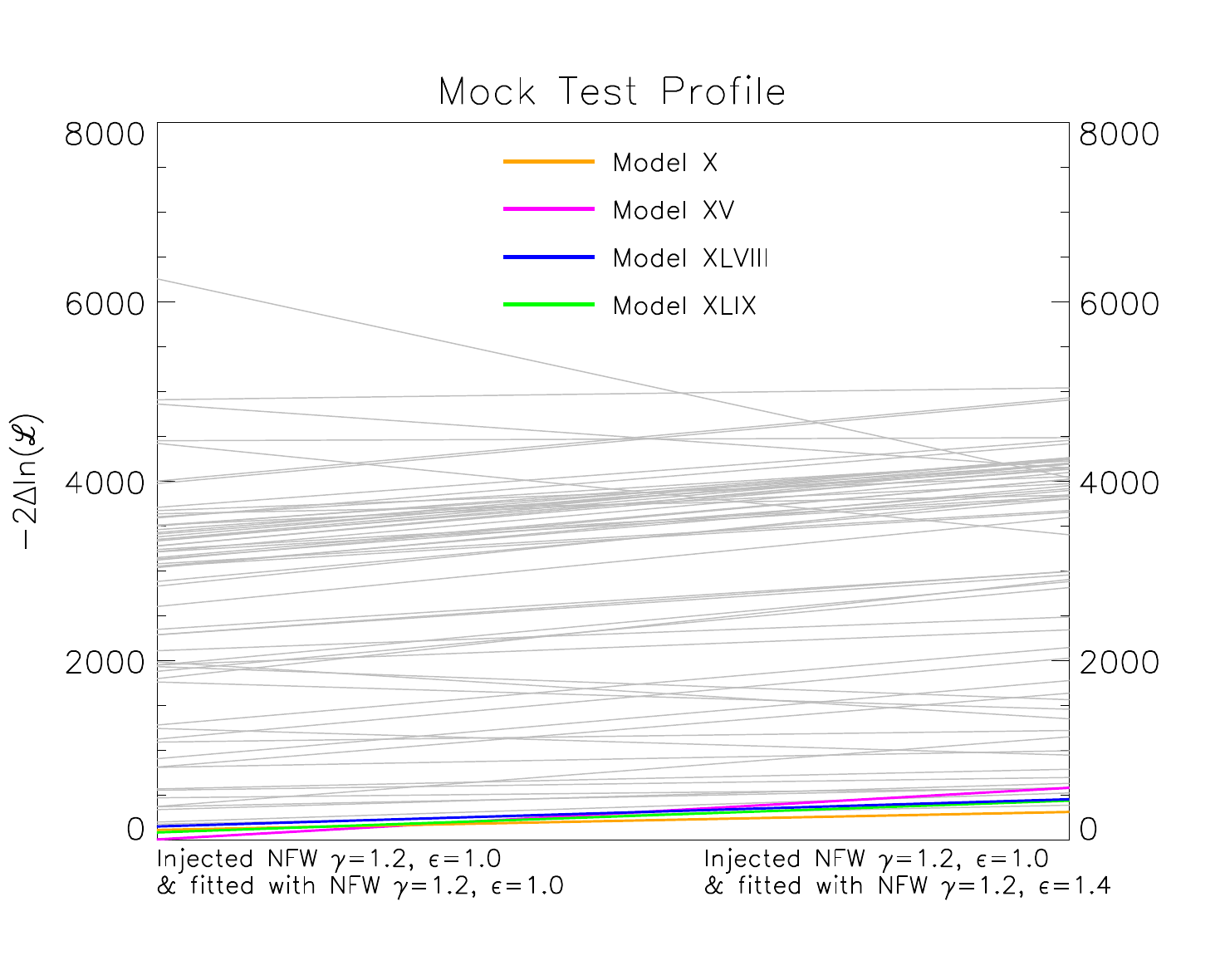}~\includegraphics[width=0.48\textwidth]{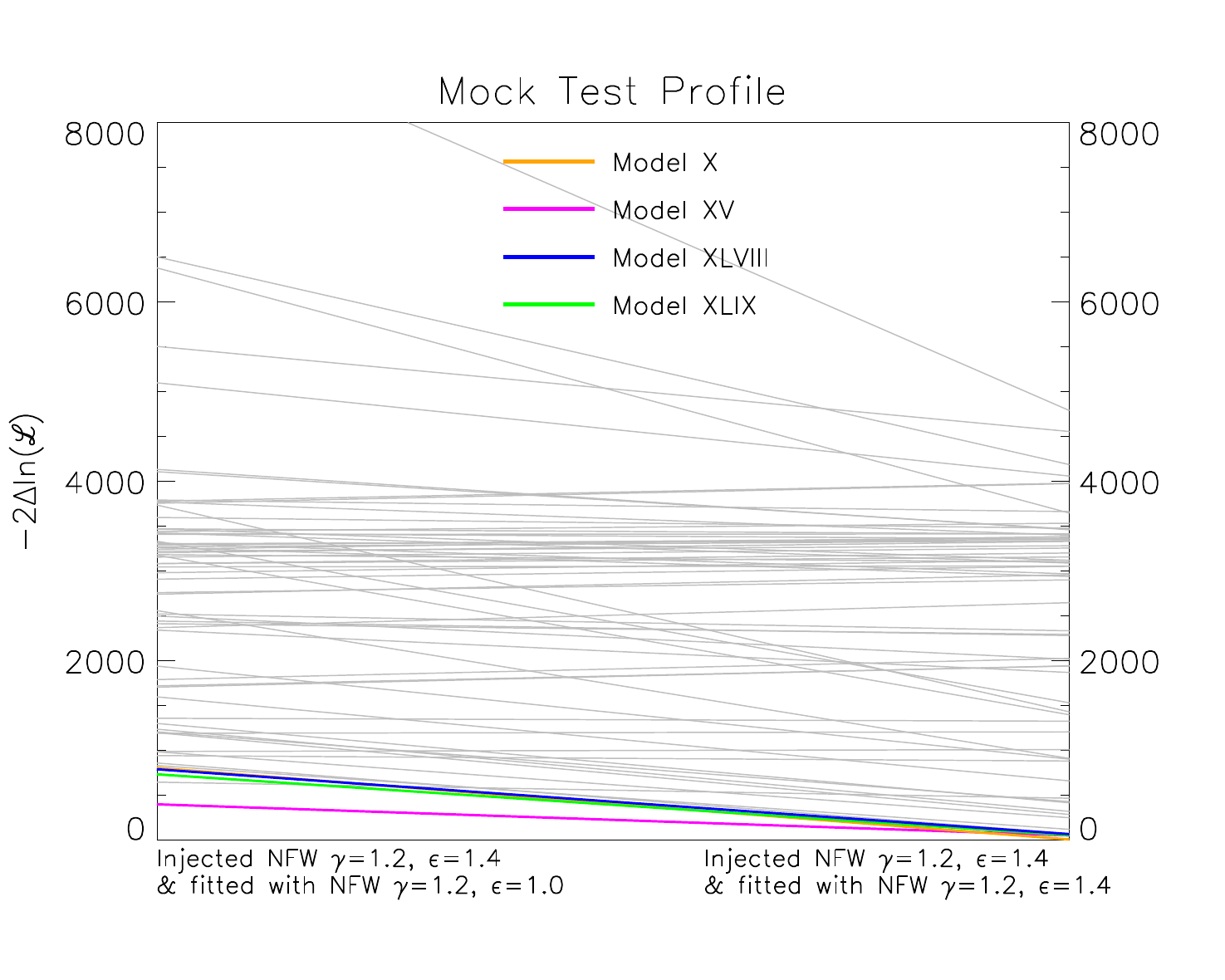}
\caption{Comparison of the difference in the total negative log-likelihood, $-2 \Delta \ln \mathcal{L}$, for injection test scenario (1) vs. scenario (2) (left), scenario (3) vs. scenario (4) (right). Colored lines represent some of the galactic diffuse emission models that give the best fit to the \textit{Fermi} data.}
\label{fig:injection_result}
\end{figure*}

The resulting energy spectra of the GCE are shown in the top two rows of Fig.~\ref{fig:injection_test}, where we apply the ``Standard 4FGLDR3 + L20'' mask. The magenta bands indicate the $2\sigma$ fit ranges for the five best-fit background galactic diffuse emission models. 
For comparison, the lower panel of Fig.~\ref{fig:injection_test} shows the GCE spectrum obtained from fitting the actual \textit{Fermi} data, applying the same mask. Upon initial examination, all the injection tests, regardless of whether the GCE template matches or mismatches, show fitted spectra that are consistent with the injected GCE spectra (shown as black lines) up to 10 GeV (except a small discrepancy in scenario (2)). 
Notably, the fluxes in these tests are approximately three times higher than those obtained from the actual \textit{Fermi} data. That should be the case since the \textit{Fermi} data already contains the GCE component before injecting the additional GCE mock component into it. 

We further look into the specifics by comparing the total negative log-likelihoods in scenario (1) vs. scenario (2) and scenario (3) vs. scenario (4). The results are shown in Fig.~\ref{fig:injection_result} in terms of the difference of total negative log-likelihoods, $-2 \Delta\ln \mathcal L$, relative to the best-fit models that include both the galactic diffuse emission backgrounds and the GCE. Our analysis shows that, in general, the cases where the GCE model matches the injected data (scenarios (1) and (4)) result in a better fit compared to their mismatched counterparts (scenarios (2) and (3)) across the majority of the 80 galactic diffuse emission background models fitted. The difference in $-2\Delta \ln \mathcal L$ ranges from approximately $50$ to $100$ for the five best-fit background models when comparing matched with mismatched scenarios. Only a few background models that yield poorer fits show a preference for the mismatched scenarios over the matched ones. 
Our results suggest that the template fitting demonstrates robustness against misinterpreting the ellipticity of the GCE and quantifies the relevant statistical penalty.

\section{The morphology of the GCE}
\label{sec:morphism}

In the subsequent sections, we provide a comprehensive list of tests focusing on the morphology of the GCE. Our investigation covers the properties of the GCE morphology, i.e., its cuspiness and 
ellipticity, across different energies and under different masks as introduced in Sec.~\ref{sec:mask} (see Fig.~\ref{fig:mask_only}).
Additionally, we evaluate and compare various alternative models for the GCE morphology proposed in the literature. 
These models encompass scenarios such as a dark matter annihilation profile, a series of stellar bulge profiles, and a combination of the dark matter annihilation profile and a stellar bulge.  Each model is examined under different masks to determine their respective goodness of fit.

\begin{figure*}
\hspace{-0.8cm}
\includegraphics[width=0.53\textwidth]{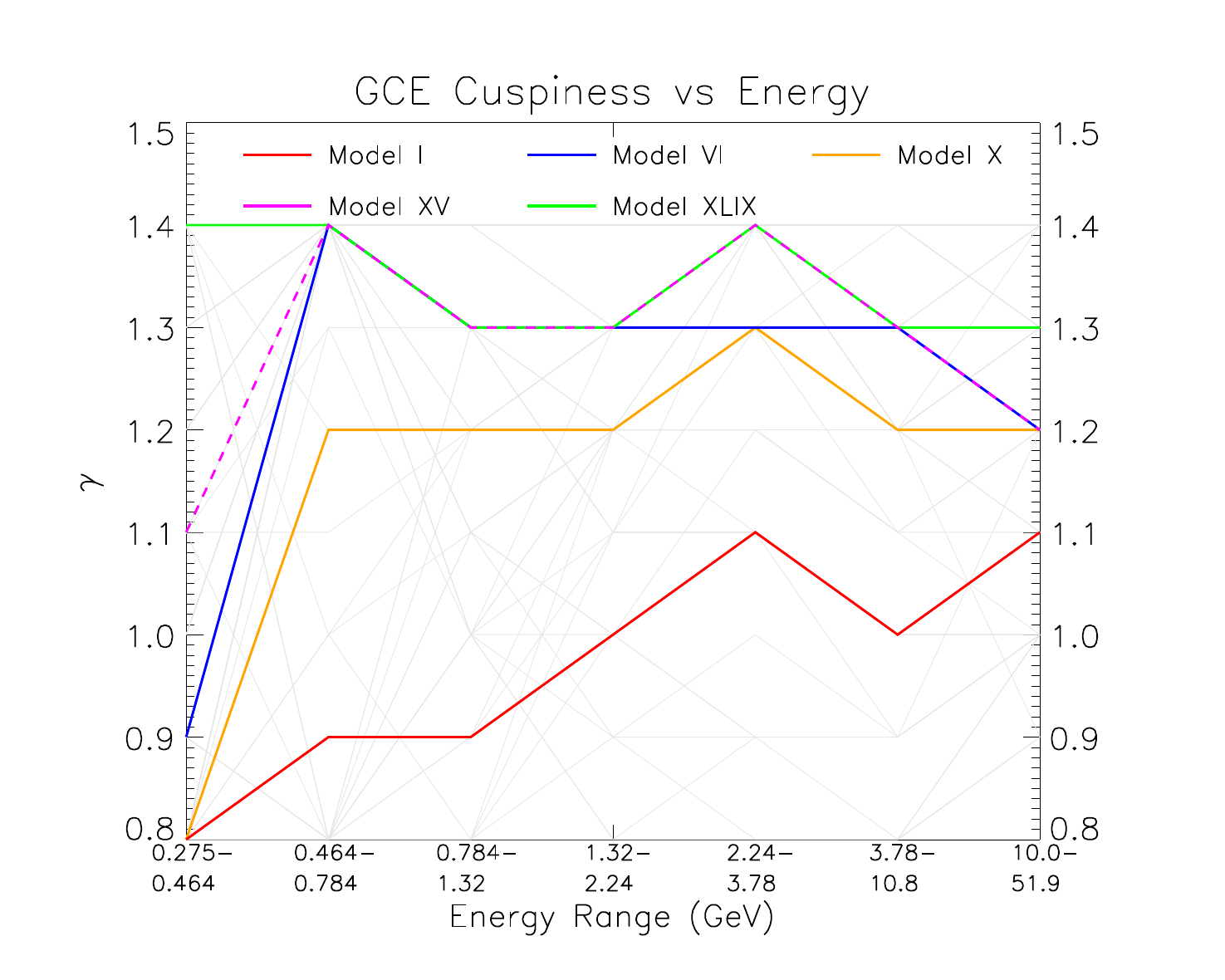}
\hspace{-0.8cm}
\includegraphics[width=0.53\textwidth]{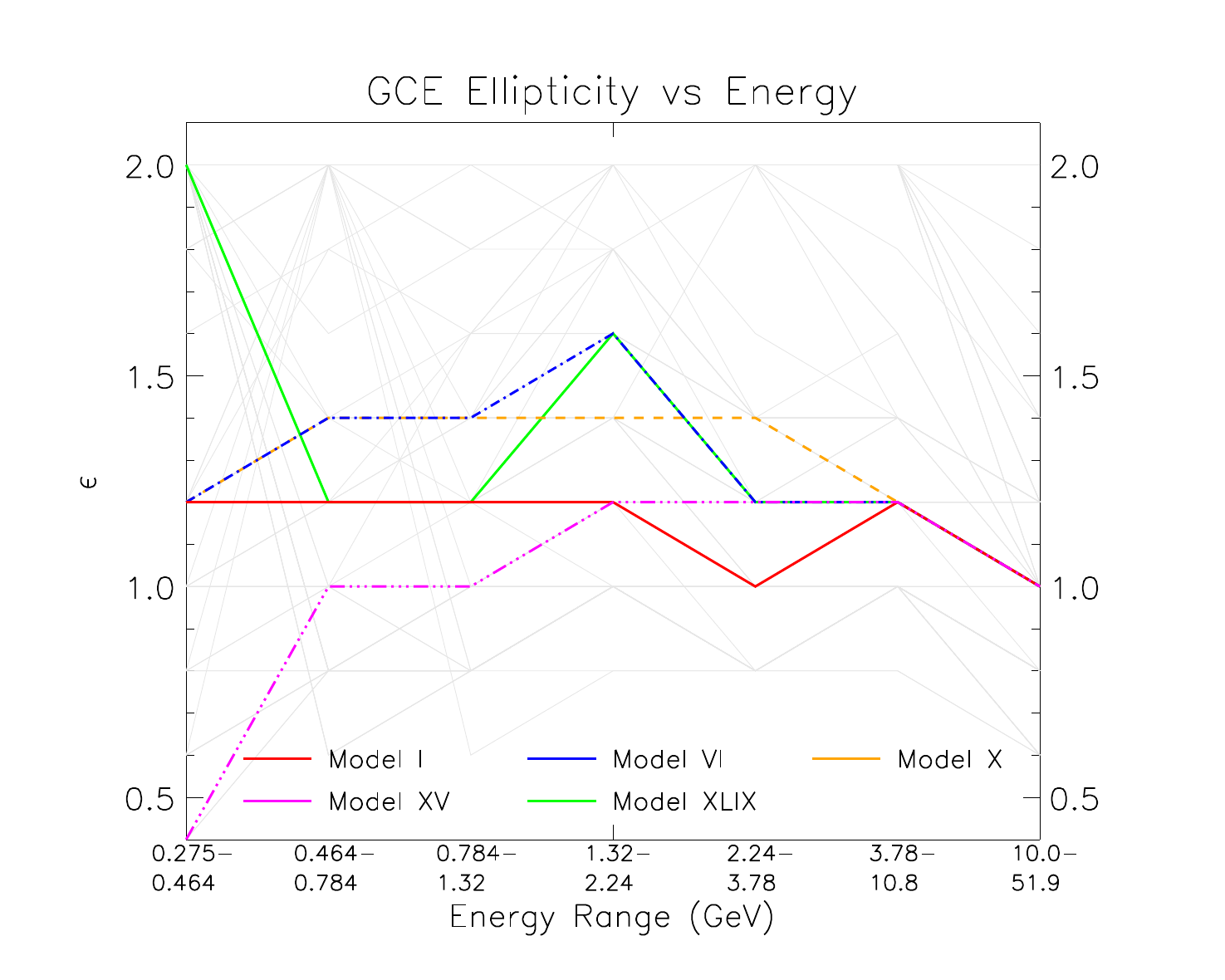}
\caption{The evolution of the fit GCE morphology with energy using the ``Standard 4FGLDR3 + L20'' mask. Left: how the inner slope (cuspiness) parameter $\gamma$ changes with energy. We show the results for all 80 galactic diffuse background models in gray lines, where we show what the best-fit $\gamma$ value was at a given energy range for each model. With the colored lines, we show the same results for some of the best overall fit models for the entire energy range. Right: Similarly, we show how the ellipticity parameter $\epsilon$ changes with energy. At low energies, due to the large masks and astrophysical background uncertainties, the GCE morphology, both in terms of its cuspiness $\gamma$ and ellipticity $\epsilon$, cannot be well constrained. However, with increasing energy as the point sources mask shrinks, the best-fit models converge to $1.1 \leq \gamma \leq 1.3$ and $\epsilon \simeq 1.0$. Some background models that provide a poor fit to the data still allow for a wider range of $\gamma$ and $\epsilon$.}
\label{fig:morphology_vs_energy}
\end{figure*}

\subsection{The dependence of the GCE morphology with energy}
\label{subsec:GCEmorphology_vs_energy}
One key question in determining the origin of the GCE 
is understanding the dependence of its morphology on energy. We expect an energy-independent morphology if the GCE originates purely from dark matter annihilation. This scenario assumes that gamma rays from the underlying 
annihilation channel are prompt emission photons rather than (ICS) photons up-scatted from high-energy electrons and positrons that interact with the interstellar medium as they propagate away from their point of origin. Dark matter particles give a dominant prompt emission component when they annihilate mainly into quarks and gluons. Instead, if the GCE is purely of stellar origin, it would consistently show an elongation along the Galactic disk ($\epsilon > 1$), regardless of energy. Such a morphology is consistent with predictions from stellar bulge profiles, where the gamma-ray emission is predominantly from the outer gap regions of MSPs and thus 
directly tracks the location of the MSPs.
Finally, the \textit{Fermi} 
bubbles might significantly affect the observed GCE emissions,
or even be related to it, being an earlier event of cosmic-ray burst activity from the GC~\cite{Carlson:2014cwa, Petrovic:2014uda, 
Cholis:2015dea}. Given the hard 
spectrum of the \textit{Fermi} bubbles, their contribution to the GCE is expected to increase with 
energy. 
In this scenario, the GCE will exhibit a more pronounced elongation perpendicular to the galactic disk at higher energies, i.e., give ellipticity parameter values of $\epsilon < 1$.

In Fig.~\ref{fig:morphology_vs_energy}, we show how the GCE morphological properties change with energy. The left panel, assuming
the GCE is spherical, shows the evolution of the cuspiness parameter $\gamma$ from Eq.~\ref{eq:DM} with energy. This parameter describes the level of contraction in the NFW profile. We remind the reader that the GCE morphology is proportional to the line-of-sight integral of $\rho(r)^{2}$. 
We use mask ``Standard 4FGLDR3 + L20'' and include results for all 80 background galactic diffuse emission models from 
\cite{Cholis:2021rpp} and highlight in color the results coming from background galactic diffuse emission models that provide the best fit to the \textit{Fermi} data. At low energies, the systematic
uncertainties of the background diffuse emission models prevent a definitive conclusion
about the value of $\gamma$. 
We remind the reader that various alternative
background models can give similar quality fits.
However, for gamma-ray 
energies above 2 GeV, there is a robust preference for the GCE to exhibit 
a cuspiness of $\gamma \geq 1$, akin to a regular 
NFW profile or a slightly adiabatically contracted version. 
Only a minority of background models at high energies favor a cuspiness of $\gamma < 1$, but these models generally have poorer 
overall fit quality for the ROI. In App.~\ref{app:GCE_Morphology_Energy_Evolution_vs_ROI}, we provide the results from the same analysis for alternative ROIs (masks). For the regular masks, our result on the cuspiness parameter being $\gamma \simeq 1.2$ remains valid. 
It should be noted that, in addition to the systematic uncertainties of the galactic diffuse emission backgrounds, accurately tracking the GCE morphology at lower energies is challenging due to the significant fraction of the ROI being removed by point source masks, as detailed in Sec.~\ref{sec:mask} and Fig.~\ref{fig:mpf}.

In the right panel of Fig.~\ref{fig:morphology_vs_energy}, again using ``Standard 4FGLDR3 + L20'' mask, we show how 
the ellipticity parameter $\epsilon$ of Eq.~\ref{eq:ellipticity} changes with energy. This parameter 
indicates the sphericity of the GCE. 
Again, we show results from all 80 background diffuse emission models of Ref.~\cite{Cholis:2021rpp} 
and highlight in color those from the best-fit background models.
Similar to the cuspiness analysis, at low energies, the combination of large 
astrophysical uncertainties of backgrounds and the small fraction of the unmasked ROI
prevents us from conclusively determining a preference for $\epsilon$.
Nonetheless, at gamma-ray energies $\gtrsim$ 3 GeV, the background models
that provide a good fit to the data consistently suggest that the GCE is \emph{approximately} spherical. In App.~\ref{app:GCE_Morphology_Energy_Evolution_vs_ROI}, we show results on the ellipticity parameter for alternative masks. An ellipticity $\epsilon \simeq 1.0-1.4$ for energies $\gtrsim$ 3 GeV  is still the case for the regular masks. The mild derivation from a perfect spherical shape ($\epsilon=1.0$) could be due to the baryonic contraction of the dark matter halo~\cite{Grand:2022olu}.

\subsection{The cuspiness and ellipticity of the GCE for alternative masks}

In this section, we test the general morphological properties of the GCE
in terms of its cuspiness and ellipticity under various mask choices.

\begin{figure*}
\centering
\includegraphics[width=0.32\textwidth]{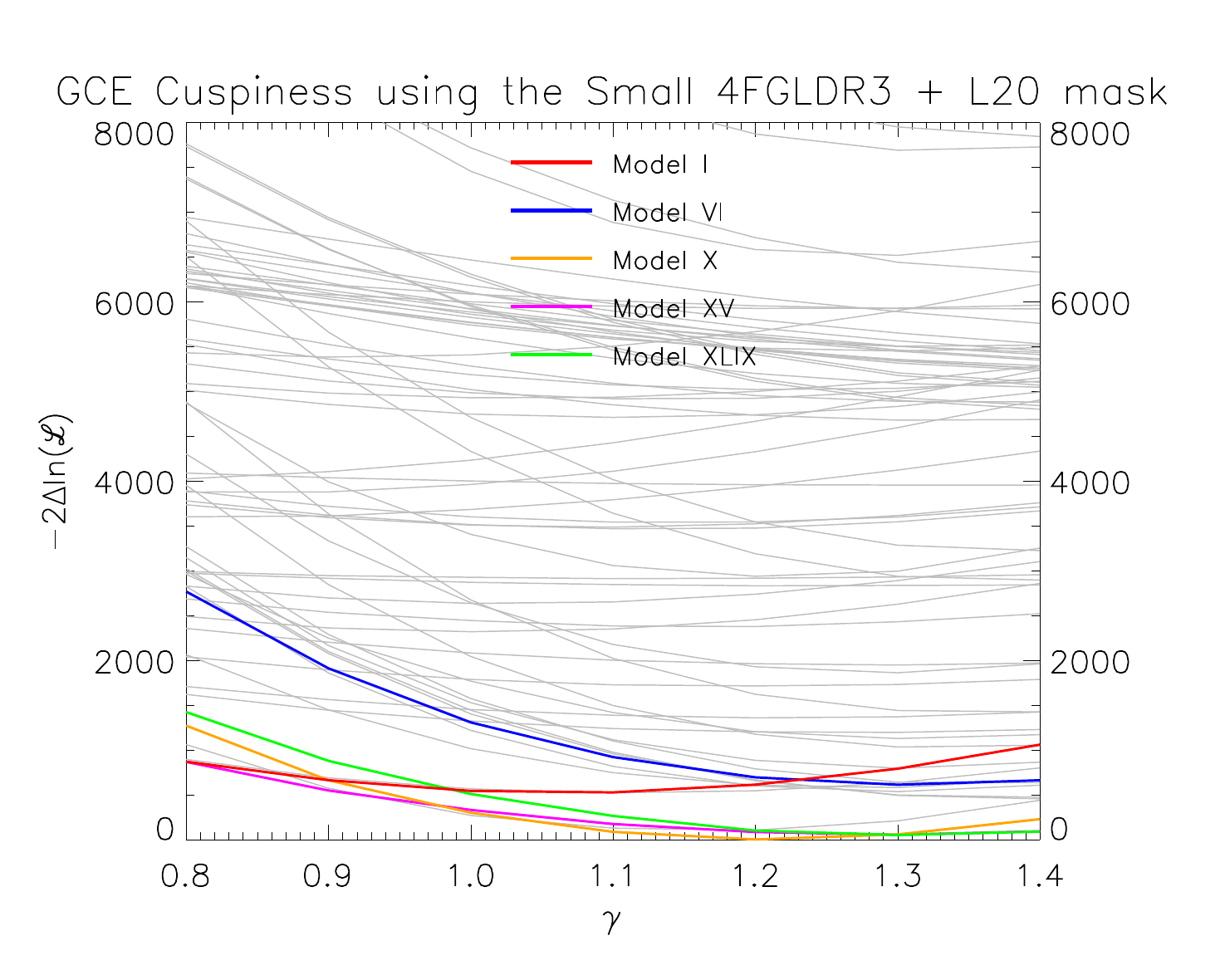}~
\includegraphics[width=0.32\textwidth]{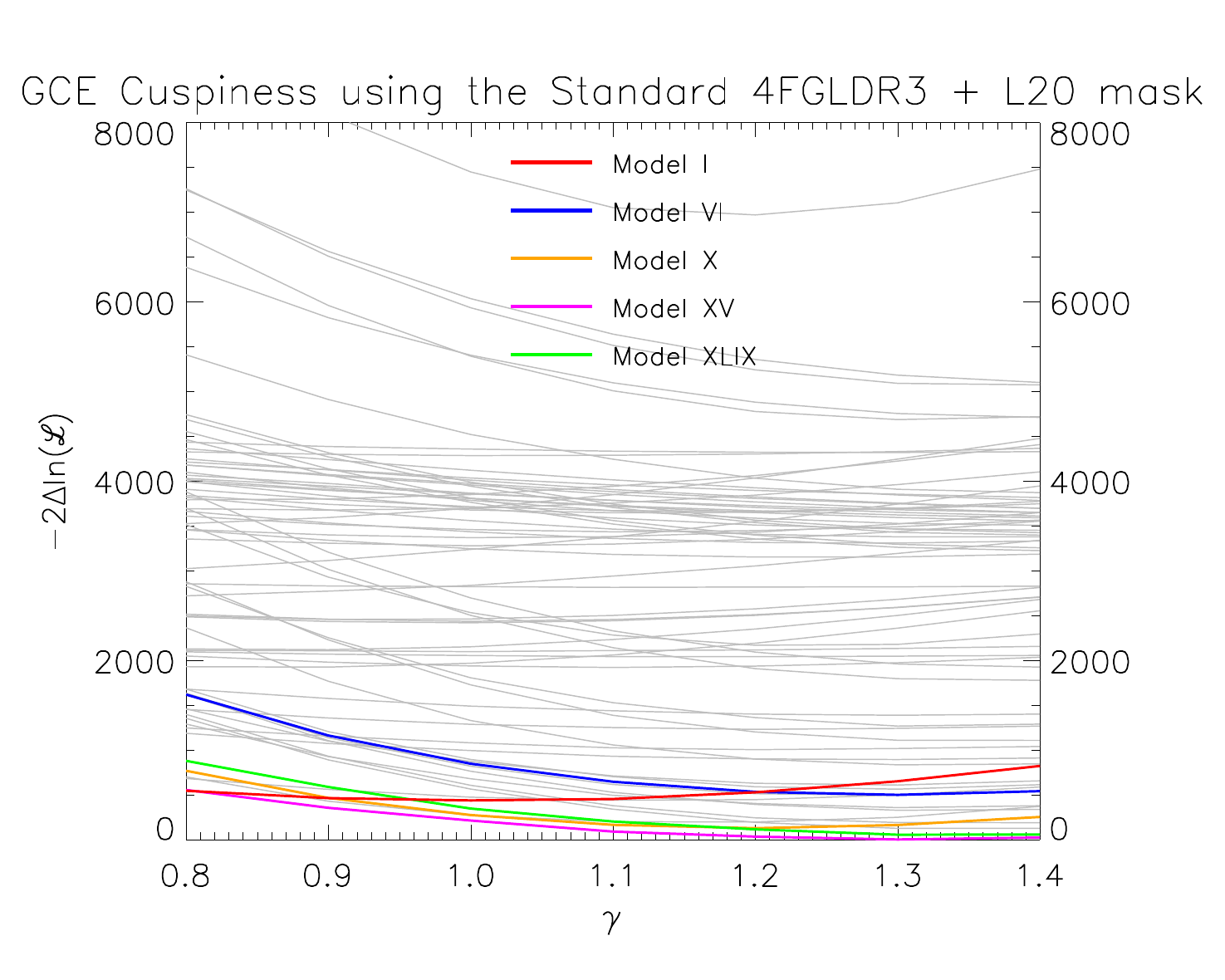}
\includegraphics[width=0.32\textwidth]{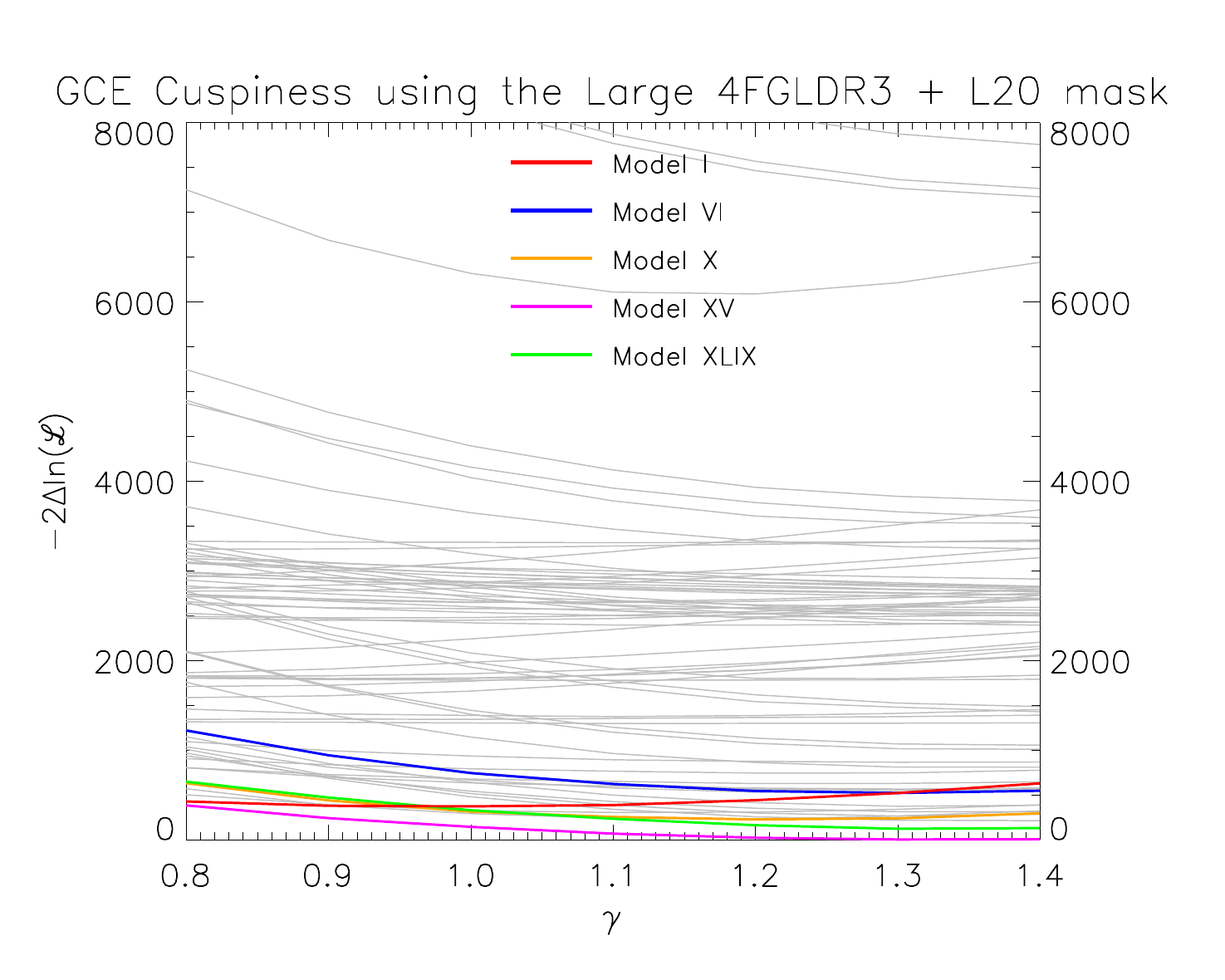}\\
\includegraphics[width=0.32\textwidth]{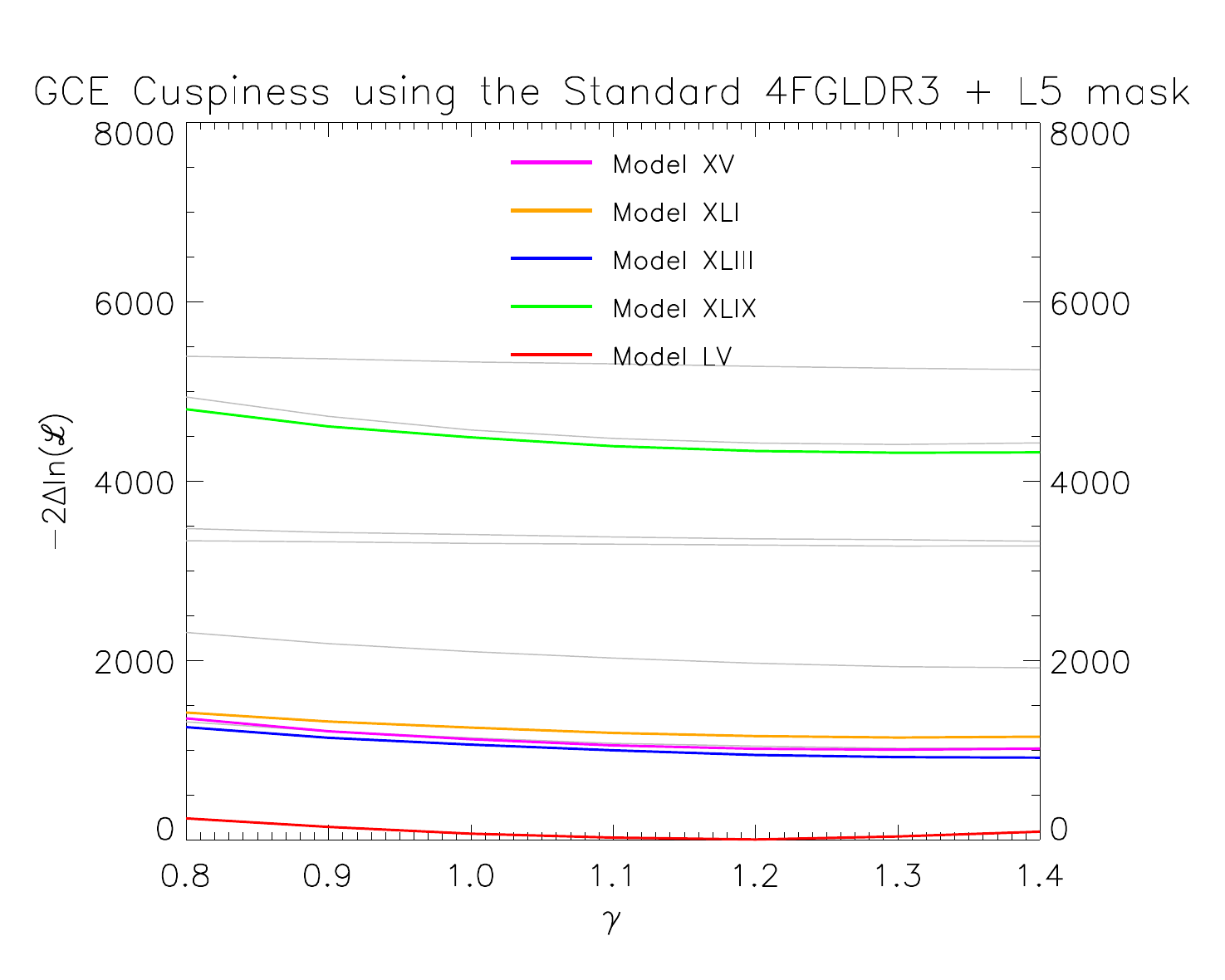}
\includegraphics[width=0.32\textwidth]{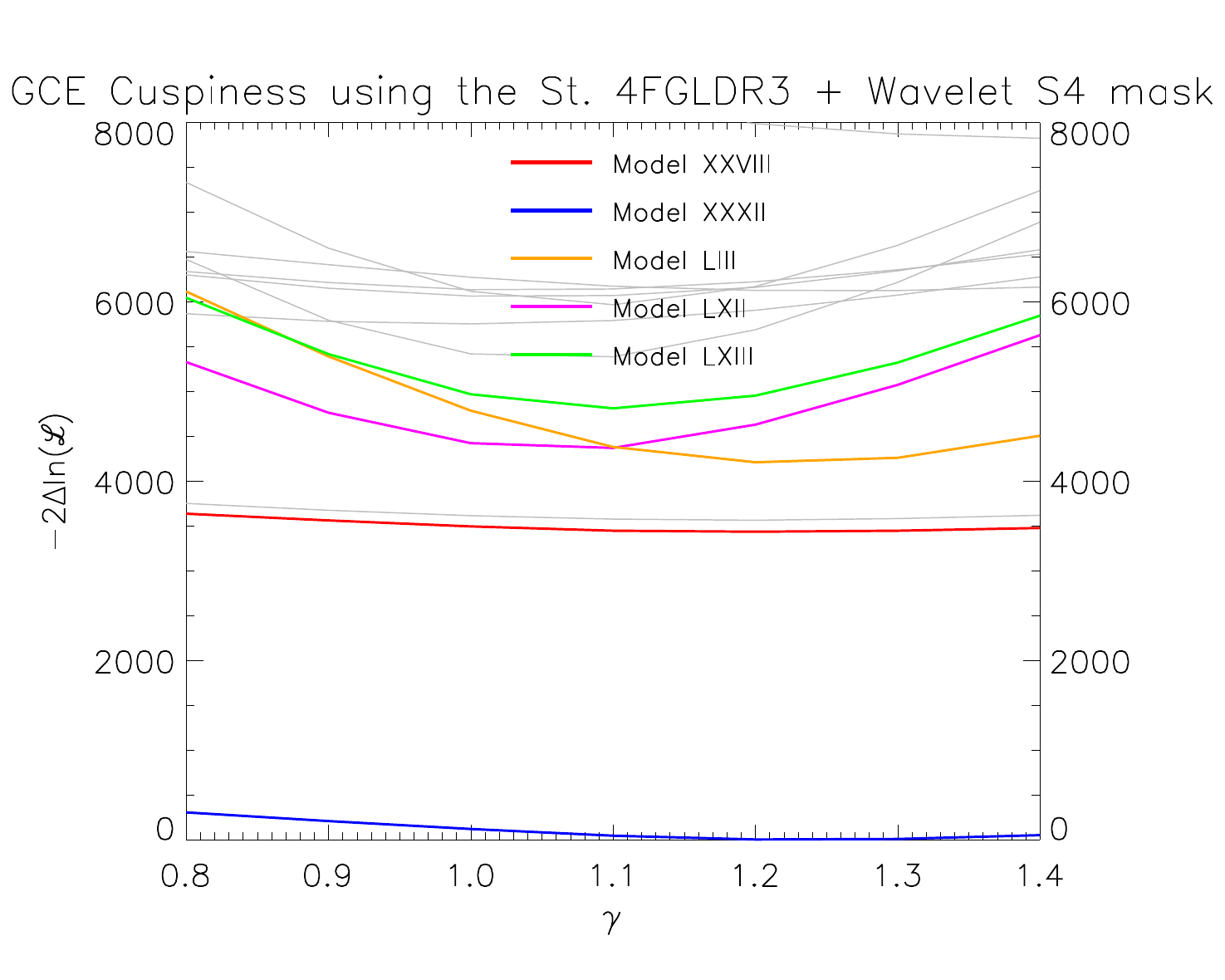}
\includegraphics[width=0.32\textwidth]{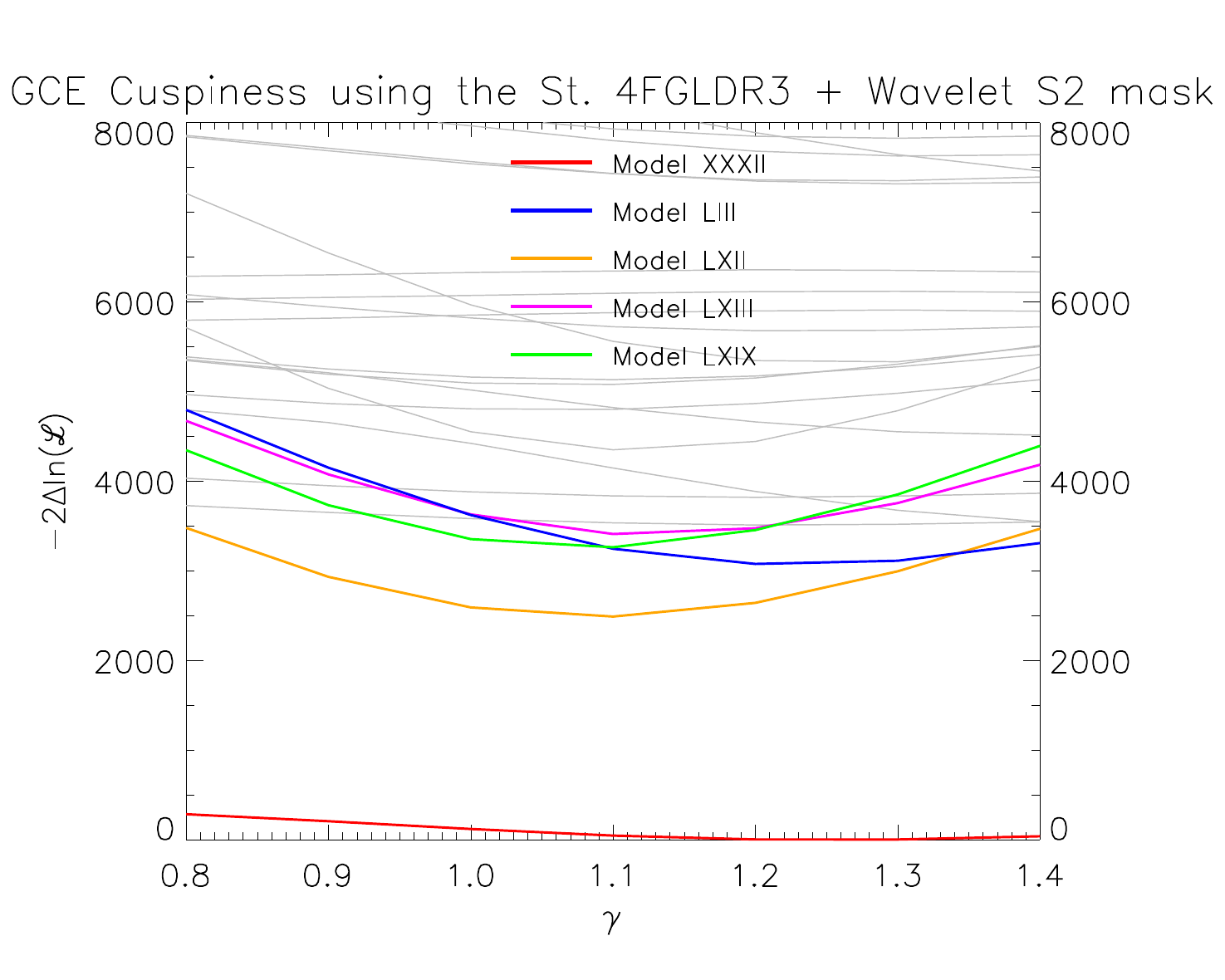}
\caption{The GCE cuspiness assuming the emission comes from annihilating 
dark matter. Log-likelihoods are evaluated by summing all energy bins. The 
dark matter profile is described by the $\gamma$ parameter of 
Eq.~\ref{eq:DM}. We show results for alternative mask choices. 
From the top left panel to the bottom right, we use masks, 
``Small 4FGLDR3 + L20'', ``Standard 4FGLDR3 + L20'', ``Large 4FGLDR3 + L20'', 
``Standard 4FGLDR3 + L5'', ``Standard 4FGLDR3 + Wavelet S4'' and ``Standard 
4FGLDR3 + Wavelet S2''. The alternative masks (ROIs) may affect the ordering between 
the diffuse emission background models that provide the best fit and the difference in $-2 
\Delta \ln \mathcal{L}$, but the common pattern is a 
systematic preference for $1.0 \leq \gamma \leq 1.4$.}
\label{fig:cuspiness_vs_mask}
\end{figure*}

Fig.~\ref{fig:cuspiness_vs_mask} examines the cuspiness $\gamma$ of the GCE.
We show the quality of fit by comparing differences in total negative log-likelihood, $-2 \Delta \ln \mathcal{L}$. This statistic reflects how well the 
combination of an astrophysical background and GCE model agrees with the 
\textit{Fermi} data for each of the six masks presented. To encompass the
 astrophysical uncertainties of backgrounds, we analyze all 80
astrophysical galactic diffuse emission models provided in Ref.~\cite{Cholis:2021rpp}. In our plots of 
$-2 \Delta \ln \mathcal{L}$, we set baseline (zero point) as the combination of background and GCE model that yields the best overall fit once summing over all energy bins. 
Lines closer to zero indicate a higher quality fit.
Since we plot the $-2 \Delta \ln \mathcal{L}$ values up to 8000, in some of the panels
(especially the lower ones of Fig.~\ref{fig:cuspiness_vs_mask}), only a subset 
of lines is visible. The absent lines are models with poorer fits. 
For each panel, we  color the combination of background and GCE model
that provides the best fit, along with four more models that provide comparatively similar quality fits. 

Direct comparison of log-likelihoods across different masks is not feasible, as the number of pixels in the remaining ROIs where fitting occurs varies. Also, the exact ranking of which combination of galactic diffuse emission background models and GCE provides the best-fit changes slightly with different ROIs (masks).
This variation is expected, mainly because regions dense in interstellar medium gas may be either revealed or masked by altering the masks, and these regions are typically associated with high gamma-ray fluxes. Nonetheless, our objective is to identify the properties of the GCE 
morphology that are robust regardless of these varying assumptions. We achieve this by searching for recurring patterns across the combination of different choices in the diffuse emission backgrounds and masks. 

In the case of Fig.~\ref{fig:cuspiness_vs_mask}, a clear pattern emerges: 
irrespective of the background model and the mask used, the GCE consistently 
shows a preference for $1.0 \leq \gamma \leq 1.4$. 
This range corresponds to the morphology of a dark matter annihilation signal, following a spatial distribution akin to a regular 
NFW profile ($\gamma =1)$ or a more contracted profile that scales as $\sim 1/r^{\gamma}$ at the inner galaxy.

\begin{figure*}[h]
\centering
\includegraphics[width=0.32\textwidth]{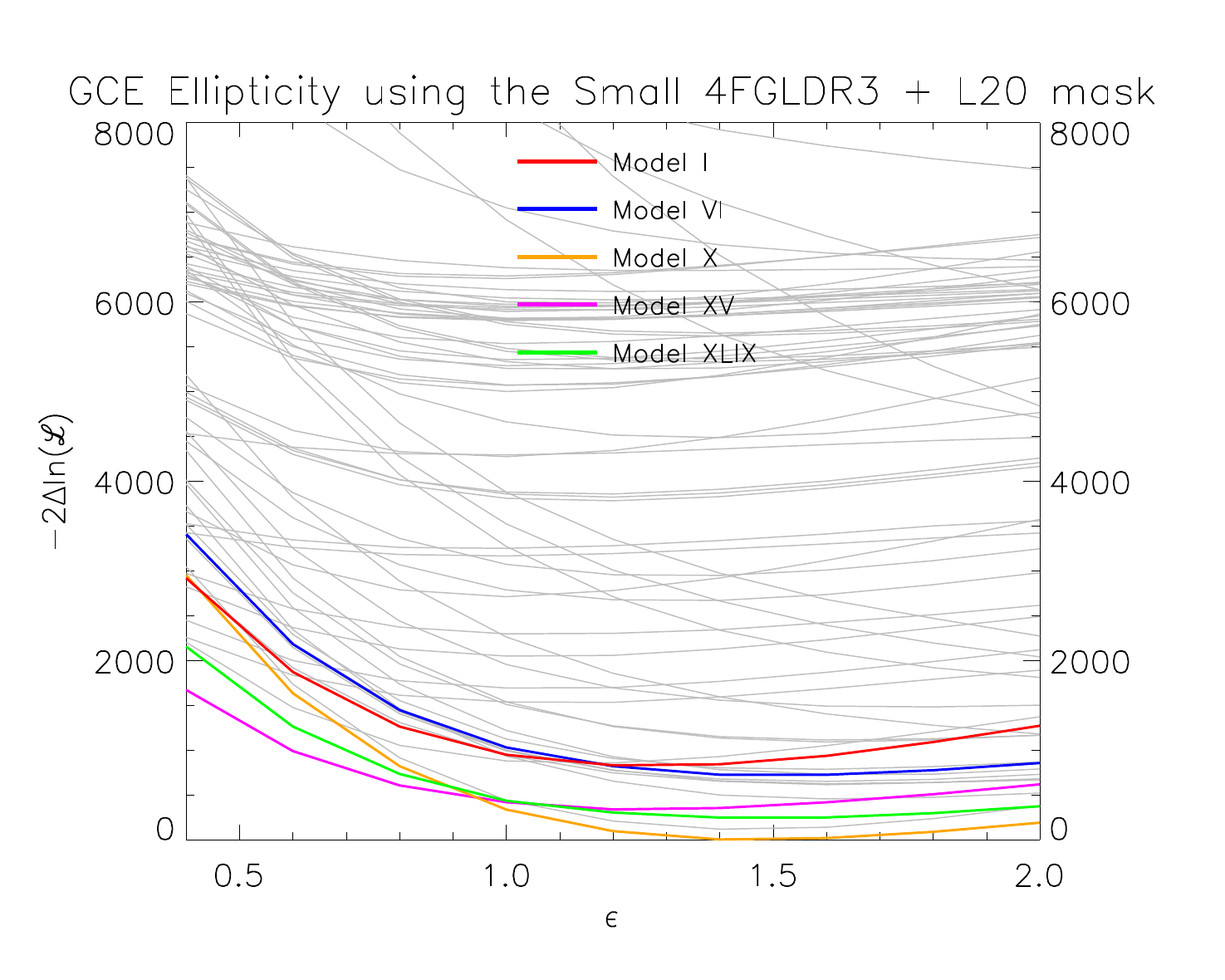}~
\includegraphics[width=0.32\textwidth]{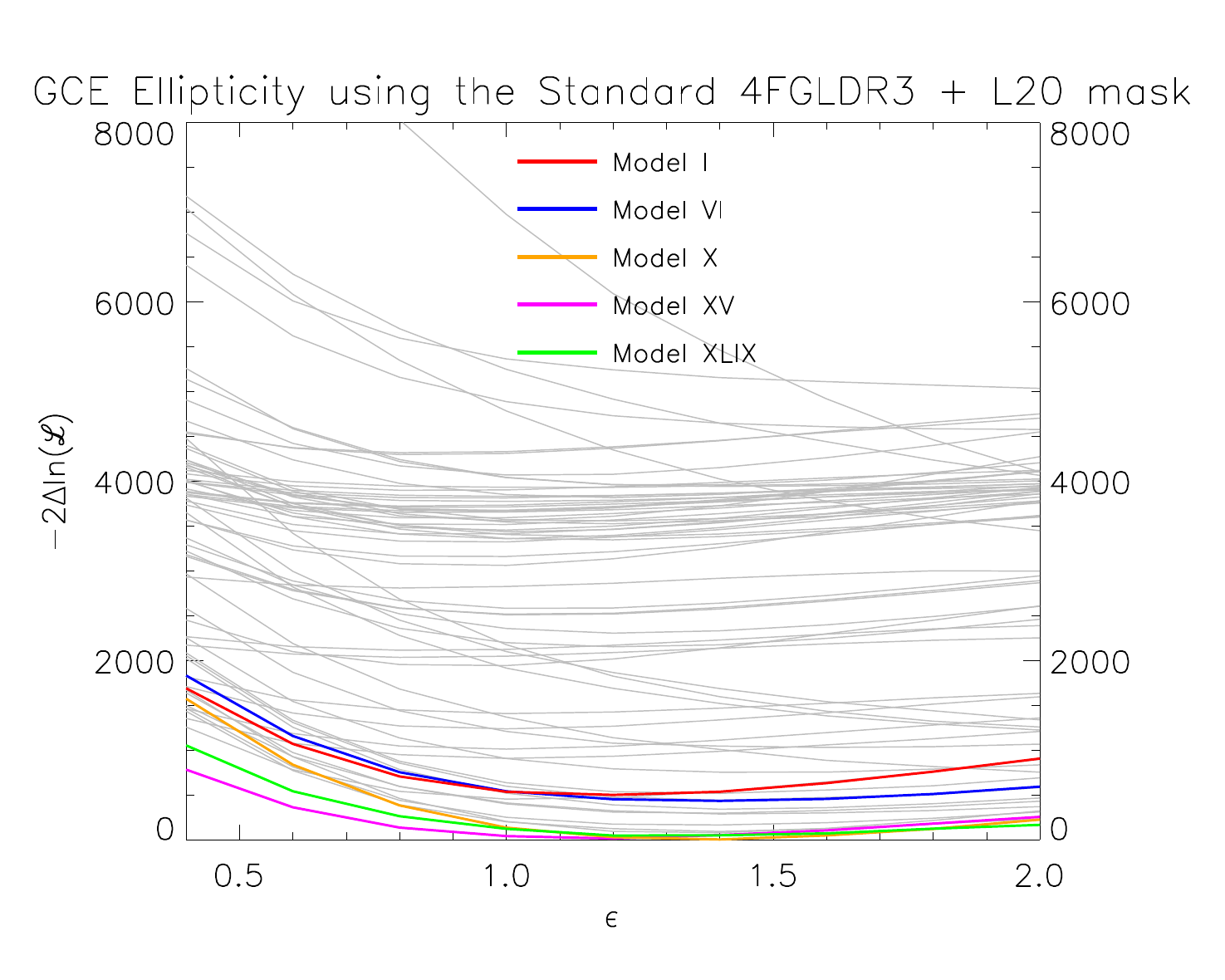}
\includegraphics[width=0.32\textwidth]{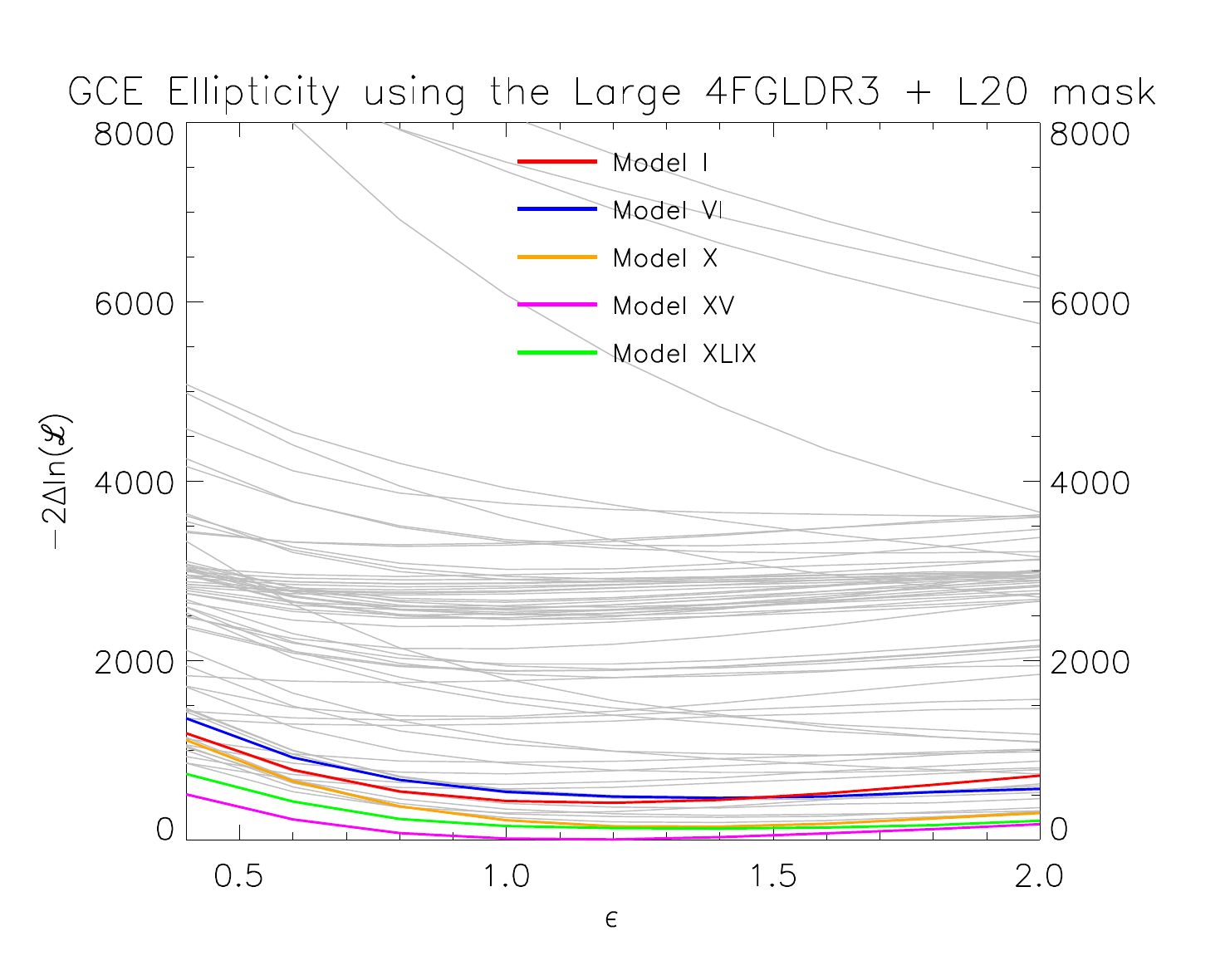}\\
\includegraphics[width=0.32\textwidth]{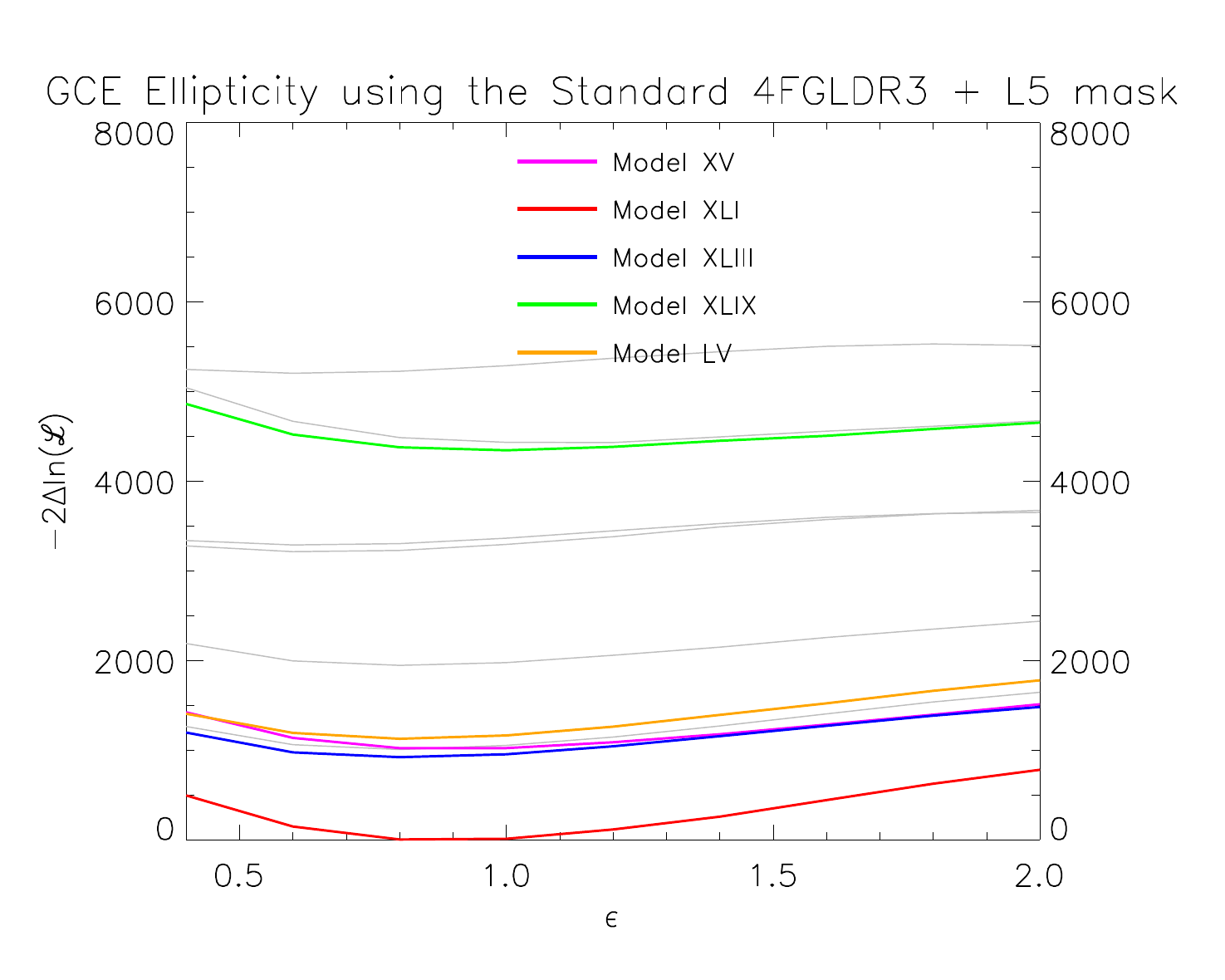}
\includegraphics[width=0.32\textwidth]{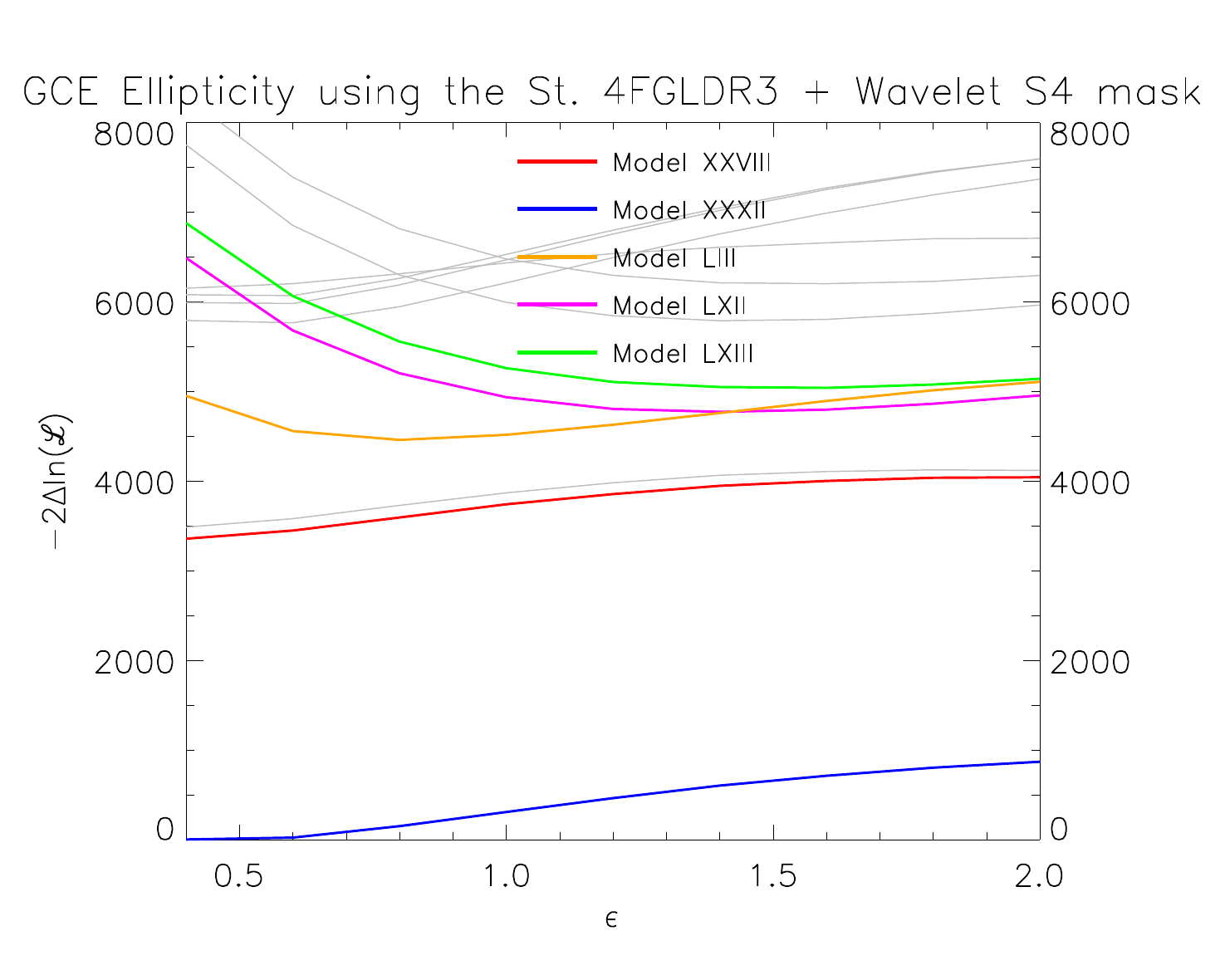}
\includegraphics[width=0.32\textwidth]{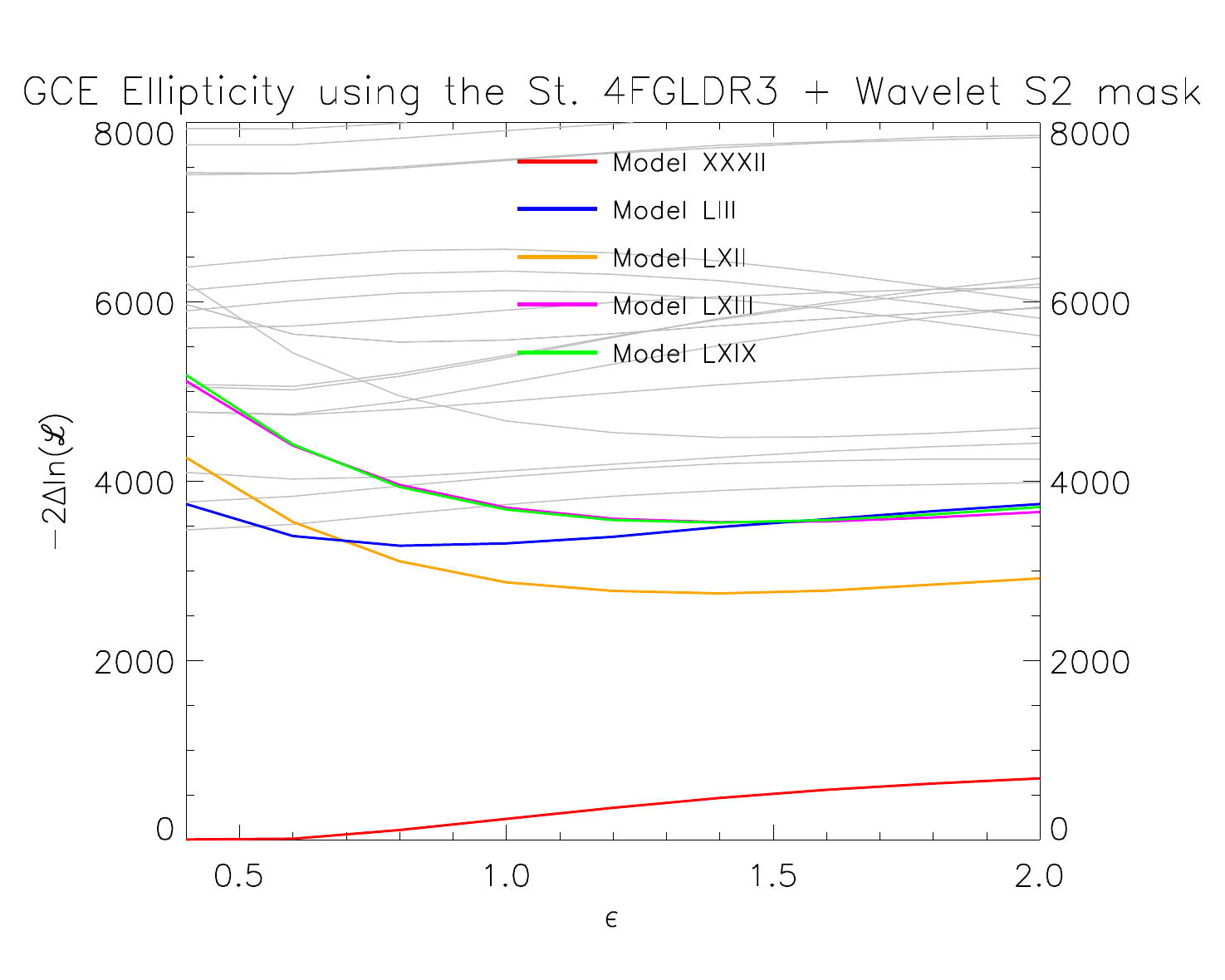}
\caption{As in Fig.~\ref{fig:cuspiness_vs_mask}, we show the GCE  
ellipticity for alternative choices of masks. Log-likelihoods are again 
evaluated by summing all energy bins. From the masks (ROIs) shown in the top panels, 
the backgrounds giving the best fit prefer $0.9 \leq \epsilon \leq 1.5$. 
Using instead as a mask the ``Standard 4FGLDR3 + L5'', we get $0.7 \leq 
\epsilon \leq 1.2$. The two wavelet-based masks of ``Standard 4FGLDR3 + Wavelet S4'' and ``Standard 4FGLDR3 + Wavelet S2'' give for a small number of 
models, but also the ones providing the best fit, a preference for  $0.4 \leq 
\epsilon \leq 0.8$.}
\label{fig:ellipticity_vs_mask}
\end{figure*}

These findings are not affected by using any of 
the remaining four masks (ROIs) of Fig.~\ref{fig:mask_only}: ``Standard 4FGLDR1 + L20'', 
``Standard 4FGLDR2 + L20'', ``Standard 4FGLDR3 + L8'', ``Standard 4FGLDR3 + Wavelet S3''. 
The ``Standard 4FGLDR1 + L20'' and  ``Standard 4FGLDR2 + L20'' masks were tested in 
Ref.~\cite{Cholis:2021rpp}, yielding the same conclusions as in this work. The ``Standard 4FGLDR3 + L8'' and  ``Standard 4FGLDR3 + Wavelet S3'' masks represent intermediate 
masking assumptions compared to those plotted here.

Fig.~\ref{fig:ellipticity_vs_mask} focuses on examining the ellipticity $\epsilon$ of the GCE.
Again, we show the quality of fit in terms of $-2 \Delta \ln \mathcal{L}$ for the
combined astrophysical background and GCE models against the 
\textit{Fermi} data, applying the same six masks as in Fig.~\ref{fig:cuspiness_vs_mask}. 
Lines closer to zero in $-2 \Delta \ln \mathcal{L}$ represent better quality fits. In each case, we color the combinations of background and GCE models that provide 
the best fit, along with four additional background models yielding similar
agreement to the observations.

The ellipticity of GCE appears to be more sensitive to the combination 
of assumptions used, compared to its cuspiness.
Under the masks shown in the top panels (``Small 4FGLDR3 + L20'', ``Standard 4FGLDR3 + L20'', 
and ``Large 4FGLDR3 + L20''), the diffuse emission background models that give the best fit prefer $0.9 \leq \epsilon \leq 1.5$. 
The ``Standard 4FGLDR3 + L5'' mask yields a similar range of $0.7 \leq \epsilon \leq 1.2$. 
However, wavelet-based masks, namely ``Standard 4FGLDR3 + Wavelet S4'' and ``Standard 4FGLDR3 + Wavelet S2'', give a range of $0.4 \leq \epsilon \leq 0.8$ for a few models that provide 
a better fit. Such values of $\epsilon$ suggest that the GCE is elongated perpendicular to the galactic disk, which would indicate the GCE comes from a burst of cosmic rays similar or the same as the one given in the \textit{Fermi} bubbles. Such an 
ellipticity may also due to a triaxial dark matter profile \cite{Diemand:2008in, Kuhlen:2008qj, 2008MNRAS.391.1685S, Dobler:2011mk}. However, this ellipticity is unlikely to result from a 
population of gamma-ray sources tracing the dense stellar regions of the Milky Way. Additionally, 
it could also be due to a non-trivial selection effect in which pixels are identified by the 
wavelet method as potential point sources.

\subsection{Comparing dark matter annihilation and stellar bulge profiles}
\label{subsec:DM_vs_Bulges}

\begin{figure*}
    \centering
    \includegraphics[width=\textwidth]{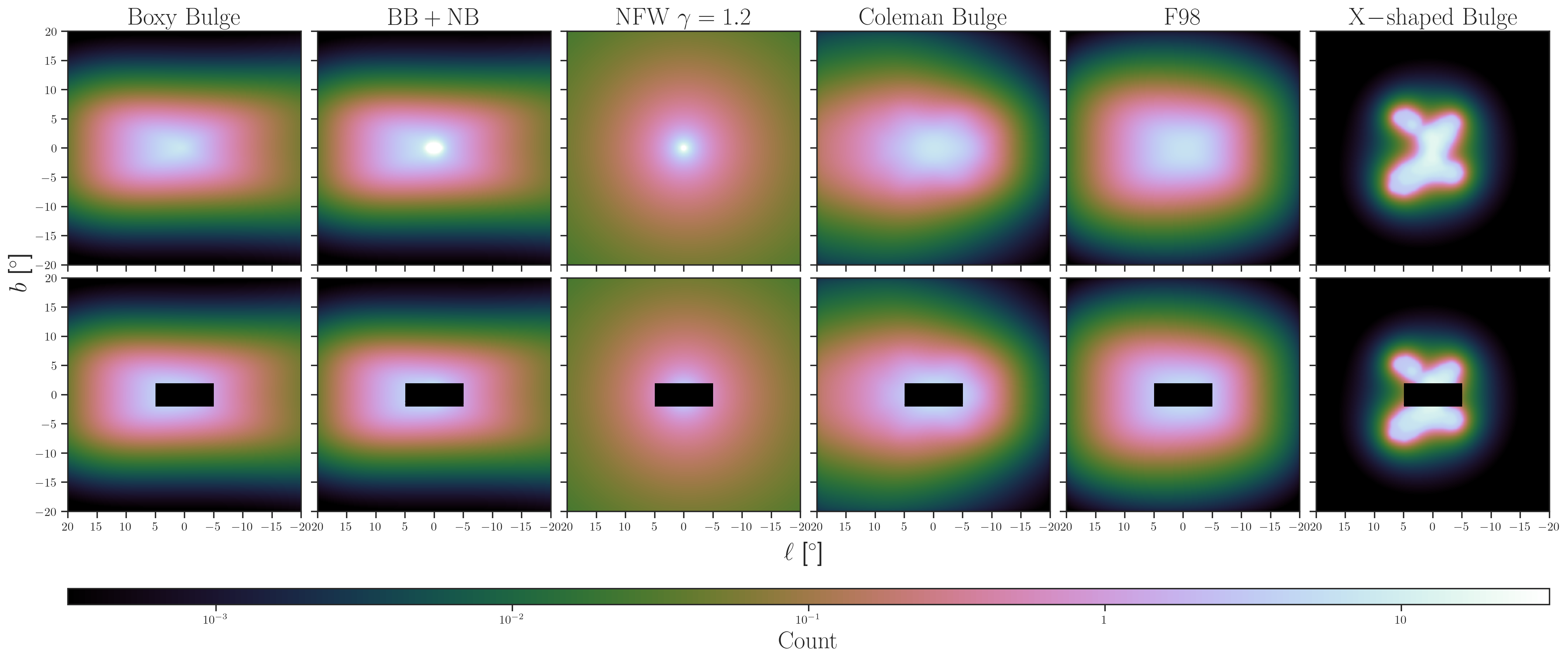}
    \caption{The ``Boxy Bulge'' (1st col.), ``BB+NB'' (2nd col.), ``Coleman Bulge'' (4th col.), ``F98'' (5th col.), and ``X-shaped Bulge'' (6th col.) profiles used in this work, together with the ``NFW $\gamma=1.2$'' profile (3rd col.), for energy bin 5 ($1.02\,\text{GeV}< E<1.32\,\text{GeV}$). The top row shows the profiles in the full $40^\circ\times 40^\circ$ ROI. The bottom row shows the same profiles with the ``L5'' disk mask, i.e., masking the region with $|\ell|<5^\circ$ and $|b|<2^\circ$.
    All the profiles are smeared by the \textit{Fermi} PSF. The color represents the photon counts of the fiducial profiles used in the template fitting.}
    \label{fig:stellar_bulges}
\end{figure*}

In this section, we compare the hypothesis of dark matter annihilation being responsible for the GCE with the alternative theory of unresolved gamma-ray point sources, such as MSPs, being the origin. The latter hypothesis suggests that the GCE, to 
some extent, follows one of the known stellar populations, such as the galactic bulge or the 
nuclear stellar cluster (nuclear bulge) at the center of the Milky Way. For the dark matter annihilation hypothesis, we adopt 
the NFW profile with $\gamma=1.2$ and $\epsilon=1.0$ (referred to as ``NFW $\gamma=1.2$''), as this choice typically provides the best fit to the gamma-ray data. We compare this with a sequence of alternative morphological models proposed in the literature for the GCE.

We compare the dark matter annihilation template and the following stellar bulge profiles:
\begin{enumerate}[(i)]
    \item The ``Boxy Bulge'' profile, based on the stellar bulge measured at the infrared \cite{2002A&A...384..112L}.
    \item A combination of the Boxy Bulge and the Nuclear Bulge ``BB + NB''. The Nuclear Bulge is measured at radio waves \cite{2002A&A...384..112L}. This combination has been proposed as the correct morphology for the GCE in Ref.~\cite{Bartels:2017vsx}.
    \item The ``Coleman Bulge'' from Ref.~\cite{Coleman:2019kax}, which has been recently suggested as an alternative for the GCE morphology \cite{OscarManojDehengPrivateComm}.
    \item The Freudenreich profile for the galactic bar, ``F98''. It is measured at the infrared~\cite{Freudenreich:1997bx}. This profile has been tested for the GCE morphology in Ref.~\cite{Macias:2016nev} and recently suggested as a possible alternative for the GCE morphology \cite{OscarManojDehengPrivateComm}.
    \item The ``X-shaped Bulge'', observed in lower energy gamma rays and suggested as having the correct morphology for the GCE in Refs.~\cite{Macias:2016nev, Macias:2019omb}.
\end{enumerate}

We showcase the morphology of the stellar bulge profiles, together with the ``NFW $\gamma=1.2$'' profile, in Fig.~ \ref{fig:stellar_bulges}. The top row shows the photon count maps, smeared by the \emph{Fermi} PSF, in the $1.02- 1.32\,\text{GeV}$ energy bin (energy bin 5) of the fiducial templates. The bottom row shows these same spatial maps after applying the ``L5'' disk mask, which covers the brightest regions across all the profiles and is the largest common masking area used in the regular masks. With the ``L5'' disk mask applied, the ``Coleman Bulge'' profile bears the greatest similarity to the spherical ``NFW $\gamma = 1.2$'' among all the stellar bugle profiles.

\begin{figure*}
\centering
\includegraphics[width=0.32\textwidth]{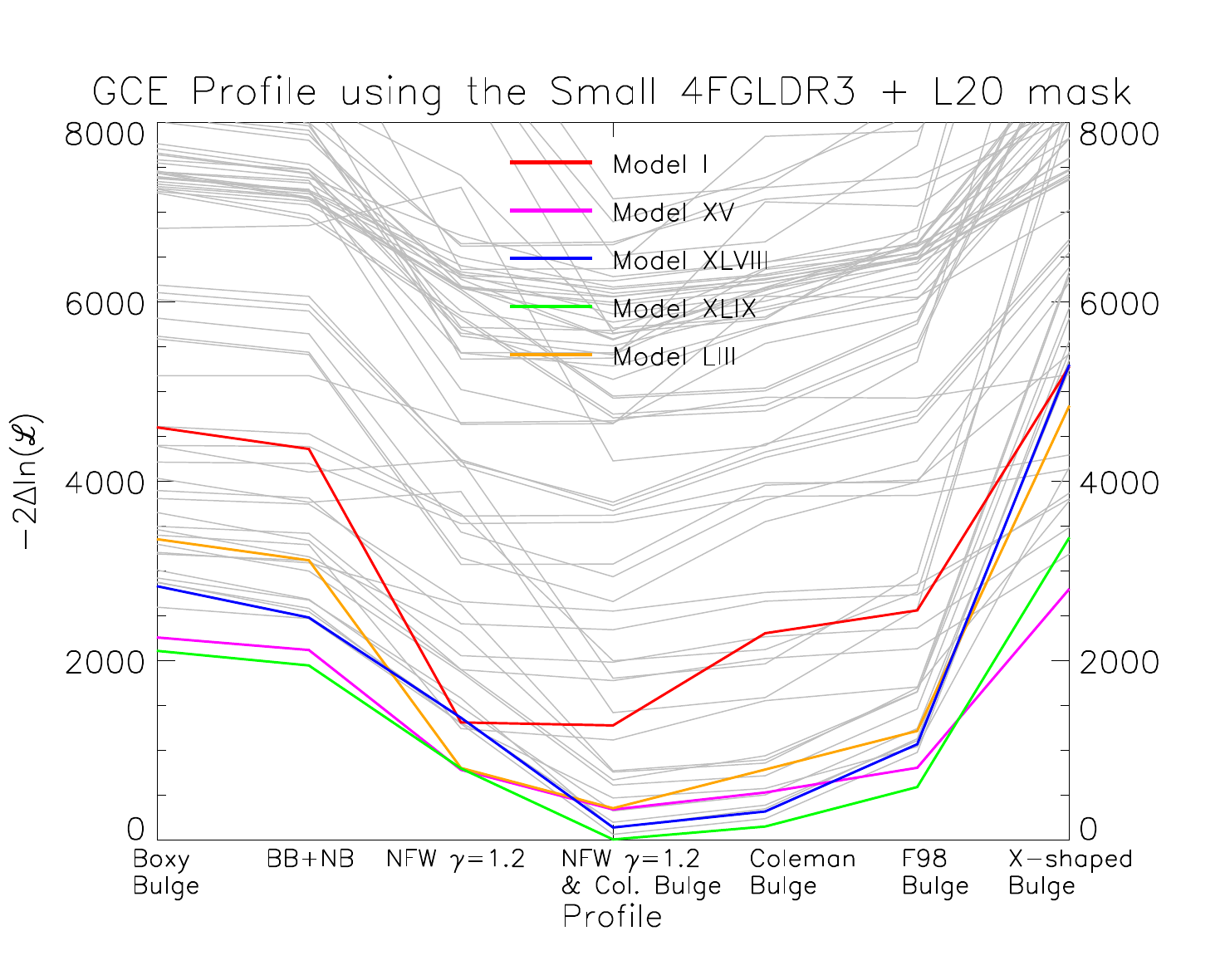}~
\includegraphics[width=0.32\textwidth]{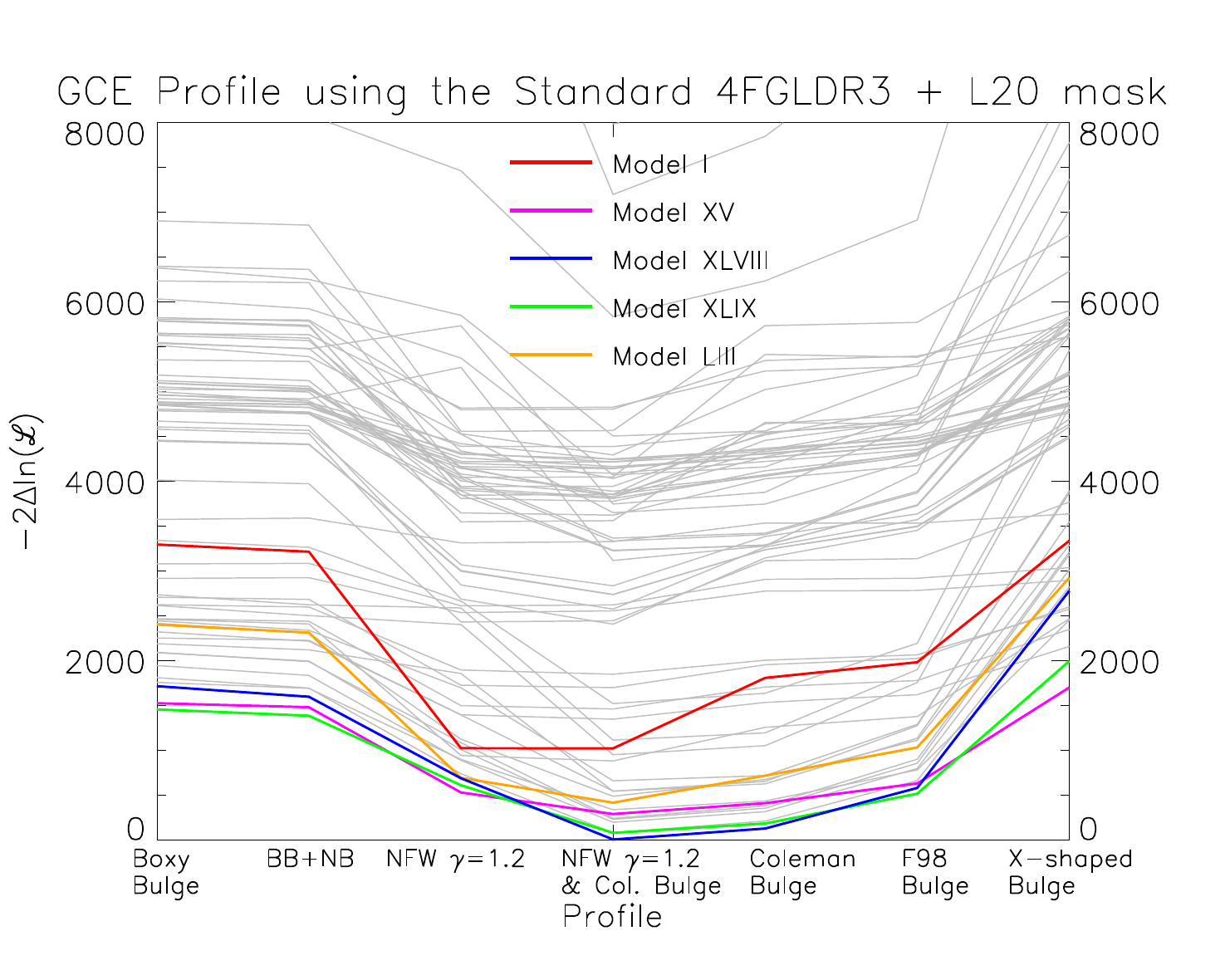}
\includegraphics[width=0.32\textwidth]{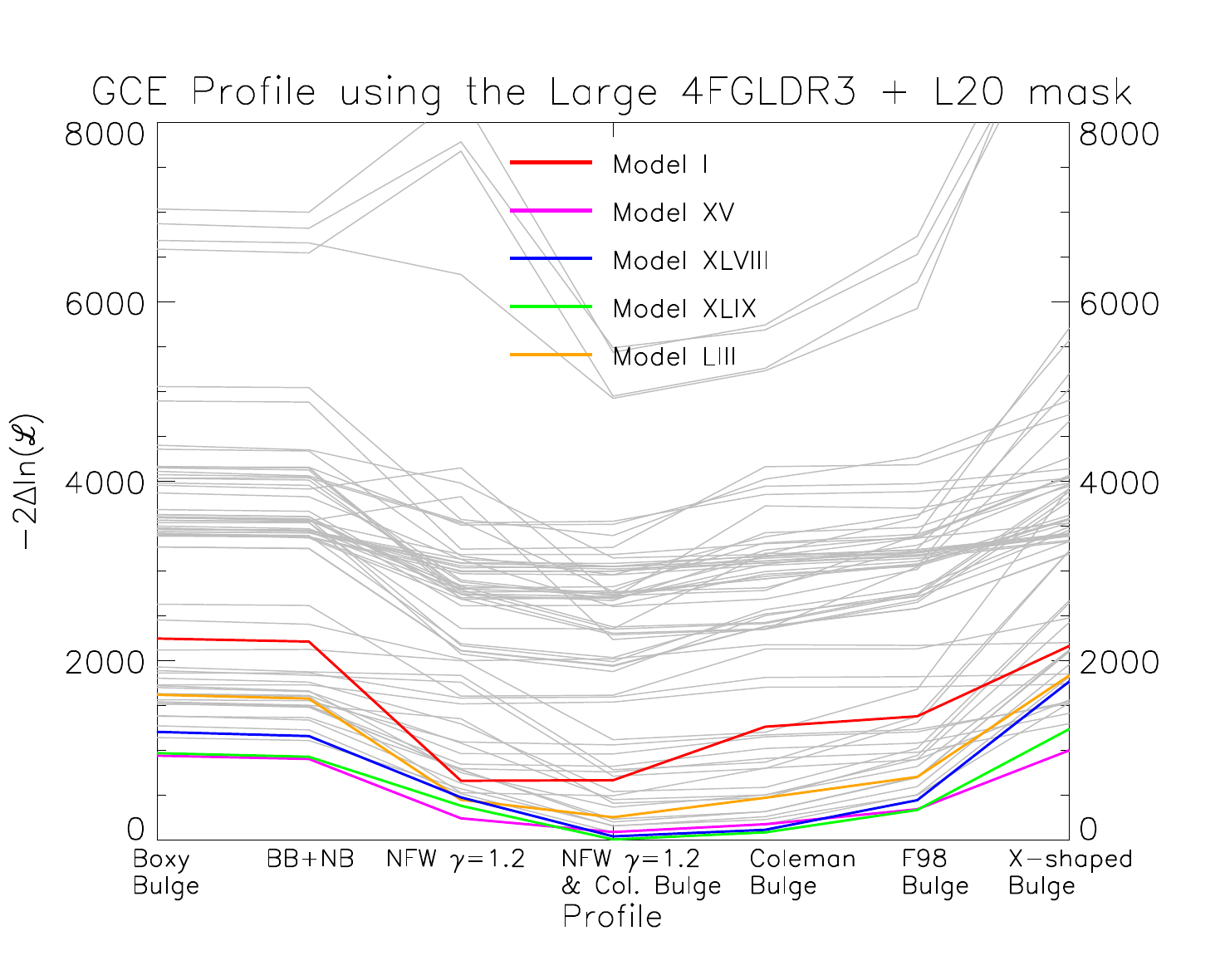}\\
\includegraphics[width=0.32\textwidth]{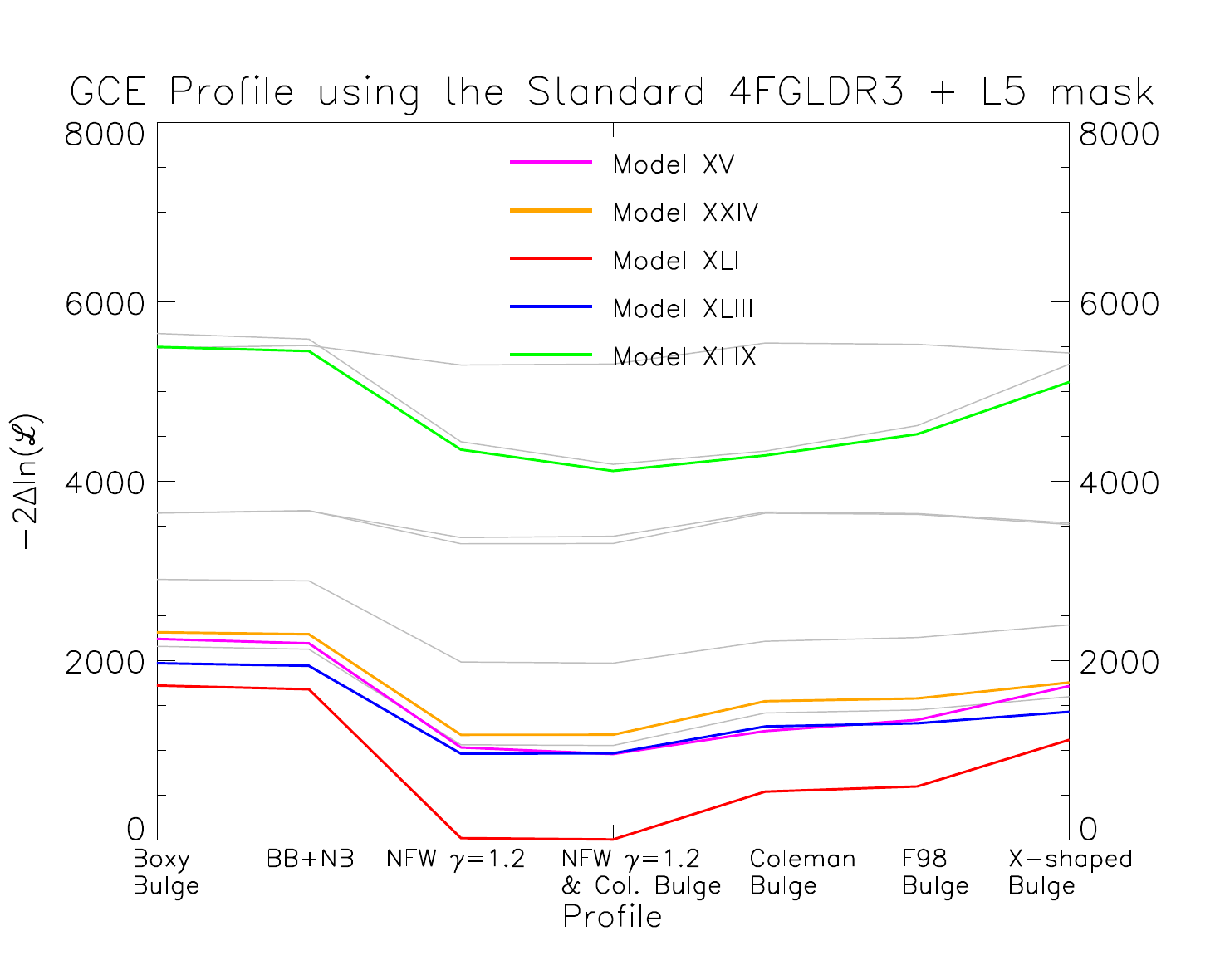}
\includegraphics[width=0.32\textwidth]{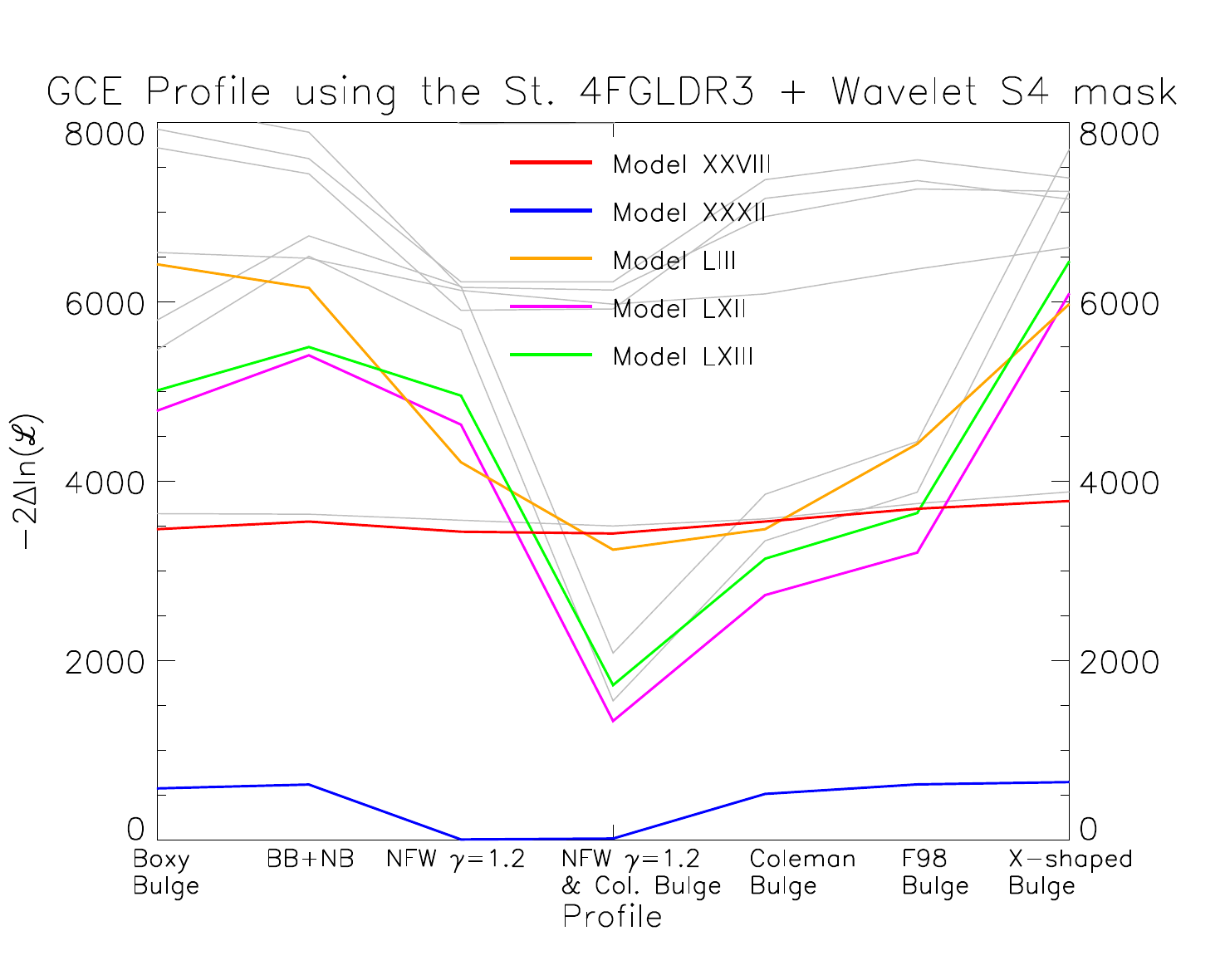}
\includegraphics[width=0.32\textwidth]{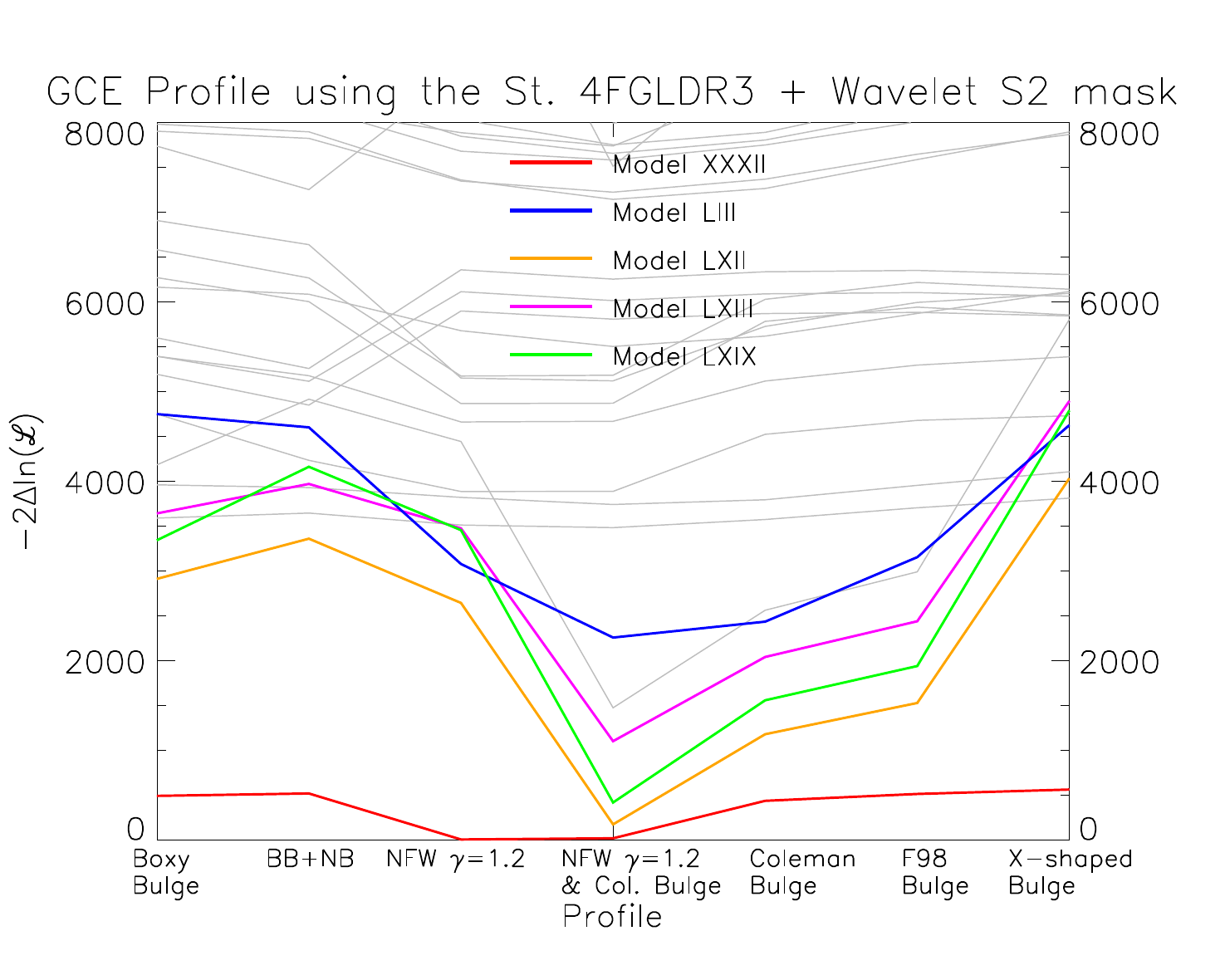}
\caption{Similarly to Figs.~\ref{fig:cuspiness_vs_mask} 
and~\ref{fig:ellipticity_vs_mask}, for alternative choices of masks
we show how the assumption that the GCE comes from dark matter annihilation (``NFW $\gamma = 1.2$'') compares to various choices of bulges for the GCE (see text for details). Log-likelihoods are again evaluated by summing all energy bins. We also include the option that the GCE originates from a combination of dark matter and a stellar population of sub-threshold point sources (``NFW $\gamma = 1.2$ $\&$ Coleman Bulge'').
From the masks (ROIs) shown in the top panels, the backgrounds giving the best fit prefer the GCE morphology to follow the ``NFW $\gamma = 1.2$ $\&$ Coleman Bulge'' choice. 
The two next best options are either the purely ``NFW $\gamma = 1.2$'' profile or the ``Coleman Bulge'' profile (see text for further details). The same result comes for the lower panel masks, with the difference that for the relevant background model used to get the best fit--which changes between masks--the pure dark matter annihilation profile for the GCE performs as well as the ``NFW $\gamma = 1.2$ $\&$ Coleman Bulge'' profile. }
\label{fig:DM_vs_Bulges}
\end{figure*}

The results from template fitting are shown in Fig.~\ref{fig:DM_vs_Bulges}. We find that the dark matter annihilation template, ``NFW $\gamma=1.2$'', generally  outperforms the ``Boxy Bulge'', ``BB+NB'' 
and the ``X-shaped Bulge'' morphologies, irrespective of the mask or the 
astrophysical background model for the galactic diffuse emissions.\footnote{For some background models the ``Boxy Bulge'' or ``BB+NB'' profiles may seem to provide a better 
than ``NFW $\gamma=1.2$''. However, these cases involve background models that yield a significantly poorer quality of fit. It highlights the importance of background model selection, as a suboptimal choice can potentially give misleading conclusions about the morphological properties of the GCE 
(see also Ref.~\cite{McDermott:2022zmq}).} 
These findings strengthen previous results from Ref.~\cite{Cholis:2021rpp}, which reached similar conclusions
regarding the GCE morphology using 
the ``Standard 4FGLDR1 + L20'' and ``Standard 4FGLDR2 + L20'' masks.

Interestingly, the ``F98'' model's performance compared to the ``NFW $\gamma=1.2$'' varies depending on the specific combination of mask and background models.
The difference in the fit quality between the two models for the GCE morphology is typically small. However, the ``NFW $\gamma=1.2$'' morphology still slightly outperforms the ``F98'' profile, particularly with regular masks such as ``Standard 4FGLDR3 + L20'', ``Small 4FGLDR3 + L20'', ``Large 4FGLDR3 + L20'', and 
``Standard 4FGLDR3 + L5'', for the best-fit background choices. The preference between ``NFW $\gamma=1.2$'' and ``F98'' varies more with wavelet-based masks, but ``NFW $\gamma=1.2$'' still outperforms the best-fit background choice.

When comparing the ``NFW $\gamma=1.2$'' to the ``Coleman Bulge'' profile, we find that the 
exact choice of mask and galactic diffuse emission background affects the preference between the two GCE templates. However, the difference in $-2 \Delta \ln \mathcal{L}$ is typically small. As shown in the fourth column of Fig.~\ref{fig:stellar_bulges}, the ``Coleman Bulge'' profile has a morphology that 
is asymmetric peanut shape, elongated along the galactic disk. However, when comparing to other stellar bulge profiles, we note that \textit{the 
``Coleman Bulge'' profile is morphologically the closest to 
a spherical NFW profile among the bulge profiles}, particularly when masking the central ``L5'' region. Such similarity makes the fit result less surprising.
We also note that which of those (if any) bulge profiles more faithfully represents the 
population of stellar progenitors to the possible 
MSPs laying below the point source threshold remains an open but important question. 

Furthermore, we explore whether the GCE could be a combination of dark matter annihilation and a stellar bulge profile, with free relative normalization for each of the 14 energy bins we study. This can be justified as even if the GCE is mostly a dark matter signal, we expect contribution from unresolved MSPs and other gamma-ray sources at some level. We test combinations of the ``NFW $\gamma = 1.2$'' profile with (1) the ``BB + NB'' profile, (2) the ``F98'' profile, and (3) the ``Coleman Bulge'' profile. 
 For the first two cases, the NFW $\gamma = 1.2$ absorbs almost all the GCE-associated emission above 0.7 GeV effectively in all background model combinations and masks. Furthermore, the difference in log-likelihood between using the combined profile and the pure ``NFW $\gamma = 1.2$'' template is small. That is in agreement with earlier results when we used the ``Standard 4FGLDR2 + L20'' mask in Refs.~\cite{Cholis:2021rpp, McDermott:2022zmq}. 
 
 However, the combination of ``NFW $\gamma = 1.2$''  with the ``Coleman Bulge'' profile (``NFW $\gamma = 1.2$ $\&$ Coleman Bulge'') presents a less definitive picture. 
 Practically in all scenarios, the combined profile ``NFW $\gamma = 1.2$ $\&$ Coleman Bulge'' performs better or as well as either the ``NFW $\gamma = 1.2$'' or the ``Coleman Bulge'' profiles alone. The ranking again varies depending on the mask and the background model. This leads us to explore further
 the amount of the GCE emission that is absorbed 
 by each of the two components (``NFW $\gamma = 1.2$'' or ``Coleman Bulge'') as a function of energy. 
 
\subsection{The GCE spectrum from dark matter annihilation and a population of point sources}
\label{subsec:DM_and_Bulges}

In this section, we focus on the case that the 
GCE is attributed to two overlapping components, each 
with its distinct morphology. One component is modeled by the spherically symmetric dark matter annihilation 
template, ``NFW $\gamma = 1.2$''.
The other is described as the ``Coleman Bulge'' profile. We are motivated to test the combined ``NFW $\gamma = 1.2$ $\&$ Coleman Bulge'' based on our findings in Sec.~\ref{subsec:DM_vs_Bulges}. This combination often provides the best fit to the gamma-ray data across various choices of background models and masks
(see also Fig.~\ref{fig:DM_vs_Bulges}).

Additionally, it is important to consider that gamma-ray point sources below the detection threshold contribute to the data, and their influence is generally captured by other templates in our fitting process. 
The isotropic diffuse emission template accounts for the extragalactic sub-threshold sources. 
The combination of galactic diffuse emission components tends to absorb some flux from the galactic 
point sources, especially those
correlated with the gas-dense regions of the 
Milky Way. By incorporating the ``Coleman Bulge'' template, we explicitly test the hypothesis that the GCE is associated morphologically with the galactic bulge. 

\begin{figure*}
\centering
\includegraphics[width=0.48\textwidth]{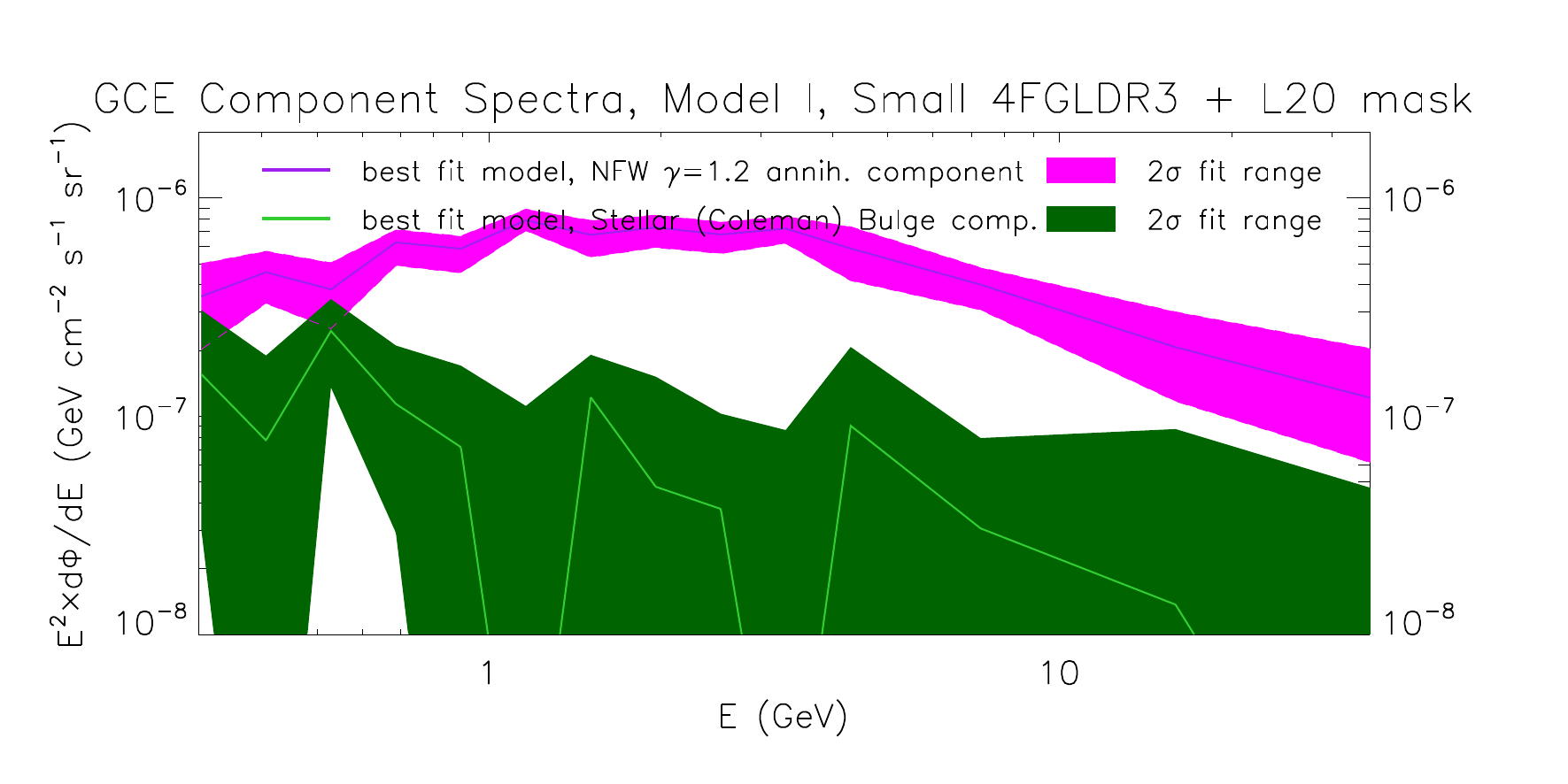}
\includegraphics[width=0.48\textwidth]{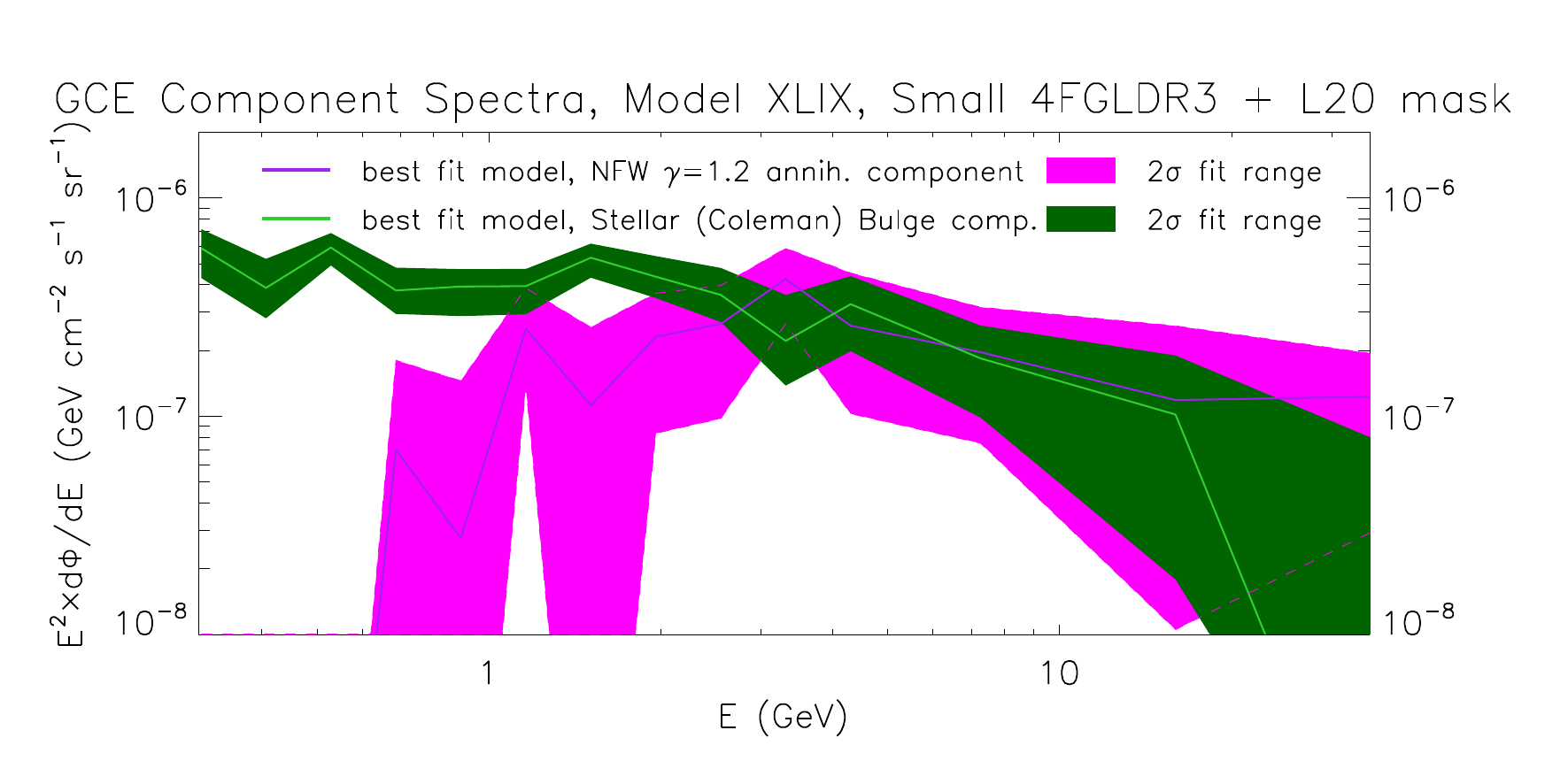}
\\
\includegraphics[width=0.48\textwidth]{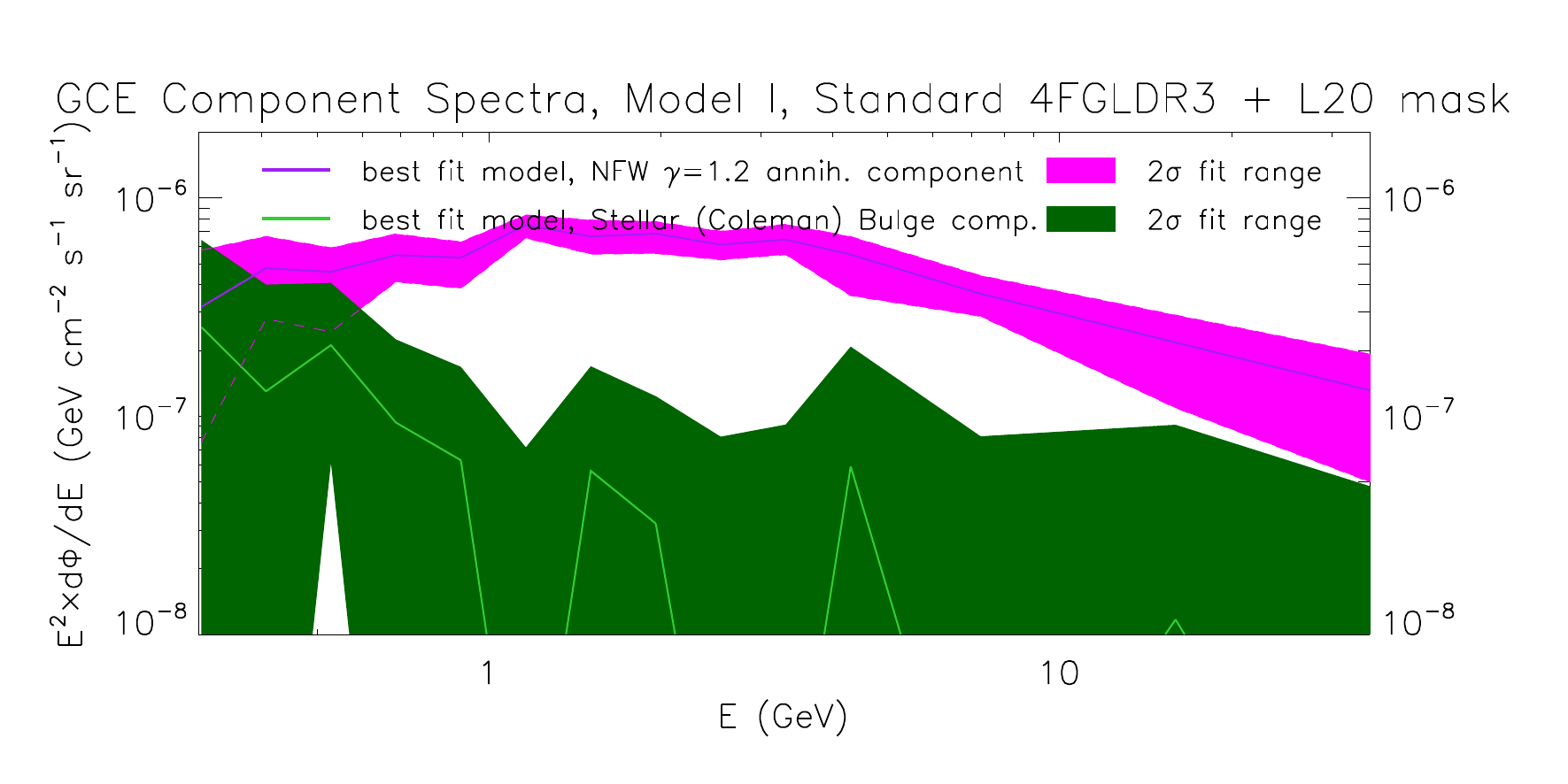}~
\includegraphics[width=0.48\textwidth]{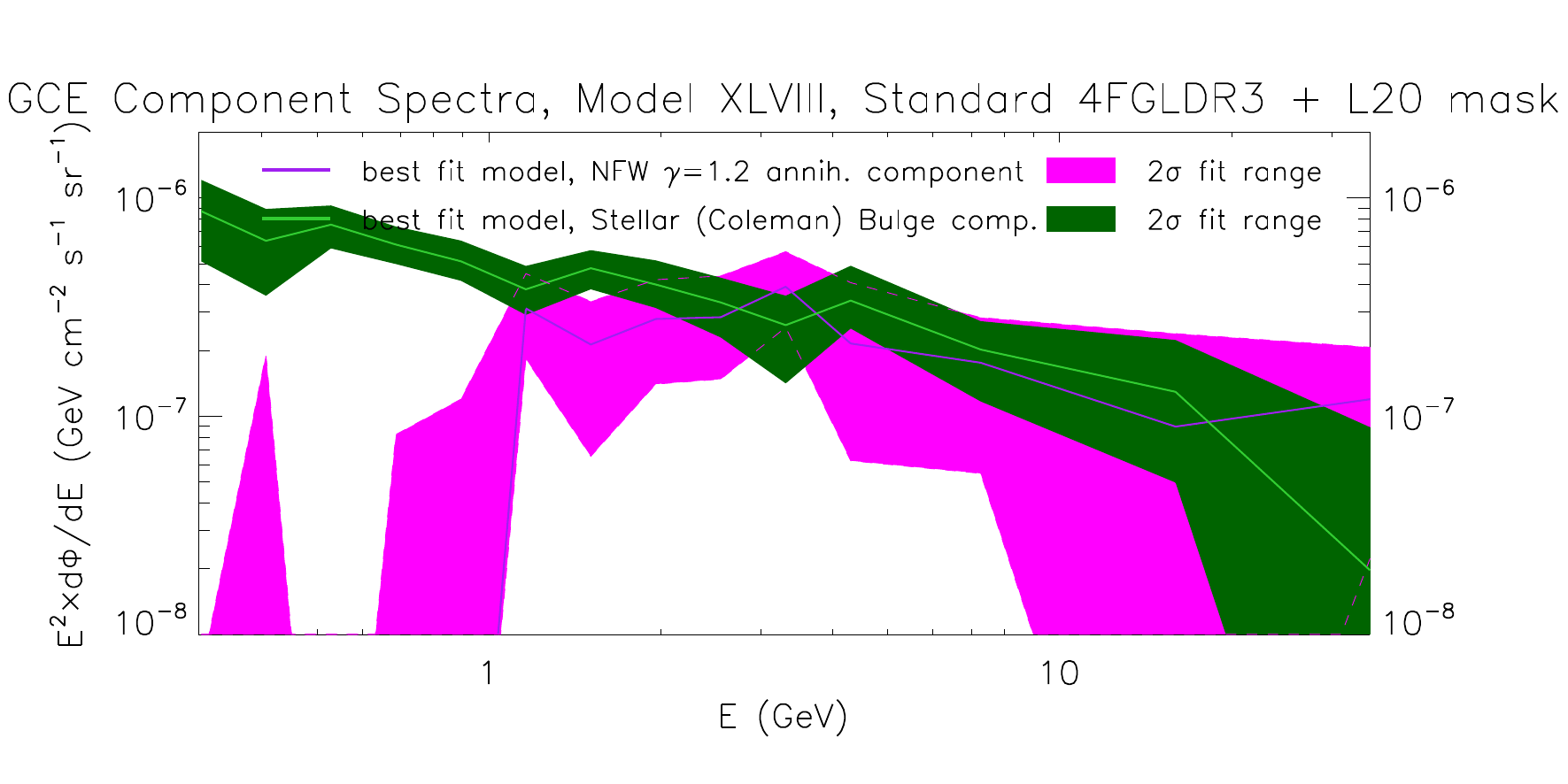}
\\
\includegraphics[width=0.48\textwidth]{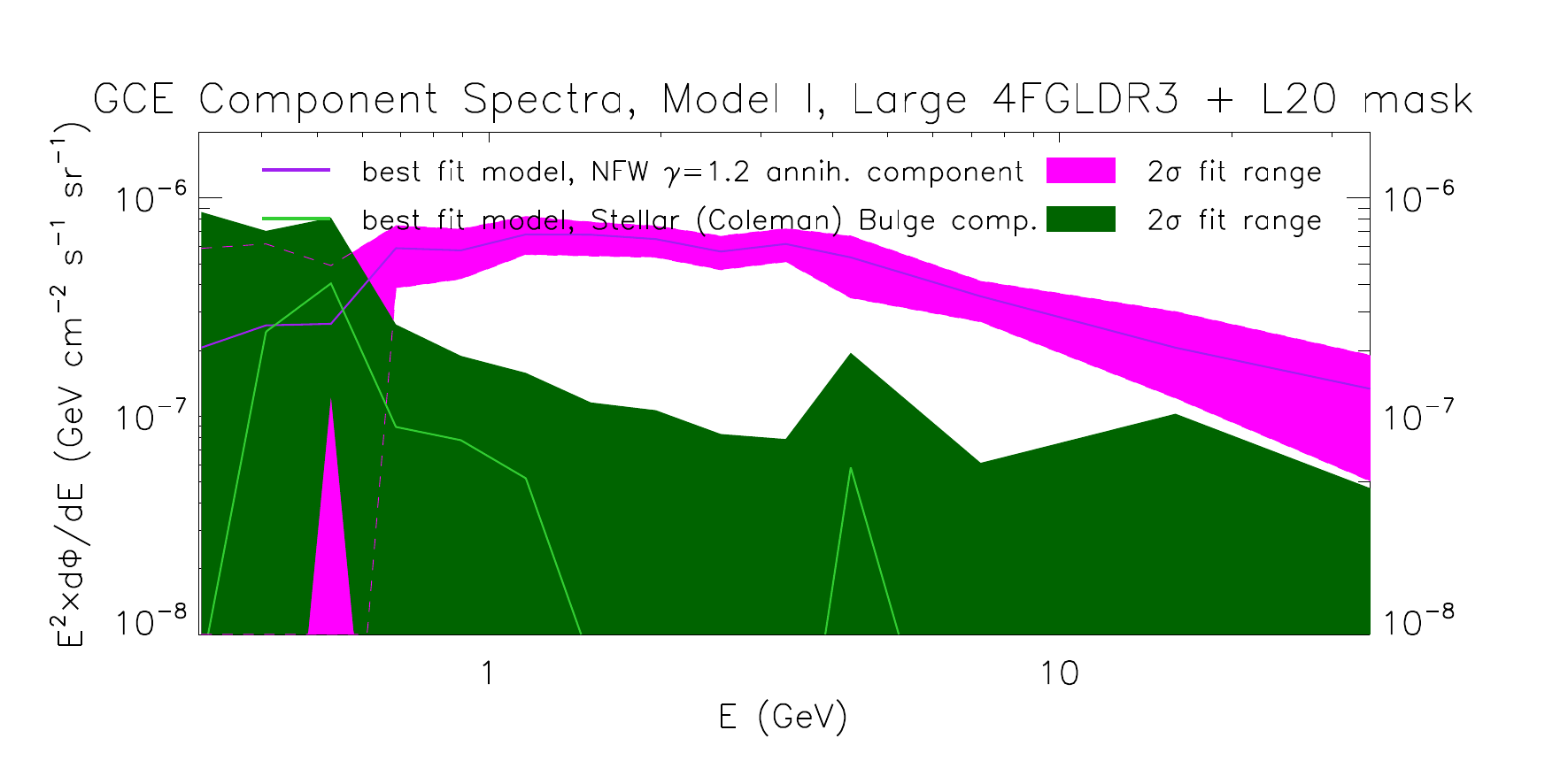}
\includegraphics[width=0.48\textwidth]{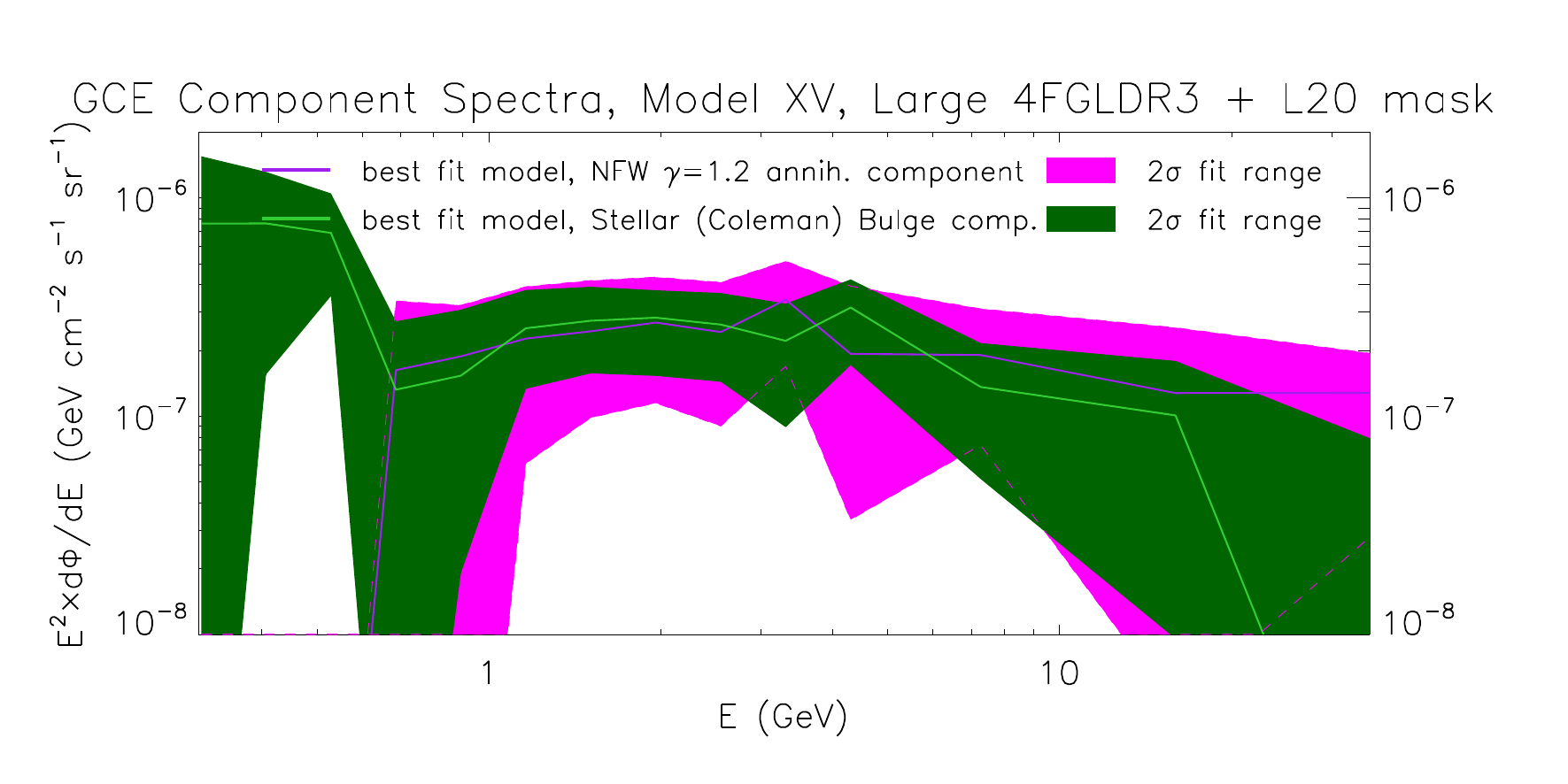}
\\
\includegraphics[width=0.48\textwidth]{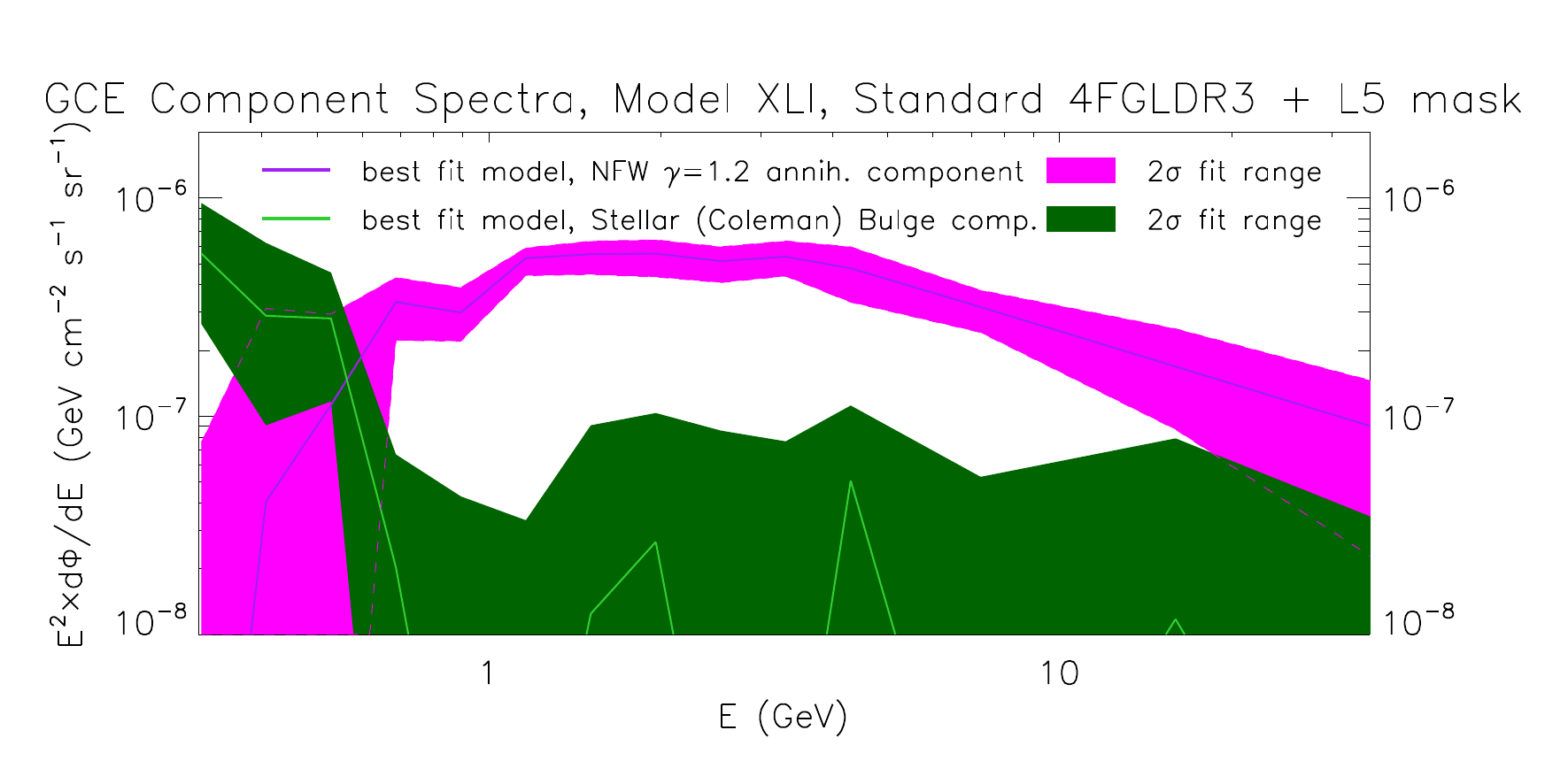}
\includegraphics[width=0.48\textwidth]{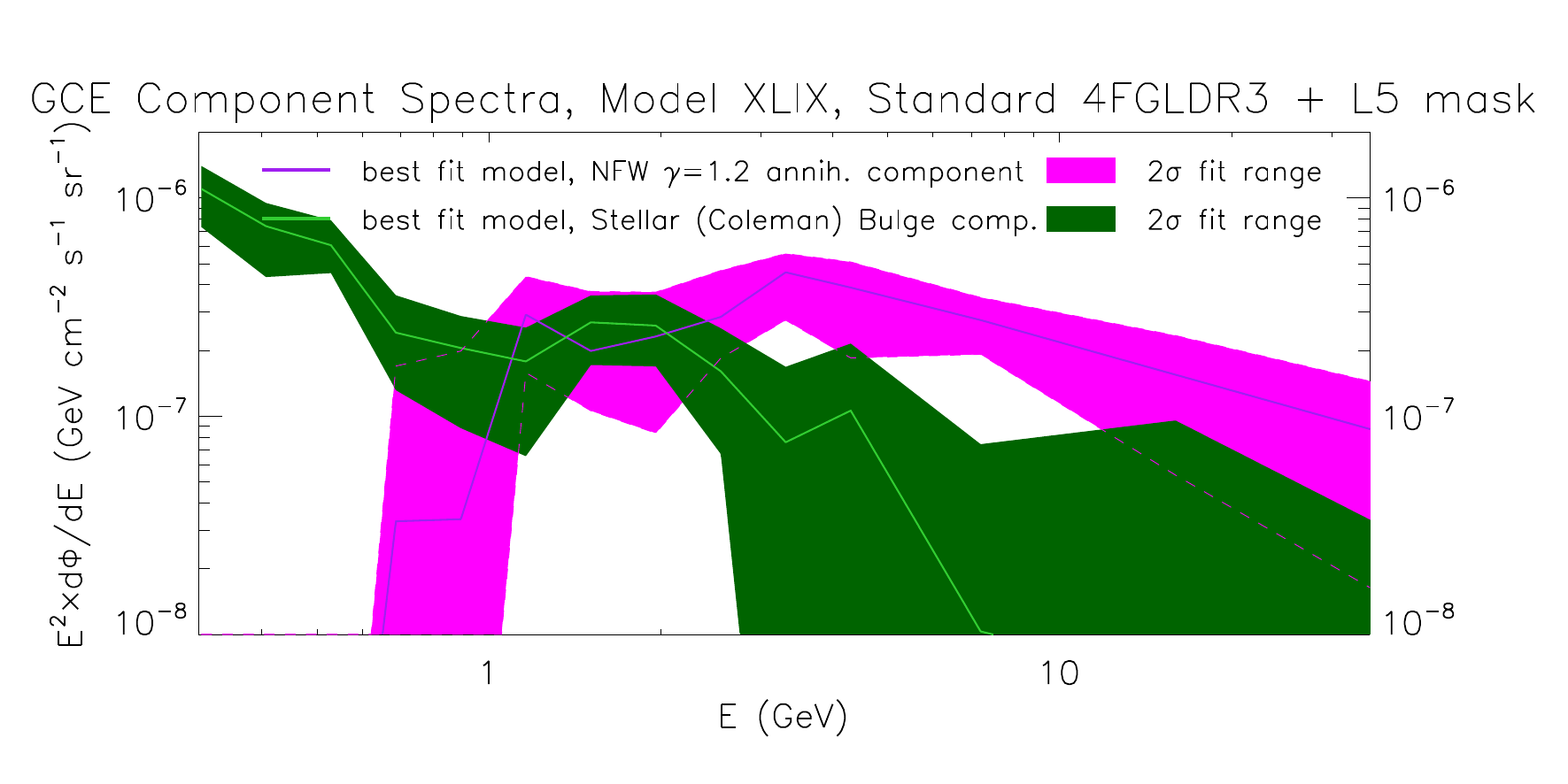}
\caption{Assuming that the GCE comes from the combination of dark matter and subthreshold gamma-ray point sources, 
we show the flux decomposition of the GCE attributed to the spherically symmetric ($\epsilon=1.0$) dark matter annihilation 
template with $\gamma = 1.2$ (purple line and magenta band) and the asymmetric and elongated along the galactic 
disk ``Coleman Bulge'' template (green line and band). 
We show the evolution with the energy of the two components contributing to the GCE. 
In each row, we show results from two different gamma-ray background models keeping the mask fixed. 
The two background models are selected to be among those providing the best agreement to the data.
From top to bottom, we use the following masks: ``Small 4FGLDR3 + L20'', ``Standard 4FGLDR3 + L20'', ``Large 4FGLDR3 + L20'' and ``Standard 4FGLDR3 + L5''.} 
\label{fig:GCE_DM_and_Bulge_Fluxes_Regular_Masks}
\end{figure*}

\begin{figure*}
\centering
\includegraphics[width=0.48\textwidth]{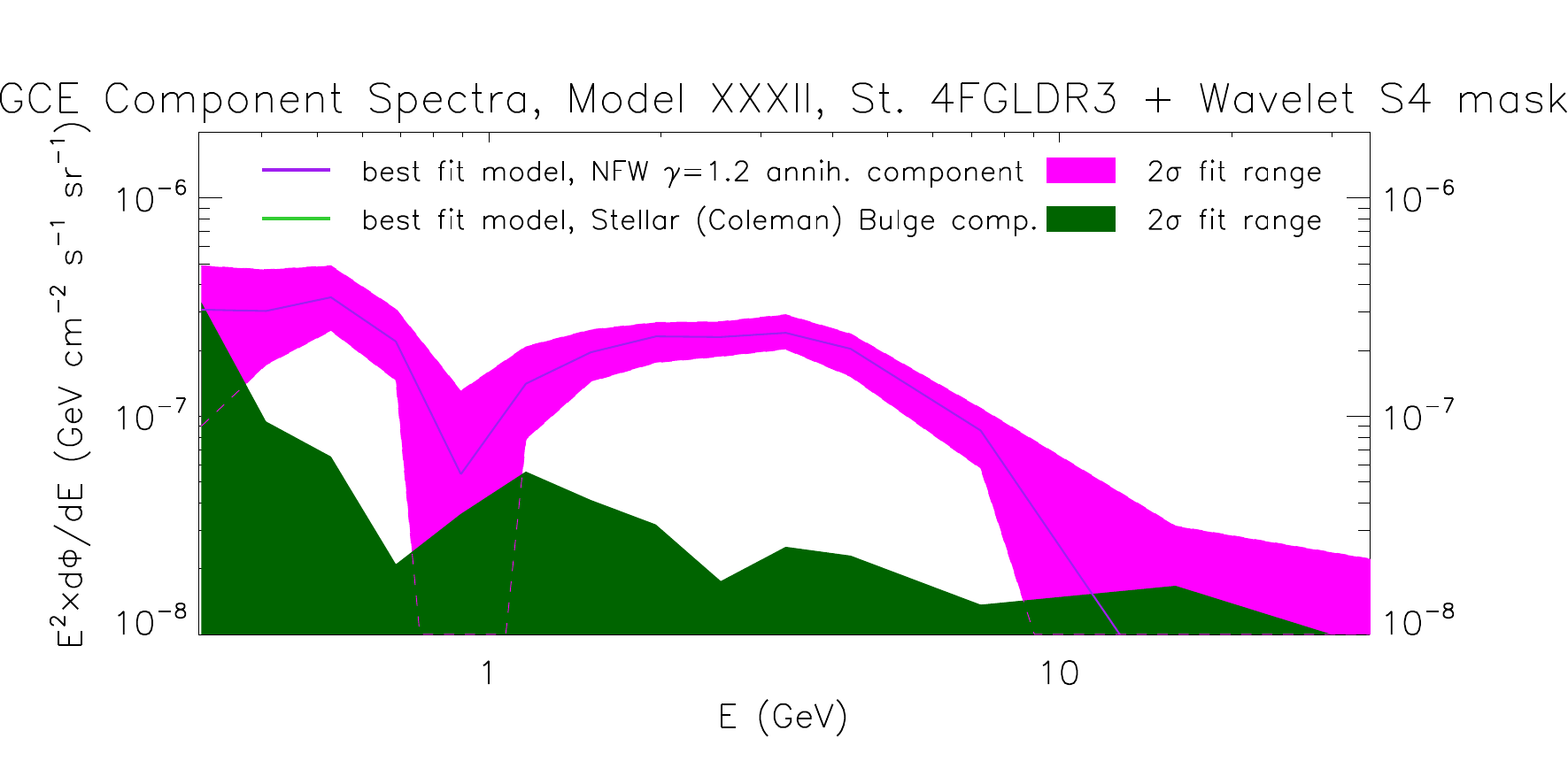}
\includegraphics[width=0.48\textwidth]{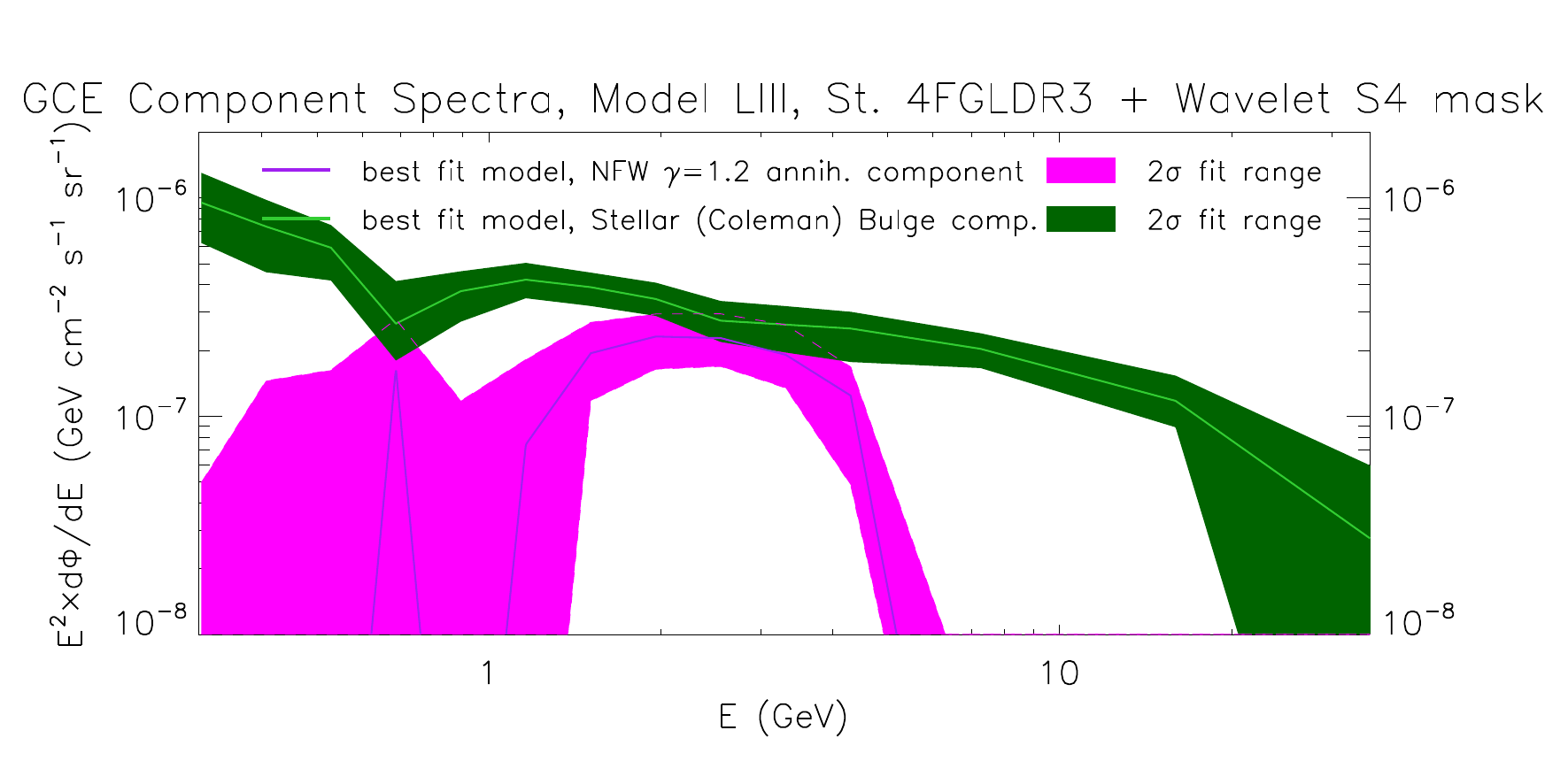}
\\
\includegraphics[width=0.48\textwidth]{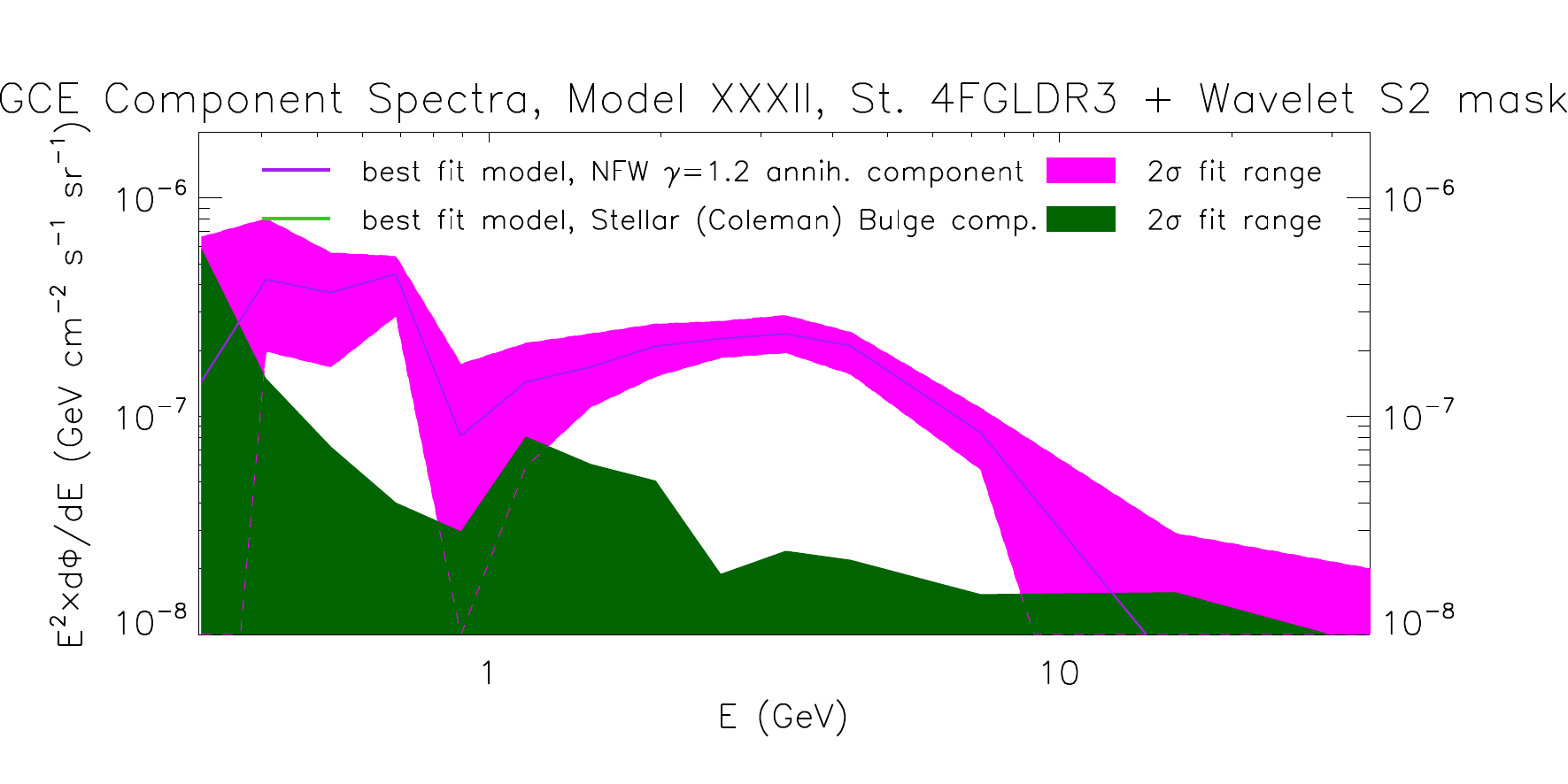}
\includegraphics[width=0.48\textwidth]{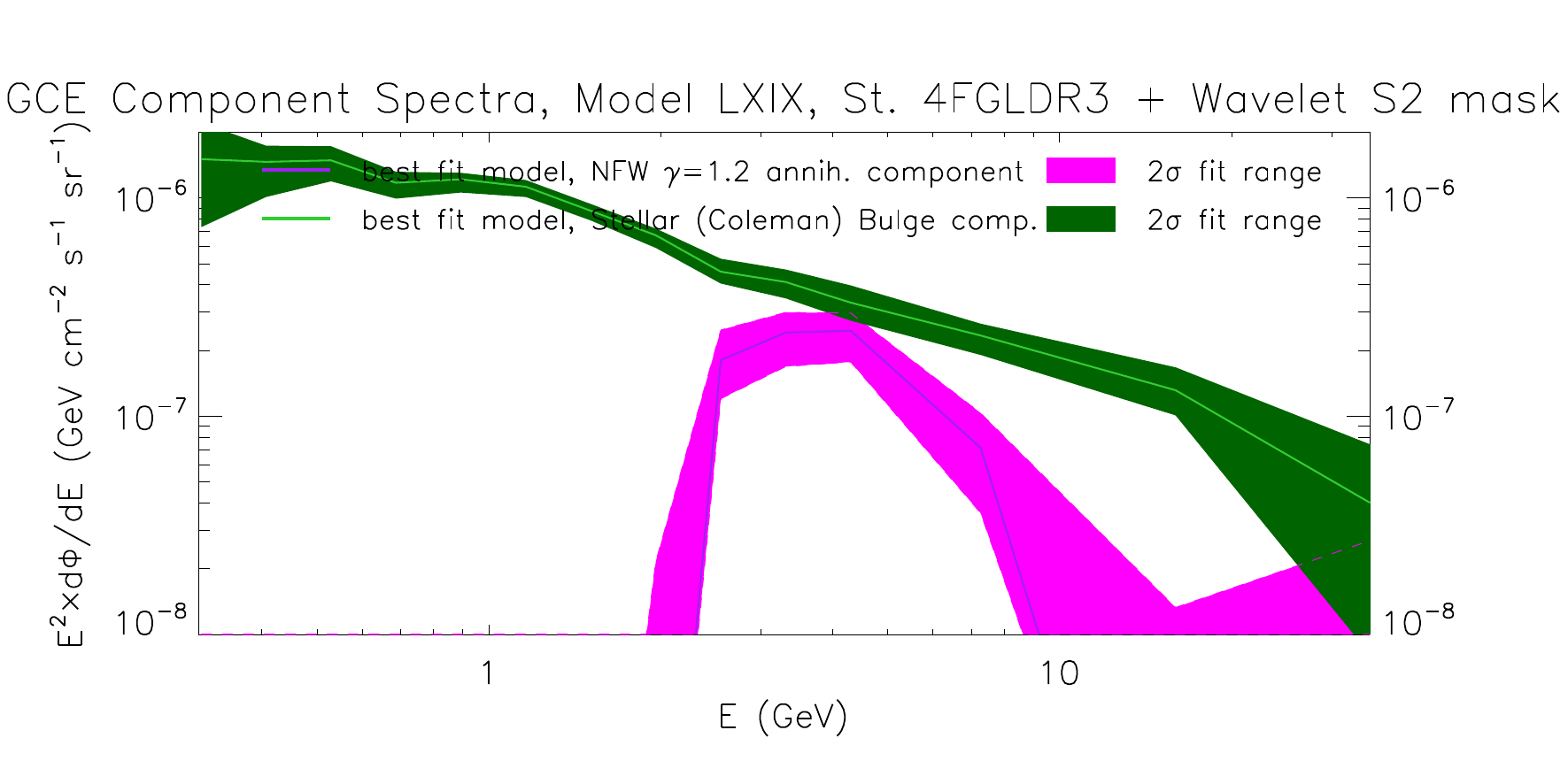}
\caption{As in Fig.~\ref{fig:GCE_DM_and_Bulge_Fluxes_Regular_Masks}, we show the flux decomposition of the GCE attributed to the dark matter annihilation 
template with $\gamma = 1.2$ and $\epsilon=1.0$ (purple line and magenta band) and the ``Coleman Bulge'' template (green line and band). We show results using masks ``Standard 4FGLDR3 + Wavelet S4'' and ``Standard 4FGLDR3 + Wavelet S2''.} 
\label{fig:GCE_DM_and_Bulge_Fluxes_Wavelet_Masks}
\end{figure*}

Figs.~\ref{fig:GCE_DM_and_Bulge_Fluxes_Regular_Masks} and~\ref{fig:GCE_DM_and_Bulge_Fluxes_Wavelet_Masks} show the flux decomposition of the GCE into these two components for the six different masks previously used in Figs.~\ref{fig:cuspiness_vs_mask}-\ref{fig:DM_vs_Bulges}. 
Each component in our study has an independently adjustable normalization for each of the 14 energy bins. 
The flux attributed to the dark matter component is shown with a purple line for the best-fit normalization and a magenta band for its 2$\sigma$ range, which are derived from the MCMC procedure that samples the coefficient parameters.
Similarly, the flux associated with the ``Coleman Bulge'' is given by a light green line for 
the best-fit normalization and a dark green band for its 2$\sigma$ range. 
For each mask, we show results from two out of the 80 background models. We pick these two
models among those providing a good fit to the data and in a manner that is intended to envelope for the reader the range of how much 
of the GCE each of the two components can absorb. 

In the analysis using regular masks, as shown in Fig.~\ref{fig:GCE_DM_and_Bulge_Fluxes_Regular_Masks}, the dark matter component consistently comprises at least 50$\%$ of the GCE emission above 0.7 GeV, and in some instances, it accounts for over 90$\%$ of the GCE emission. 
Moreover, even after accounting for systematic uncertainties, the spectrum of the dark matter component is roughly the same as what has been proposed in previous works 
as~\cite{Hooper:2010mq, 2009arXiv0912.3828V, Abazajian:2010zy, Hooper:2011ti, Hooper:2013rwa, Gordon:2013vta, Abazajian:2014fta, Daylan:2014rsa, Calore:2014xka, Zhou:2014lva, TheFermi-LAT:2015kwa, Linden:2016rcf, Cholis:2021rpp}.
This suggests that, within a factor of $\simeq 2$ adjustments in the dark matter cross section and mass, 
the existing explanations for the dark matter origin of the GCE remain valid, even considering 
all tested background diffuse emission model systematics, mask choices, and the potential independent contribution of a galactic bulge component. 

It is also notable that even when the ``Coleman Bulge'' component contributes maximally (as seen in the right panels of Fig.~\ref{fig:GCE_DM_and_Bulge_Fluxes_Regular_Masks}), 
its spectrum is \textit{distinctly different} from that typically associated with MSPs \cite{Abazajian:2010zy, Abazajian:2014fta, Cholis:2014noa, Bartels:2015aea}. Instead, it more closely resembles the 
gamma-ray spectrum from the regular population of
galactic point sources \cite{Fermi-LAT:2022byn}, which includes MSPs but also contains the emission 
from younger gamma-ray pulsars, including pulsar wind nebulae, sources in globular clusters, binaries, 
supernova remnants, and star-forming regions. 
This suggests that the actual contribution of regular MSPs to the GCE may be relatively minor (see e.g. \cite{Hooper:2013nhl, Cholis:2014lta, Zhong:2019ycb, Leane:2019xiy}). Results from the ``Standard 4FGLDR1 + L20'', ``Standard 4FGLDR2 + L20'' and ``Standard 4FGLDR3 + L8'' masks, lead to the same conclusions.   In App.~\ref{app:GCE_Components_Statistical_Preference_vs_Energy}, we provide additional information on the statistical preference for either the dark matter or the ``Coleman Bulge'' components of the GCE as a function of energy.

Fig.~\ref{fig:GCE_DM_and_Bulge_Fluxes_Wavelet_Masks}, shows the results obtained using our wavelet-based masks ``Standard 4FGLDR3 + Wavelet S2'' and ``Standard 4FGLDR3 + Wavelet S4''. The  GCE spectra derived from these masks are distinctively different from those derived when portions of the galactic disk are masked. 
Depending on the specific case, the dark matter component or the ``Coleman Bulge'' may dominate the GCE emission. 
However, it is important to note that the gamma-ray spectrum associated with the ``Coleman Bulge'', again, does not resemble the known spectra of MSPs. 
This discrepancy suggests a significant contribution from unresolved point sources along the disk.
Results from the ``Standard 4FGLDR3 + Wavelet S3'' mask agree with those of the other two wavelet-based masks. 

\begin{figure*}
\includegraphics[width=0.48\textwidth]{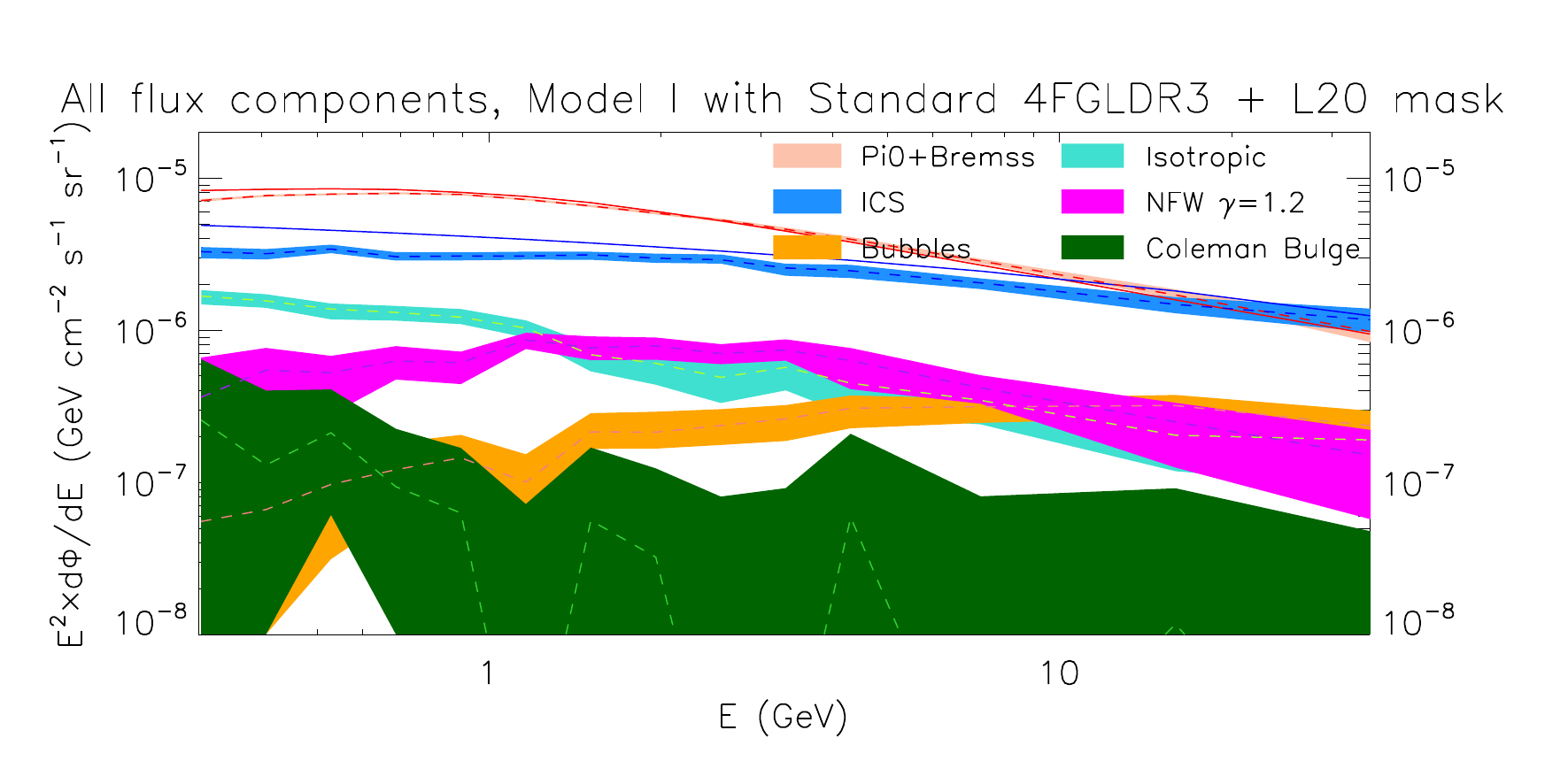}
\includegraphics[width=0.48\textwidth]{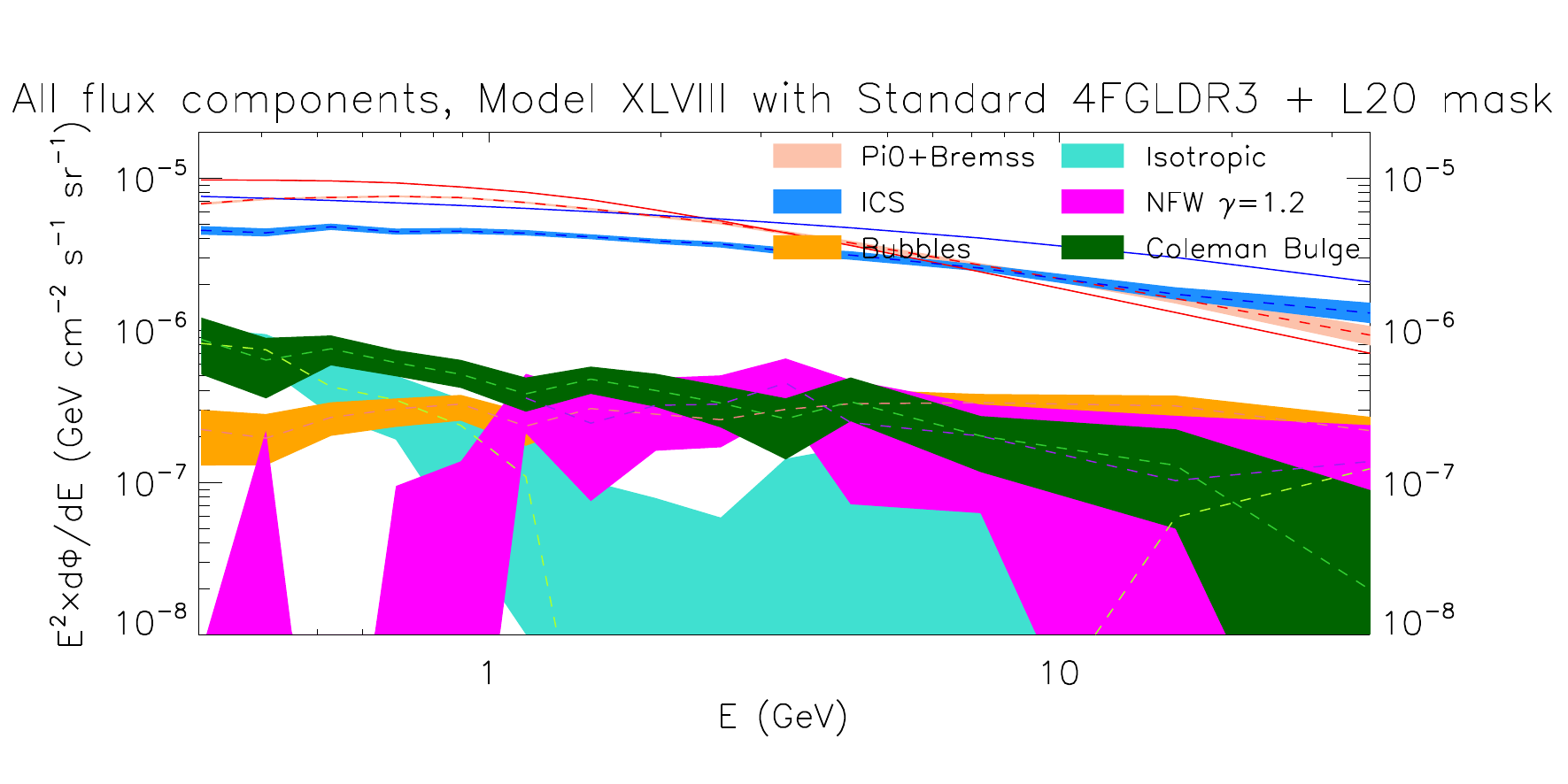}
\caption{For the $40^{\circ} \times 40^{\circ}$ ROI with the ``Standard 4FGLDR3 + L20'' mask, we provide all galactic diffuse emission components. Those are the Pi0+Bremsstrahlung in orange lines and band, the ICS in blue lines and band, the isotropic in turquoise, \textit{Fermi} bubbles in orange, the dark matter component in purple line and magenta band, and the ``Coleman Bulge'' component in green line and band. Left panel: We use galactic diffuse emission model I of Ref.~\cite{Cholis:2021rpp}. Right panel: We instead use the model XLVIII  from the same reference (see text for further details).} 
\label{fig:All_spectra}
\end{figure*}

Finally, in Fig.~\ref{fig:All_spectra}, as a point of reference for the reader, we provide a complete decomposition of gamma-ray emission from the 
$40^{\circ} \times 40^{\circ}$ ROI for one of the masks and two galactic diffuse emission background
models. The solid lines represent the original assumptions of the Pi0+Bremsstrahlung and the ICS before fitting each energy bin. The dashed lines show the fitted spectra for each component, along with their corresponding 2$\sigma$ ranges as colored bands.
On the left panel of Fig.~\ref{fig:All_spectra}, we use the galactic diffuse emission model I of Ref.~\cite{Cholis:2021rpp}, where the dark matter template dominates the GCE. On the right panel, we use model XLVIII from the same reference instead, in which case the dark matter template contributes approximately half of the GCE flux above 1 GeV. Note that in the case of the right panel, the flux component associated with the ``Coleman Bulge'' profile does not have the hard spectrum followed by an exponential cut-off expected by MSPs. Instead, the spectrum associated with the ``Coleman Bulge'' is that expected from the population of all galactic gamma-ray point sources. 
Furthermore, the inferred isotropic emission component 
deviates from the true isotropic emission calculated at high latitudes. 
On the left panel, the ``Coleman Bulge'' template component has a similar spectrum to that on the right but an order of magnitude dimmer.
We find such spectral patterns indicative of the ``Coleman Bulge'' template are likely some combination of extragalactic and galactic point sources and not an unresolved millisecond pulsar 
population. 

\section{Discussions and Conclusions}
\label{sec:conclusions}

Masking is an essential step for the template-fitting analysis of the GCE, crucial for minimizing the influence of the galactic disk and the known point sources on the GCE emission, which are challenging to model accurately. In this study, we constructed a variety of masks based on the information of the point sources and the disk. These include regular masks, which combine a band-shaped disk mask with point source masks based on the \textit{Fermi} 4FGL catalogs, and wavelet-based masks, which combine masks for high-significance peaks identified by the wavelet method with masks for point sources also based on 4FGL. For the regular masks, we varied the band length for the disk mask and the radii for the point source masks.

We then extensively investigated the impact of masking on the GCE morphology analysis. Through injection tests, we demonstrated that our analysis method, the template fitting, is sensitive and robust against the ellipticity of the GCE. Through systematic analyses, we further showed that the choice of masks does not significantly impact the inferred cuspiness and ellipticity of the GCE, assuming the GCE follows contracted, squared, and integrated NFW profiles, a common template to model dark matter annihilation. This robustness in the characteristics of the GCE is evident at an energy bin-by-energy bin level, except for the lowest energy bins, due to limited statistics after masking and large astrophysical background uncertainties.

We also explored various hypotheses about the GCE's morphology based on different stellar bulge profiles, including ``Boxy Bulge'', ``BB+NB'', ``Coleman Bulge'', ``F98'', and ``X-shaped Bulge''. We compared these results to dark matter annihilation (``NFW $\gamma=1.2$''). Across different choices of masks and background diffuse emission models, we found that the ``NFW $\gamma =1.2$'' and the ``Coleman Bulge'' profiles offer the best fits to the GCE, with the latter profile giving slightly more often a better fit to the gamma-ray data, but with a typically small difference in terms of  $-2\Delta \ln \mathcal L$. 

Additionally, we tested combinations of dark matter annihilation with the ``BB+NB'', the ``F98'', and the ``Coleman Bulge''. Among those combinations, we found that ``NFW $\gamma =1.2$ \& Coleman Bulge'' offers the best fit across different choices of masks and background emission models. It also consistently fits better than ``NFW $\gamma=1.2$'' or ``Coleman Bulge'' alone. This prompted further exploration of the contributions from the two components as a function of energy under different masks. We found that the dark matter component consistently comprises a significant portion of the GCE for $E>0.7\,\text{GeV}$. Although, in some cases, the ``Coleman Bulge'' can be an important component, the corresponding spectra distinctly differ from that typically associated with MSPs, resembling more the spectra of the regular galactic point sources. Besides, when the ``Coleman Bulge'' is an important component, the resulting spectrum associated with the isotropic emission deviates from the measurement at high latitudes.
The fact that the ``Coleman Bulge'' does not yield a MSP-like spectrum is the case across both regular and wavelet-based masks. All regular masks still give a prominent spherically symmetric GCE component.

Our results using wavelet-based masks occasionally show a different GCE morphology and spectrum from those obtained with regular masks, possibly due to too much contribution from other galactic point sources.  However, the template fitting based on wavelet-based masks generally maintains a consistent ranking of GCE hypotheses (dark matter annihilation, stellar bulges, and combinations) as those based on regular masks. The wavelet-based masks do accentuate the differences in $-2\Delta \ln \mathcal L$ between the best-fit background models and poorer-fit ones. Further exploration with a broader pool of background models may offer more suitable candidates for template fitting under wavelet-based masks.

\acknowledgments{
We thank Samuel D.~McDermott for participating in the early stage of the project.
We thank Matthew Buckley, Dan Hooper, Manoj Kaplinghat, Tracy Slatyer, Deheng Song, and Yitian Sun for valuable discussions
 and Oscar Macias and Deheng Song for providing us with the F98 and Coleman Bulge templates. 
 This work was motivated in part by discussions during the Galactic Center Excess workshop hosted by the New High Energy Physics Center at Rutgers University.
We also acknowledge the use of \texttt{GALPROP} \cite{galprop}. This research uses computational resources at the University of Chicago’s Research Computing Center. We thank Edward W. Kolb for providing access to the resources.
YZ acknowledges the Aspen Center for Physics for its hospitality
during the completion of this study, which is supported by the
National Science Foundation under Grant PHY-1607611. YZ
is supported by the Kavli Institute for Cosmological Physics
at the University of Chicago through an endowment from the
Kavli Foundation and its founder Fred Kavli.
IC acknowledges that this material is based upon work supported by the U.S. Department of Energy, Office of Science, Office of High Energy Physics, under Award No. DE-SC0022352. 
}

\begin{appendix}

\section{The GCE Morphology's  Energy Evolution for alternative masks}
\label{app:GCE_Morphology_Energy_Evolution_vs_ROI}

In Sec.~\ref{subsec:GCEmorphology_vs_energy} of the main text, we presented for the ``Standard 4FGLDR3 + L20'' mask the evolution with energy of the GCE's cuspiness and ellipticity. Here, we show results for the other five masks, for which figures are presented in the main text. Those are the, ``Small 4FGLDR3 + L20'', ``Large 4FGLDR3 + L20'', ``Standard 4FGLDR3 + L5'', ``Standard 4FGLDR3 + Wavelet S4'' and ``Standard 4FGLDR3 + Wavelet S2'' masks. For ease of comparison, we also include the results of ``Standard 4FGLDR3 + L20''.   

\begin{figure*}
\centering
\includegraphics[width=0.32\textwidth]{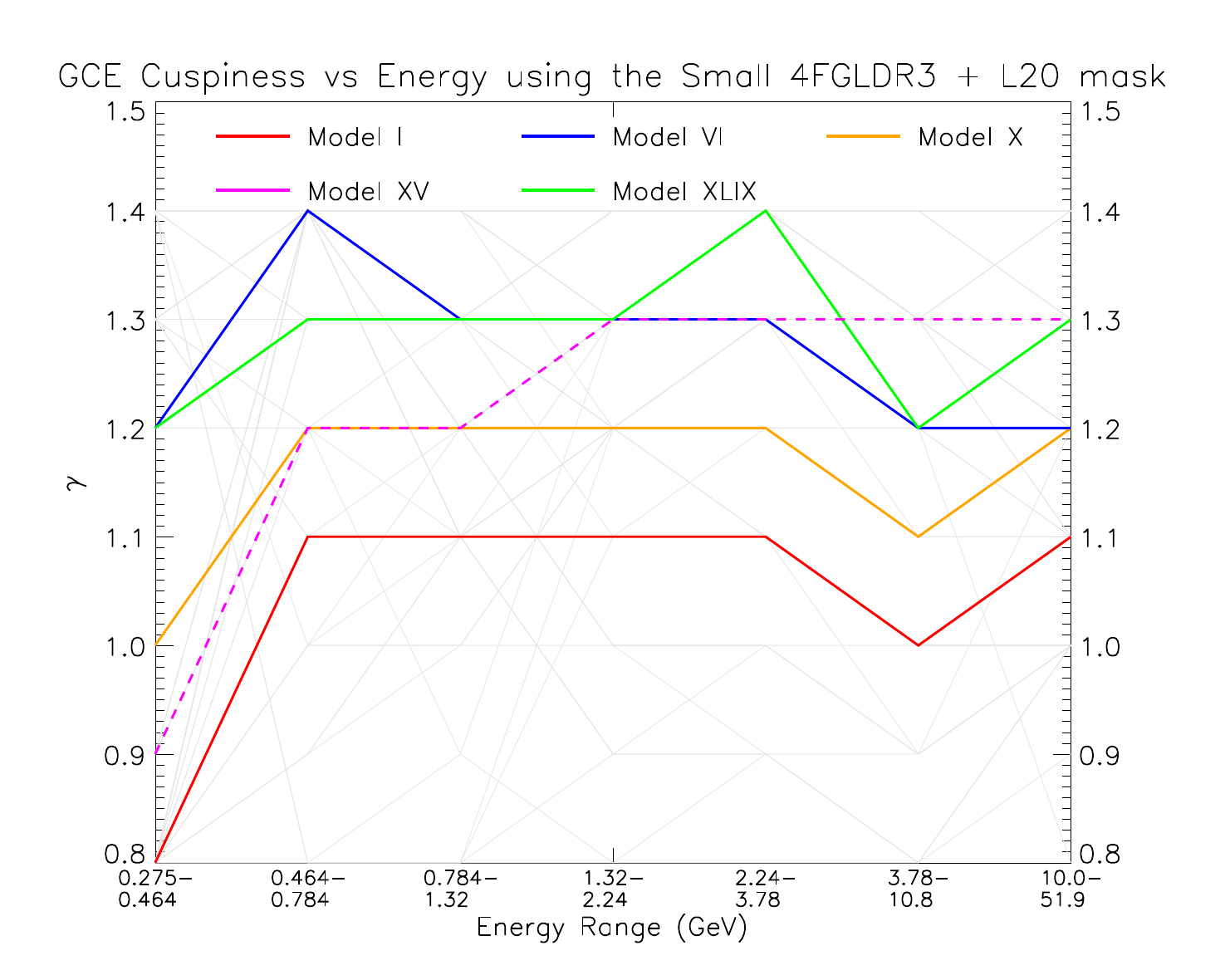}
\includegraphics[width=0.32\textwidth]{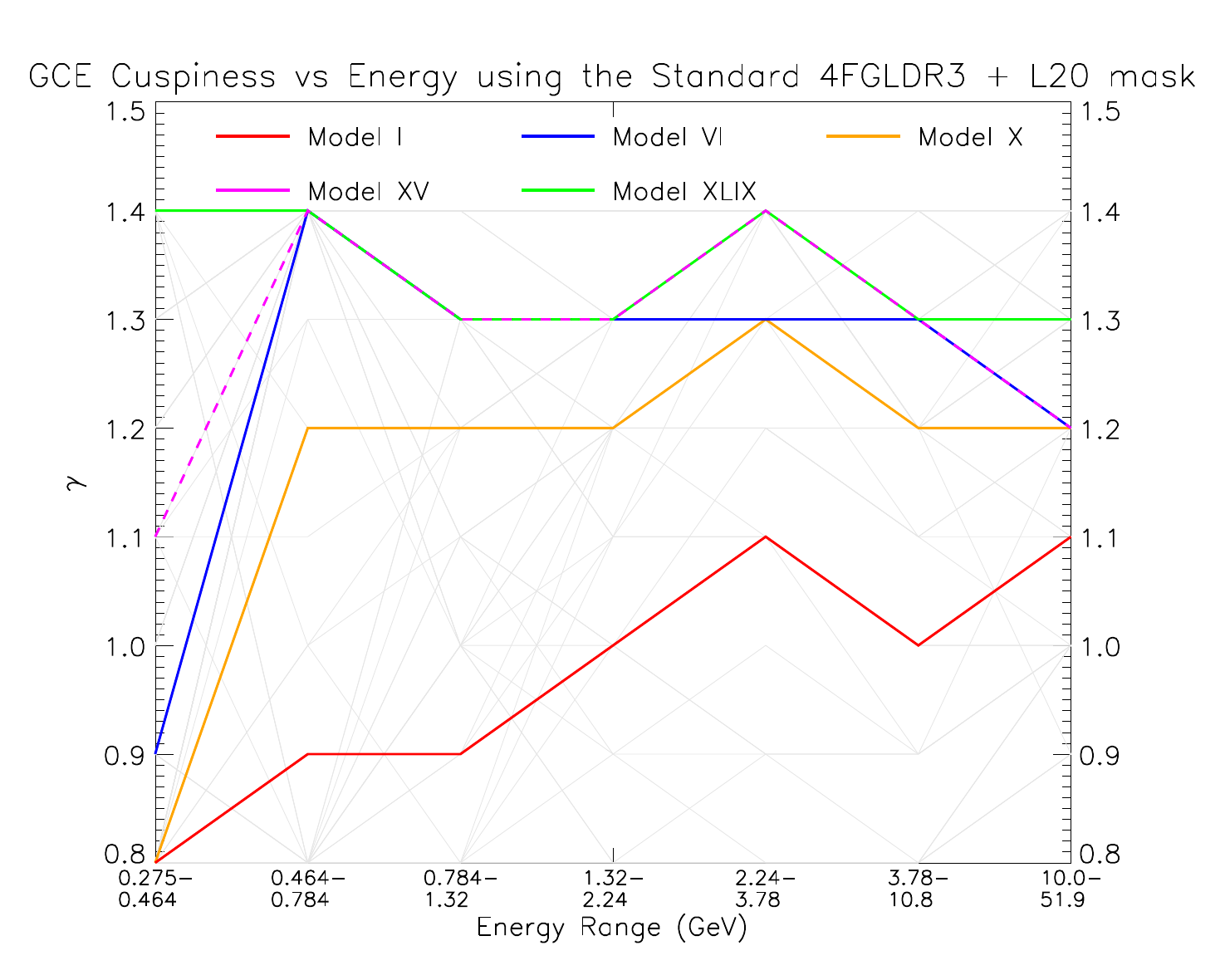}
\includegraphics[width=0.32\textwidth]{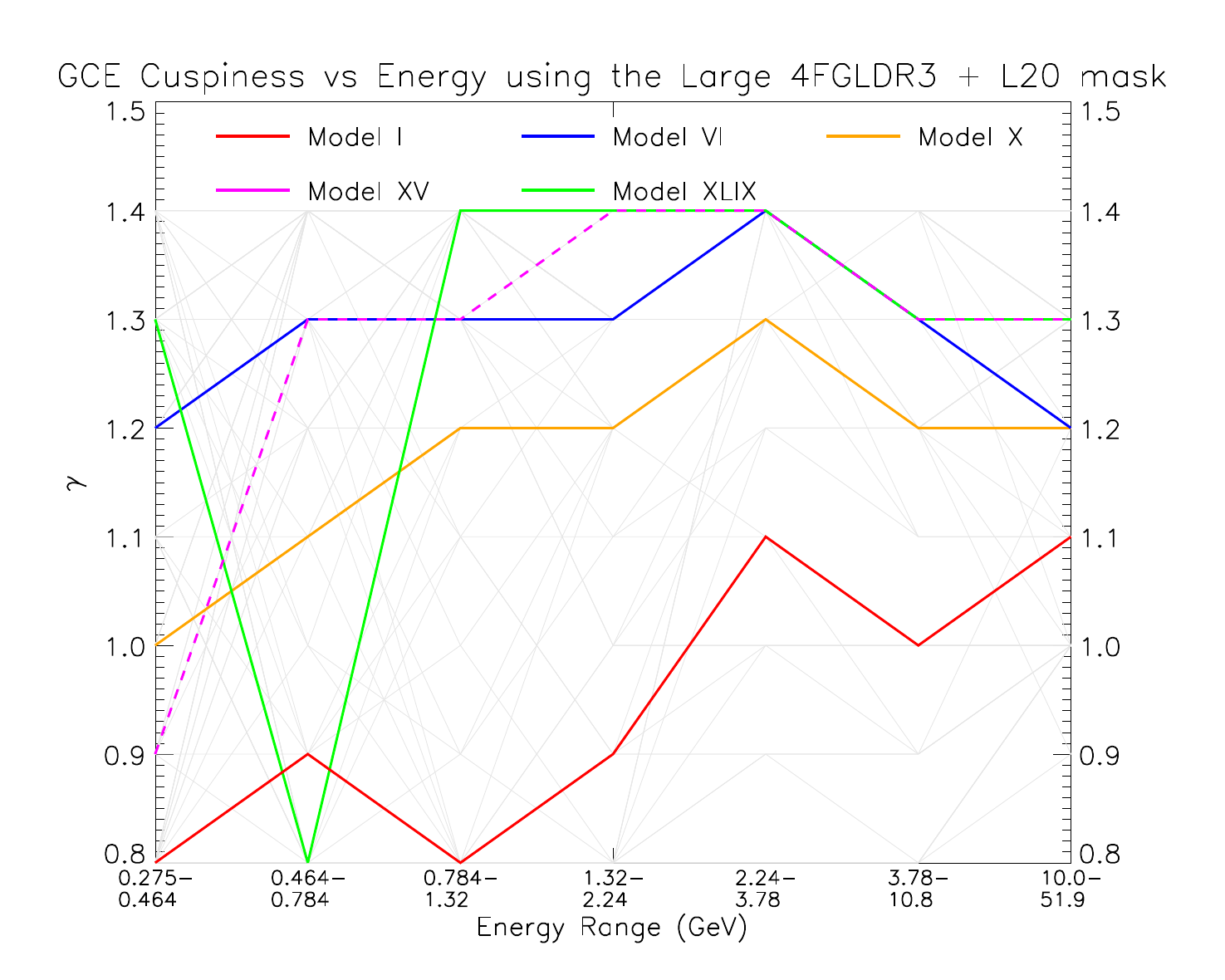}
\\
\includegraphics[width=0.32\textwidth]{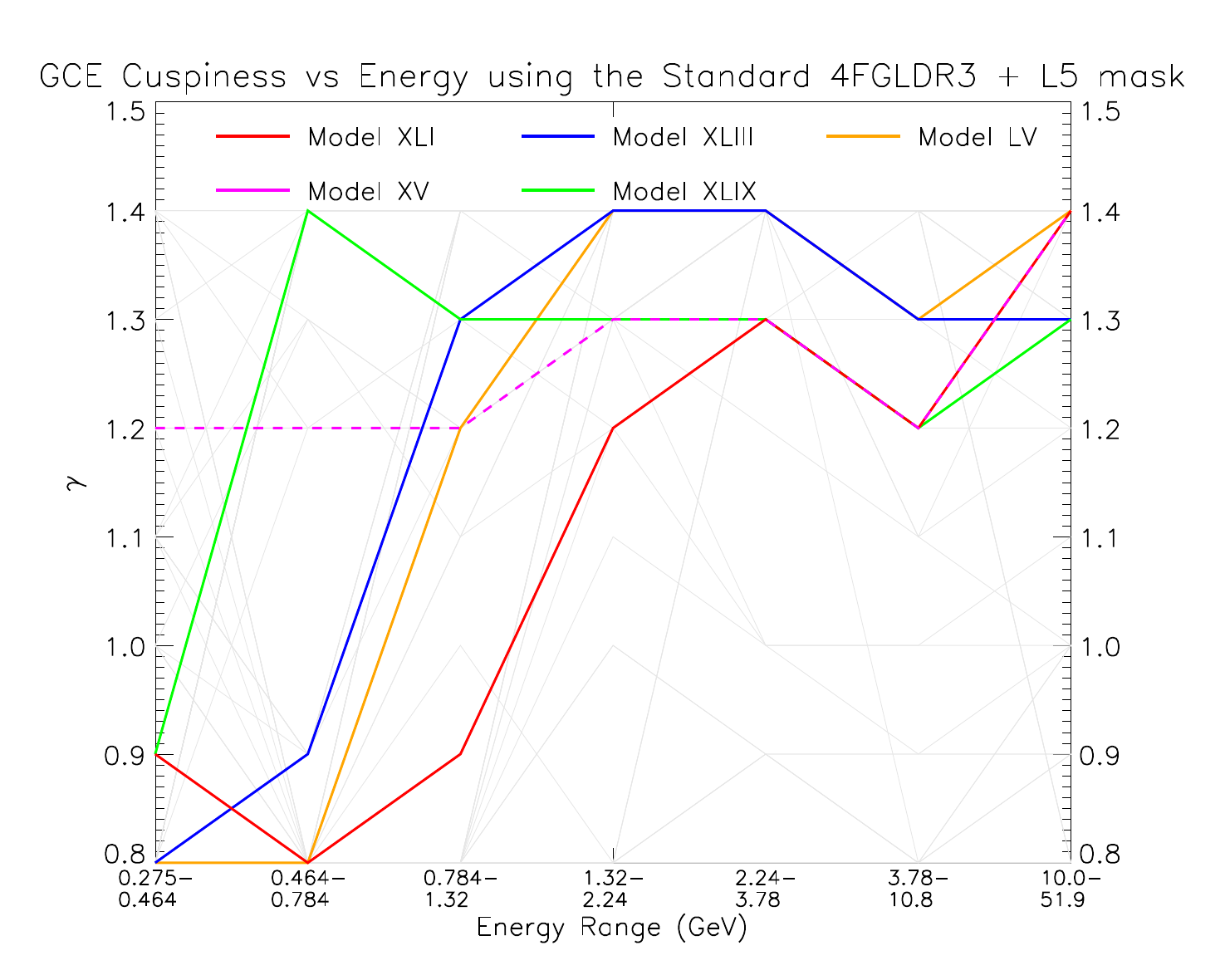}
\includegraphics[width=0.32\textwidth]{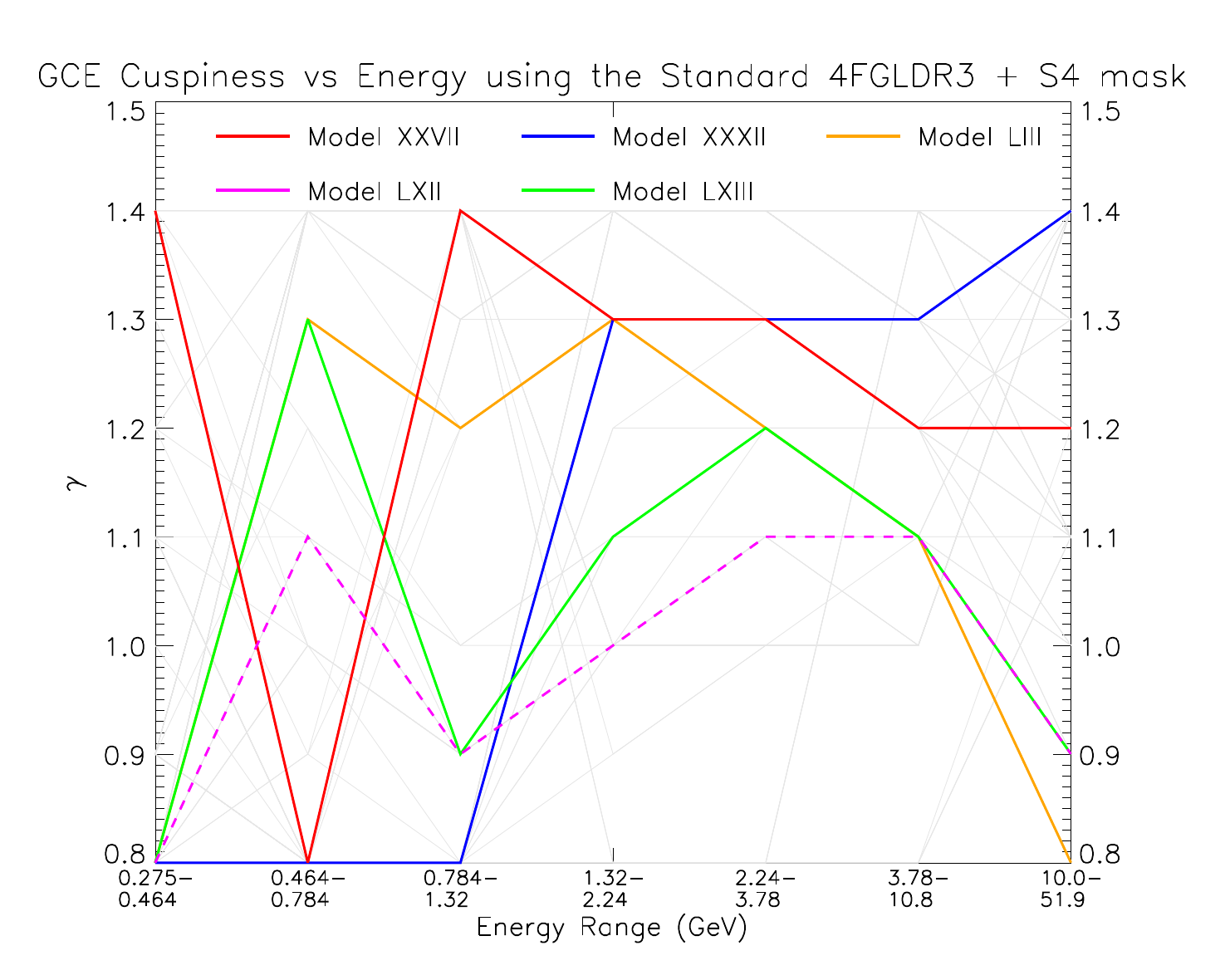}
\includegraphics[width=0.32\textwidth]{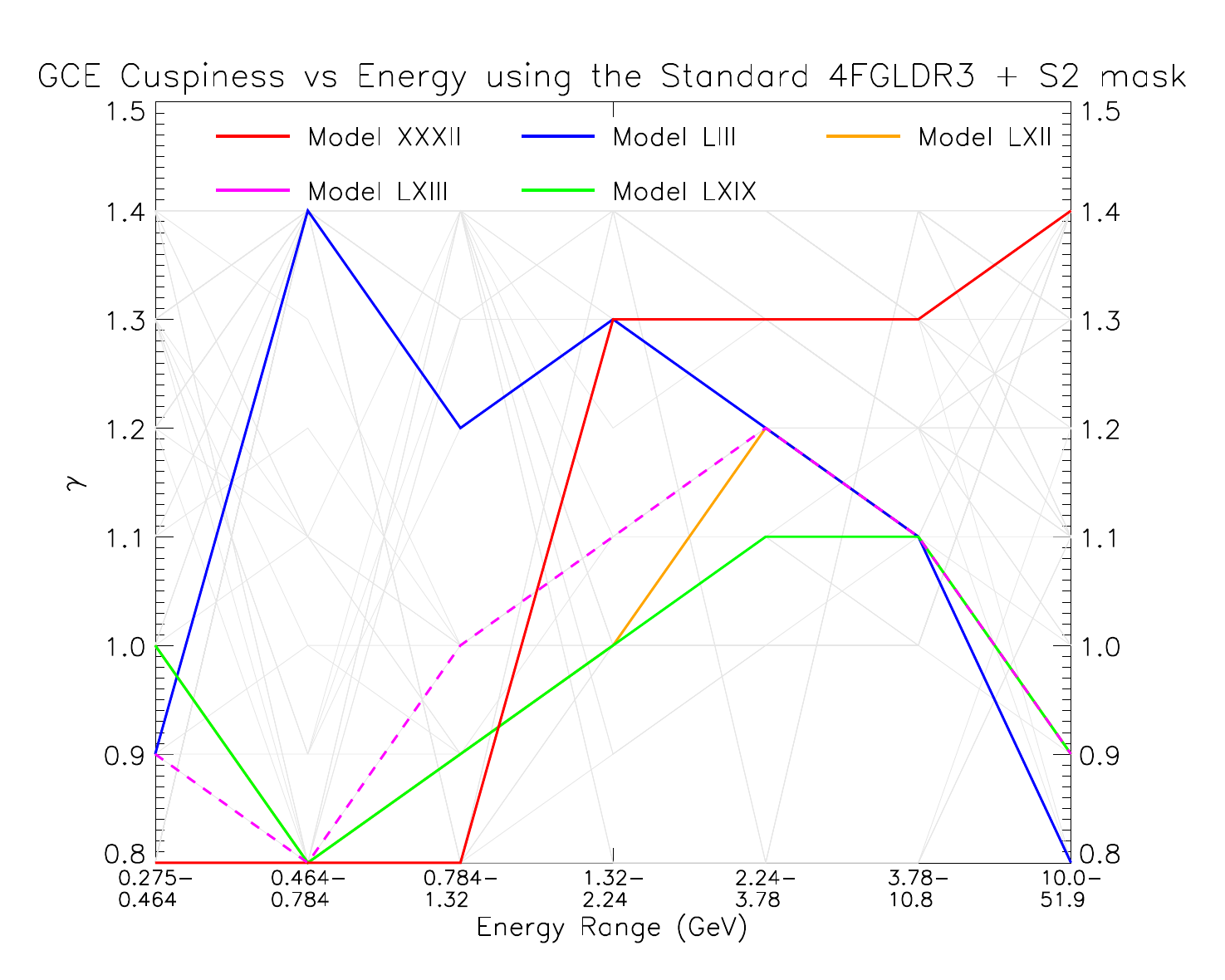}
\caption{As in the left panel of Figure~\ref{fig:morphology_vs_energy}, the evolution of the cuspiness of the GCE with energy for a sequence of masks (see text for more details).} 
\label{fig:Cuspiness_Energy_Evolution_Alt_Masks}
\end{figure*}

In Fig.~\ref{fig:Cuspiness_Energy_Evolution_Alt_Masks}, 
we show the cuspiness evolution with energy for these six masks. Using the regular masks (first four panels) and focusing on the background diffuse emission models that give the best fits (colored lines), the cuspiness converges to values of $1.1 \leq \gamma \leq 1.4$ for energies 3 GeV or higher. Only the wavelet-based masks provide inconclusive results on the GCE cuspiness. As described in the main text, we think that to be the result of the wavelet-based masks allowing too much contribution of other galactic (non-MSP) point sources, whose spectrum has nothing to do with the GCE emission. 

\begin{figure*}
\centering
\includegraphics[width=0.32\textwidth]{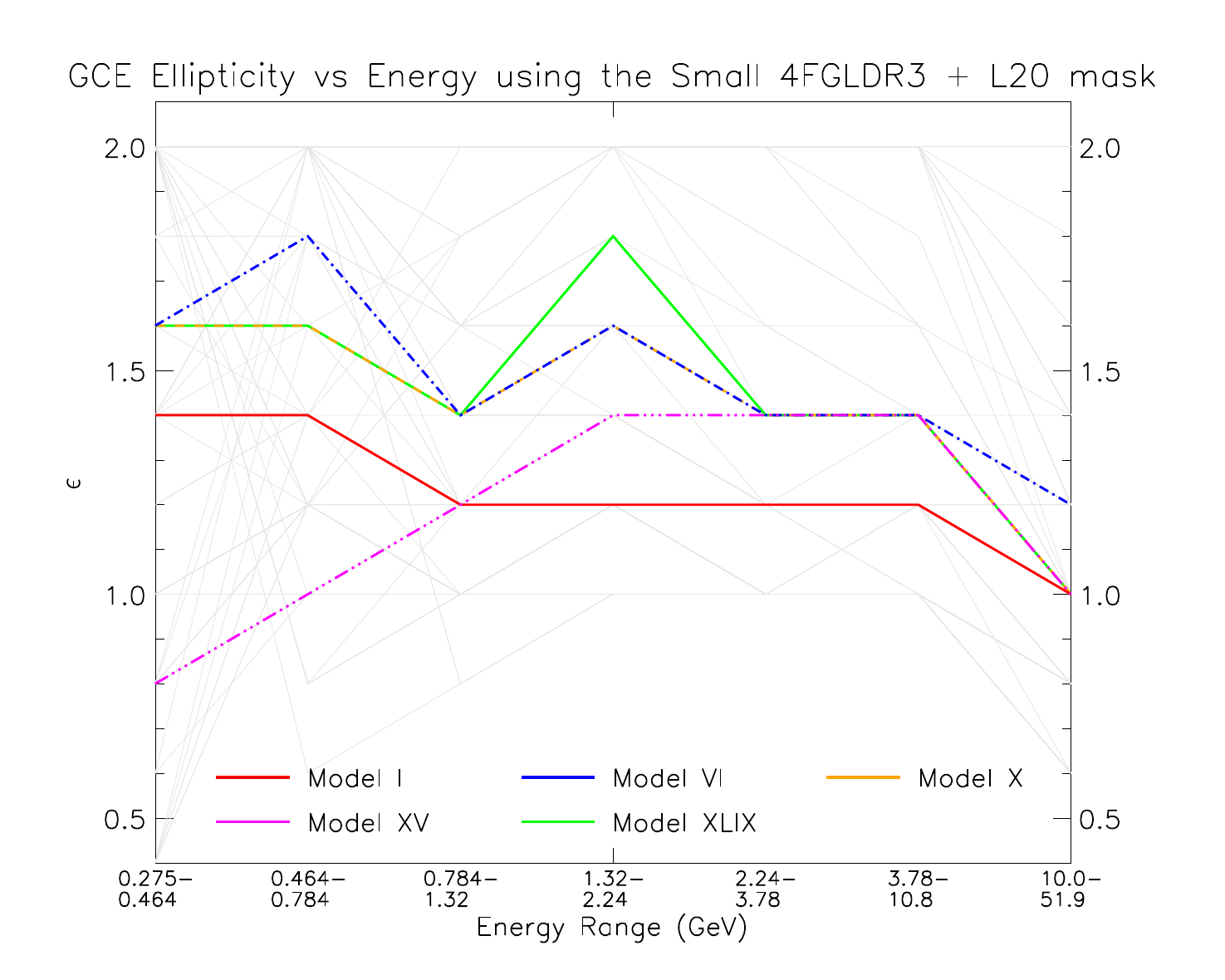}
\includegraphics[width=0.32\textwidth]{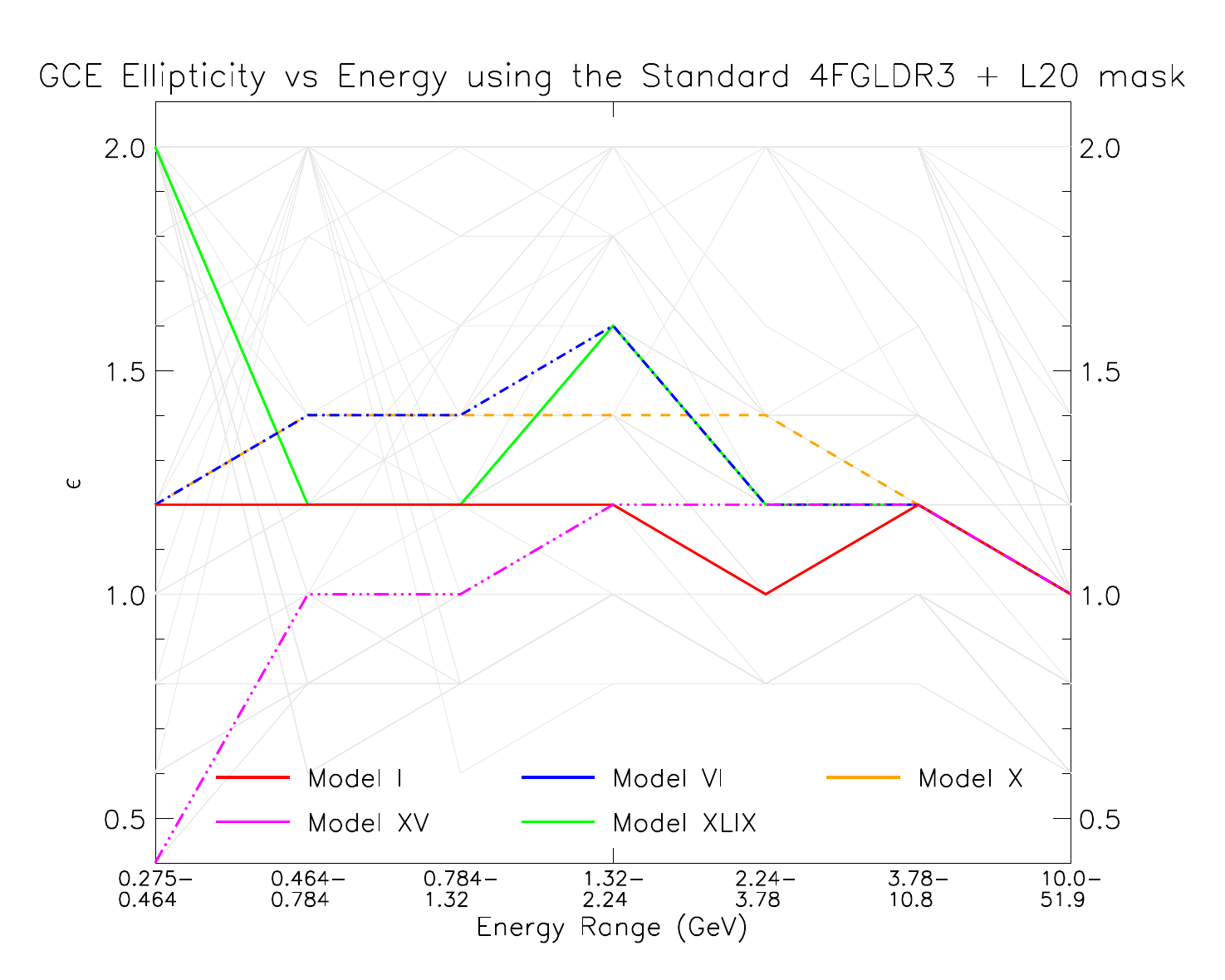}
\includegraphics[width=0.32\textwidth]{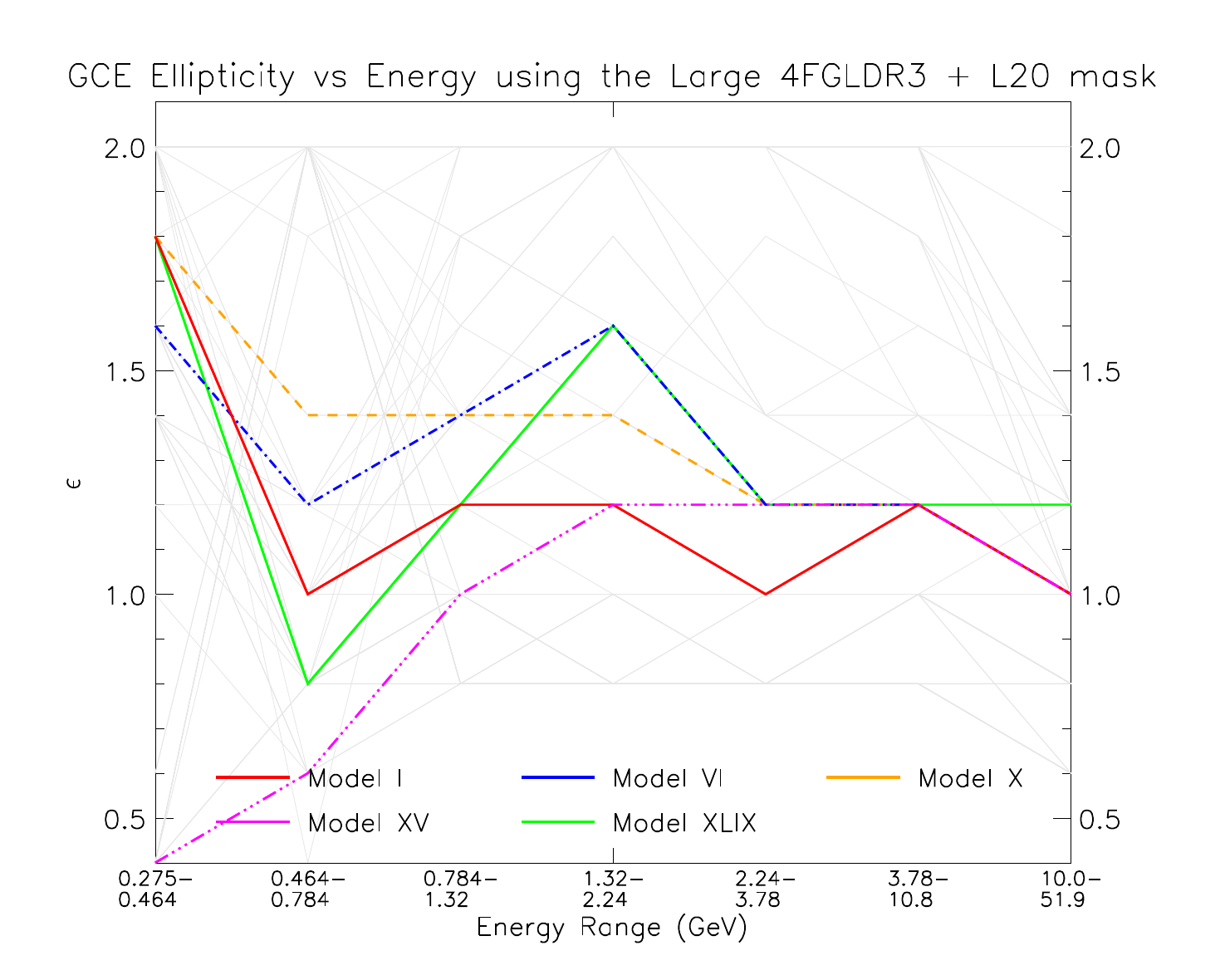}
\\
\includegraphics[width=0.32\textwidth]{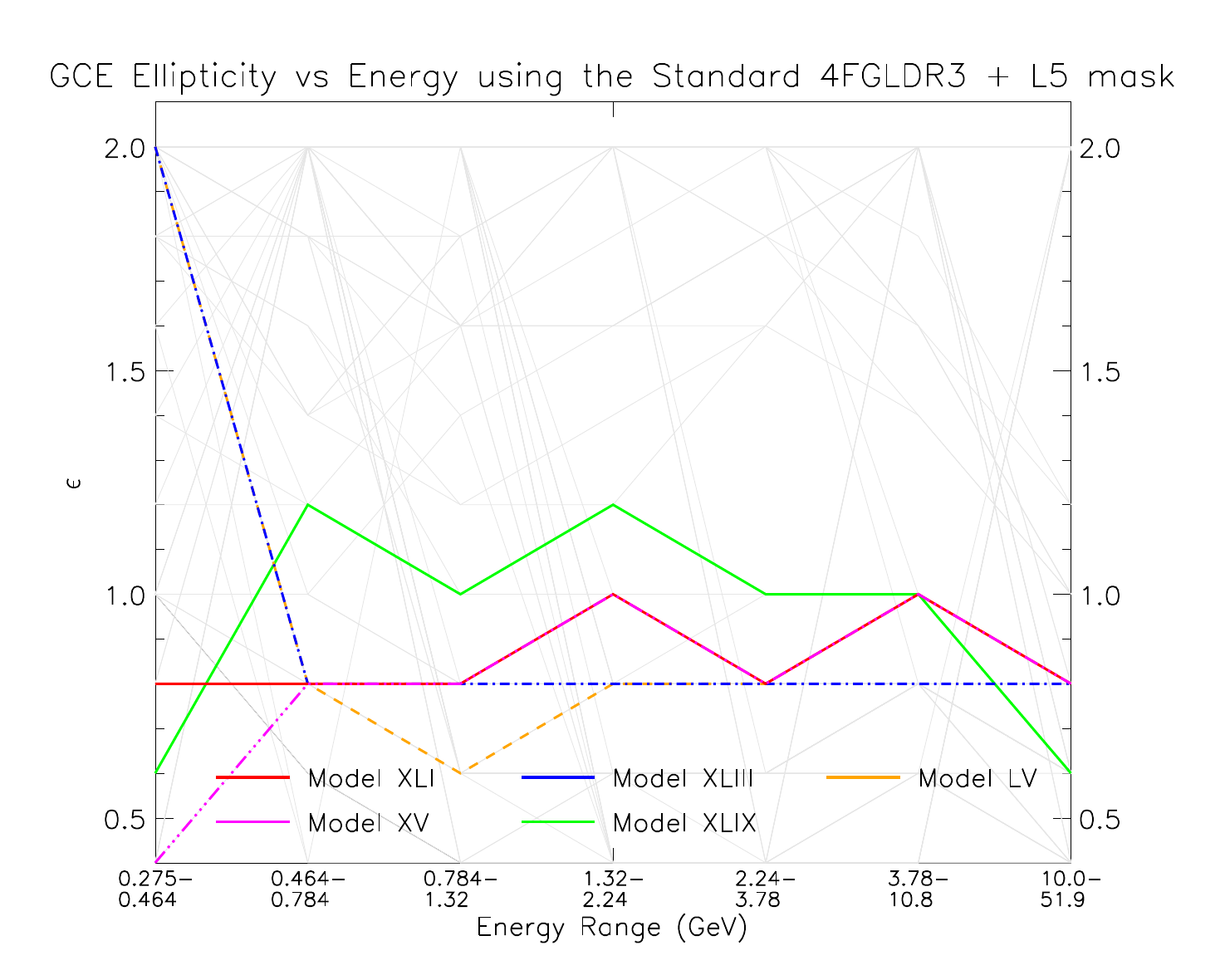}
\includegraphics[width=0.32\textwidth]{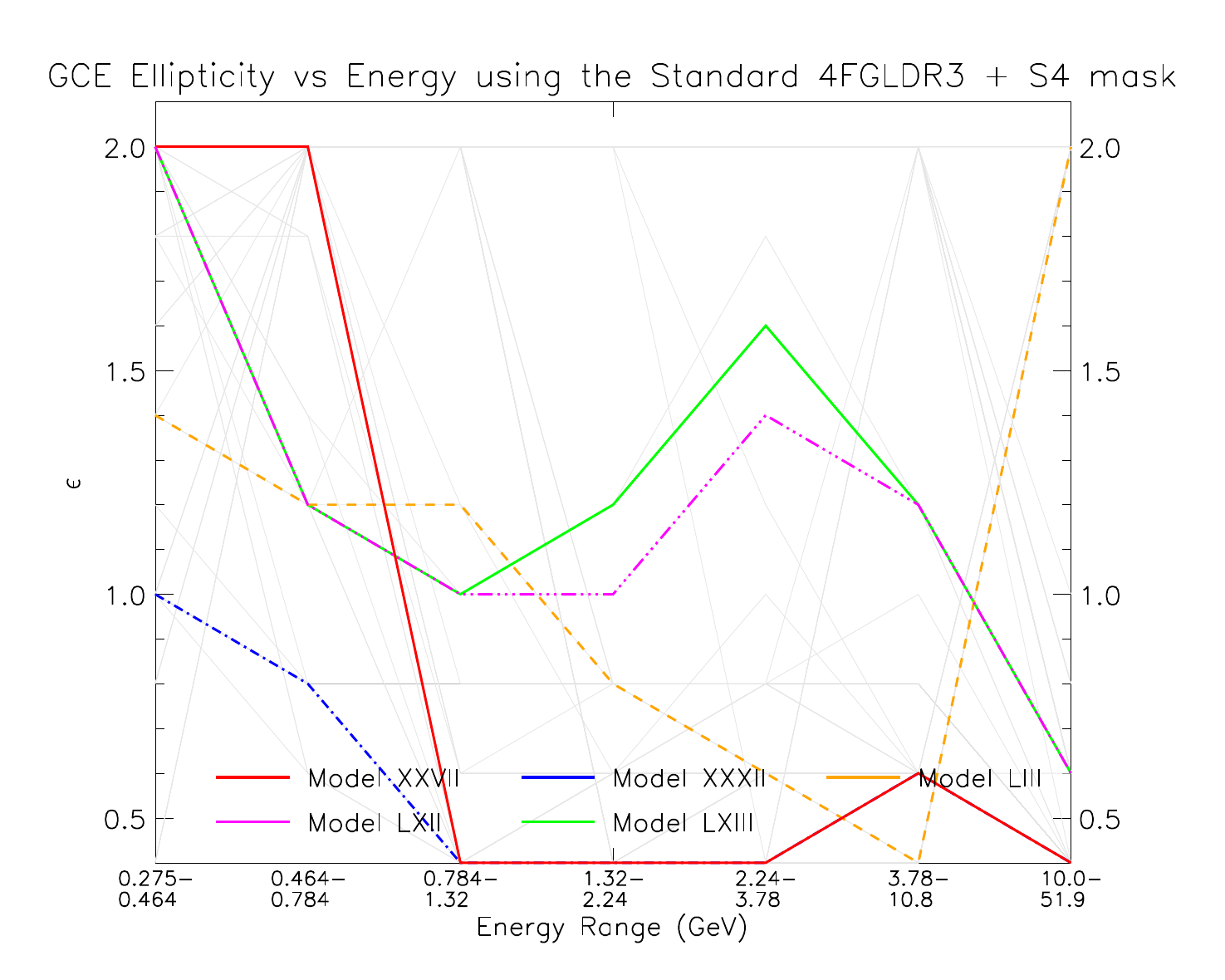}
\includegraphics[width=0.32\textwidth]{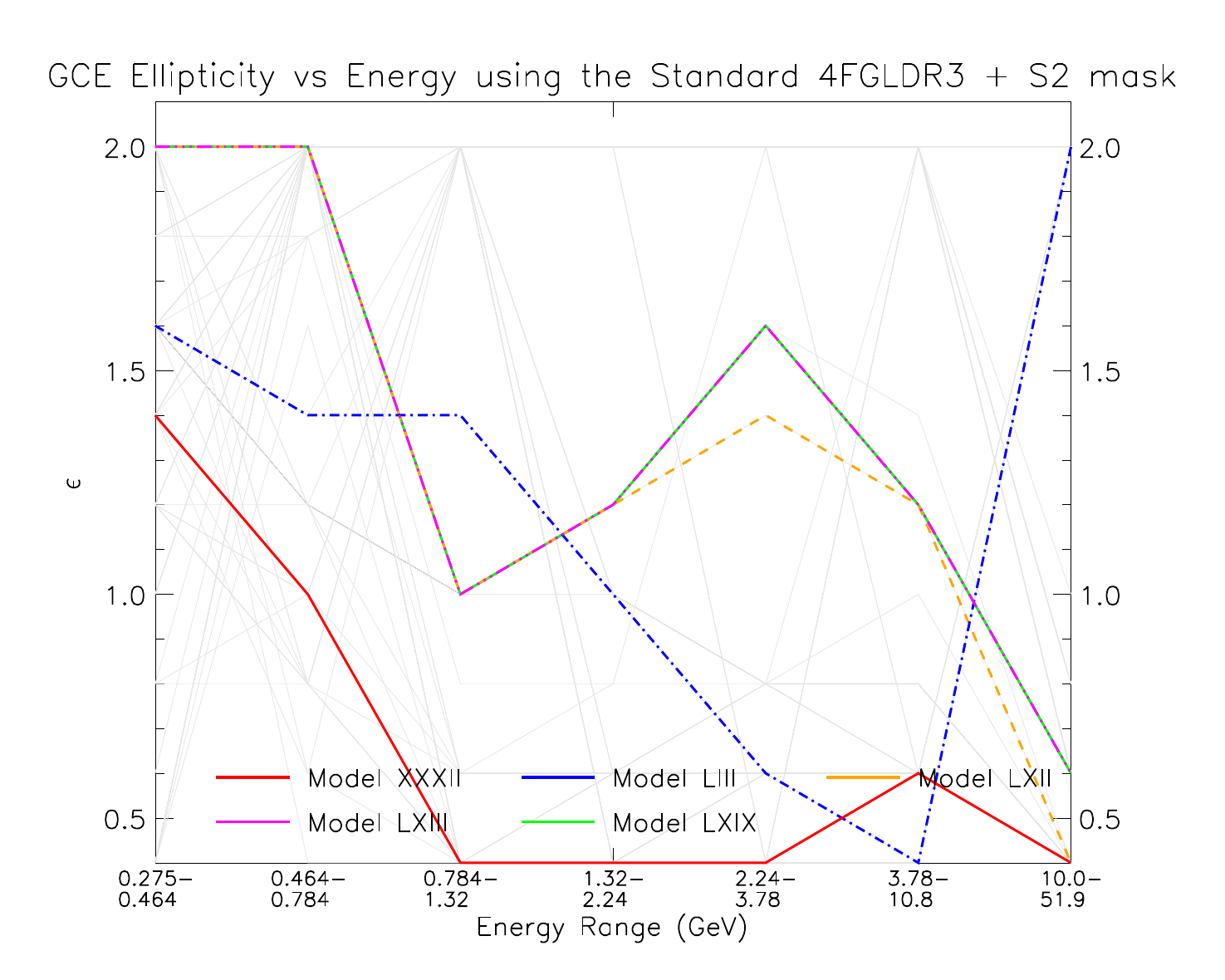}
\caption{As in the right panel of Figure~\ref{fig:morphology_vs_energy}, the evolution of the ellipticity of the GCE with energy for a sequence of masks (see text for more details).} 
\label{fig:Ellipticity_Energy_Evolution_Alt_Masks}
\end{figure*}

In Fig.~\ref{fig:Ellipticity_Energy_Evolution_Alt_Masks}, 
we show the ellipticity evolution with energy for the same six masks. Again, the regular masks, in combination with the best-fit background models, converge to values of $0.8 \leq \epsilon \leq 1.2$ for energies $>3$ GeV. Instead, the wavelet-based masks provide inconclusive results on the GCE ellipticity. 

\section{The energy evolution of the statistical preference for the GCE components}
\label{app:GCE_Components_Statistical_Preference_vs_Energy}

In the main text, in Secs.~\ref{subsec:DM_vs_Bulges} and ~\ref{subsec:DM_and_Bulges} 
and Figs.~\ref{fig:DM_vs_Bulges}, ~\ref{fig:GCE_DM_and_Bulge_Fluxes_Regular_Masks} and~\ref{fig:GCE_DM_and_Bulge_Fluxes_Wavelet_Masks} 
we investigated the possibility that the GCE is a combination of both a component following the dark matter annihilation profile 
and a component that traces the ``Coleman Bulge'' profile.
As we showed in the main text, the component tracing the dark matter annihilation profile is always present, with a normalization that for energies $E>0.7\,\text{GeV}$ accounts for $\simeq 50-90\%$ of the total GCE flux. 
Only when we use the ``Standard 4FGLD3 + Wavelet S2'' mask, we find a lower flux associated with the dark matter annihilation profile.  

\begin{figure*}
\centering
\includegraphics[width=0.48\textwidth]{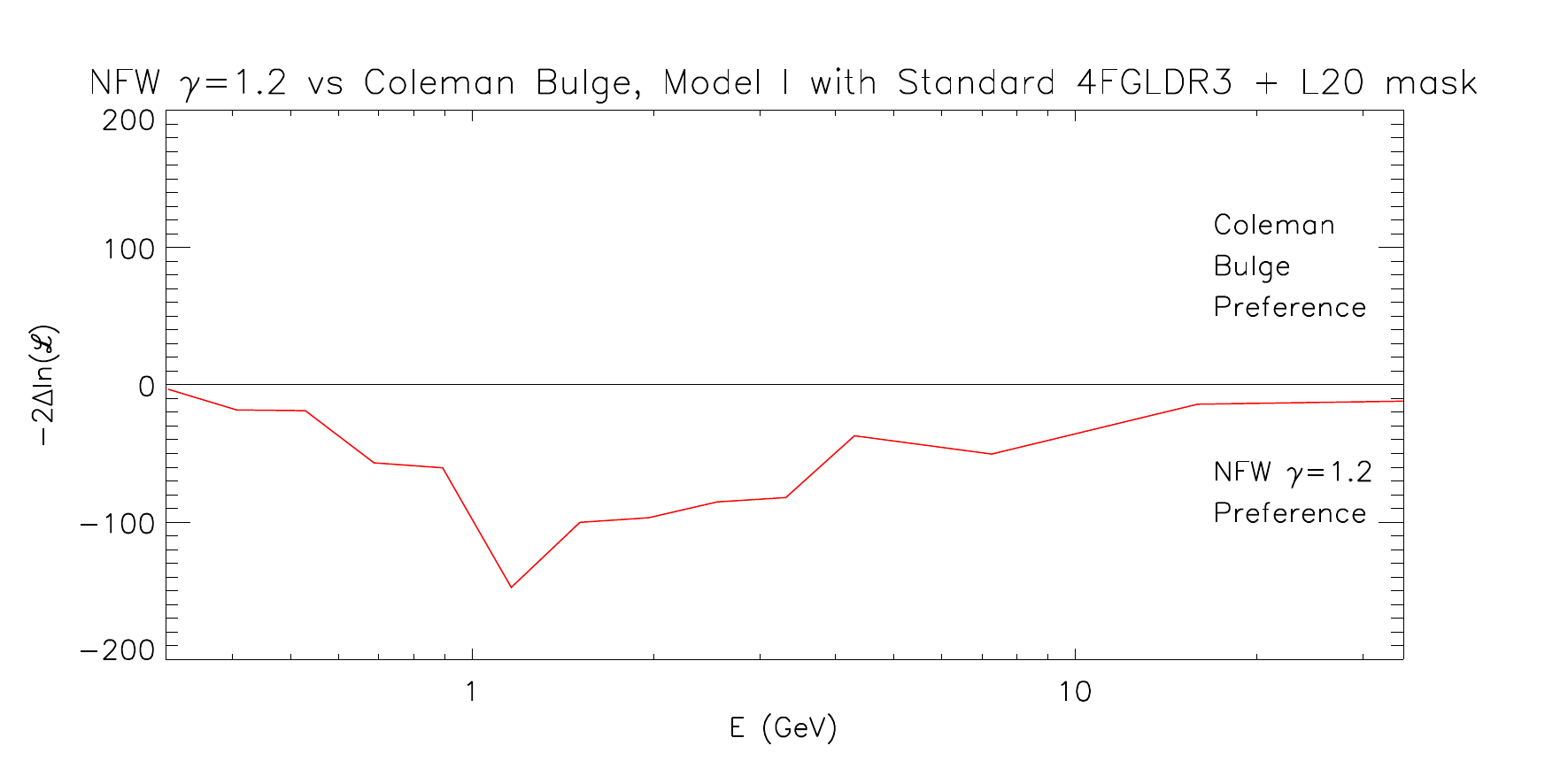}
\includegraphics[width=0.48\textwidth]{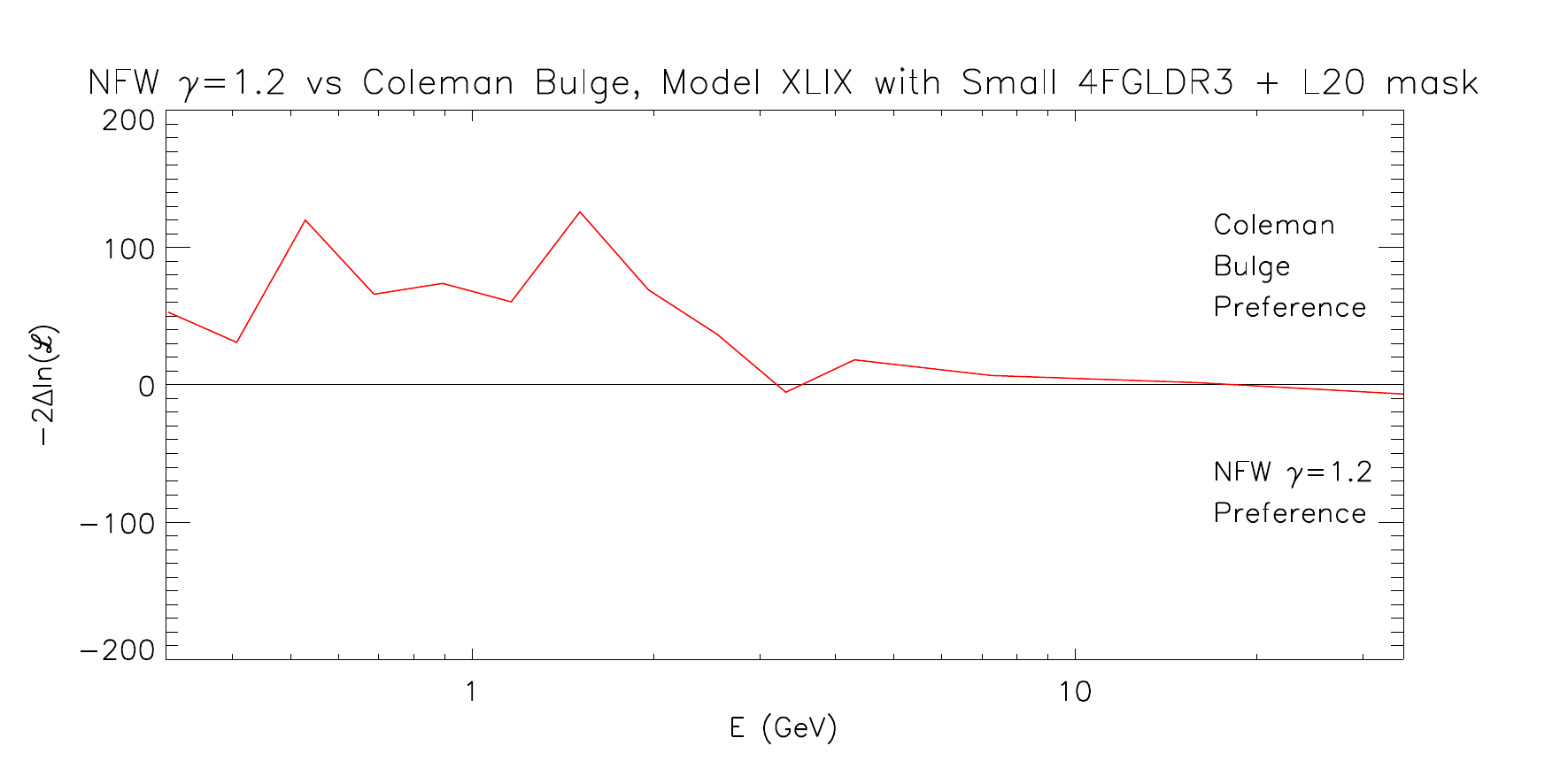}
\\
\includegraphics[width=0.48\textwidth]{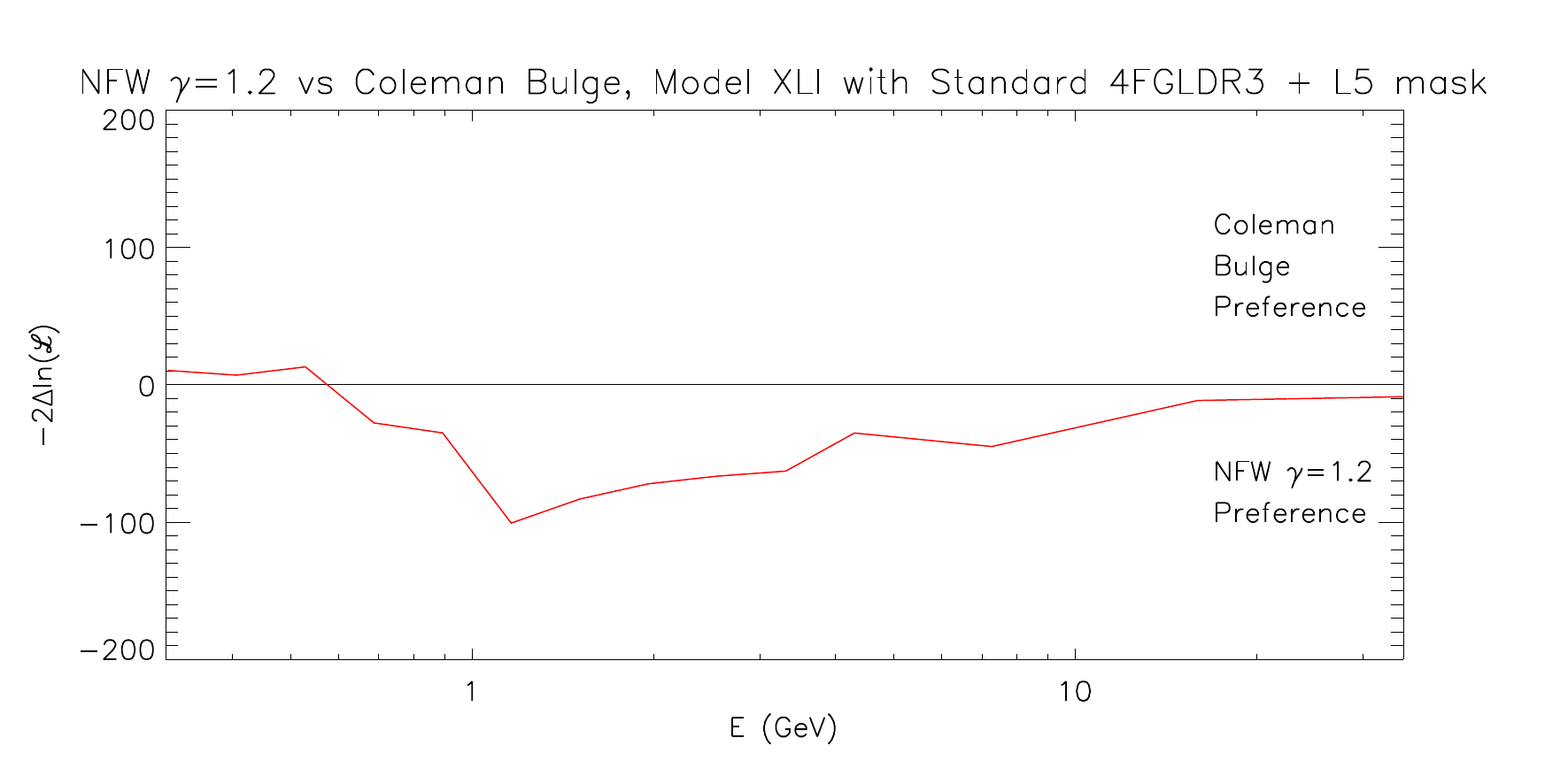}
\includegraphics[width=0.48\textwidth]{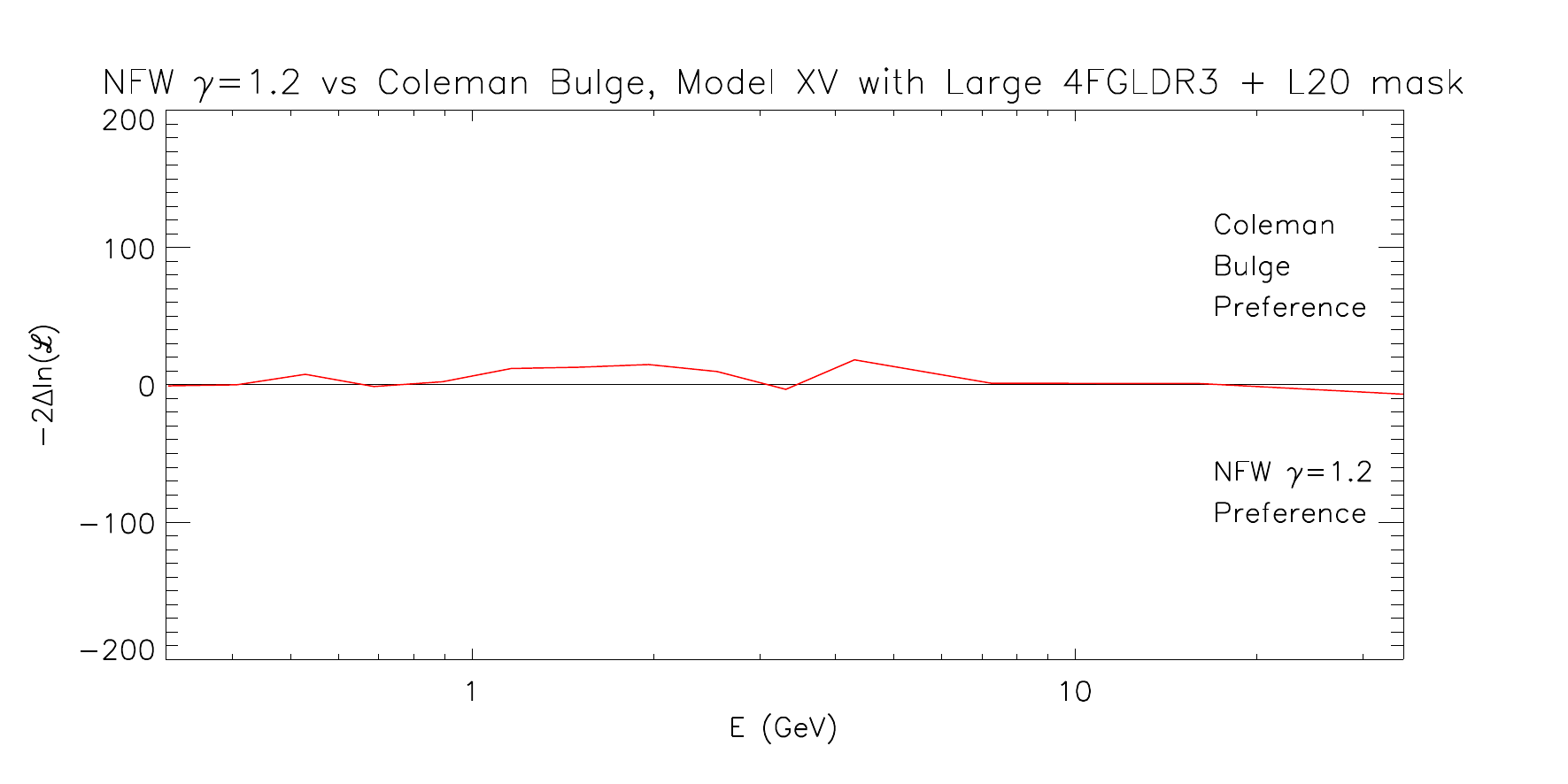}
\\
\includegraphics[width=0.48\textwidth]{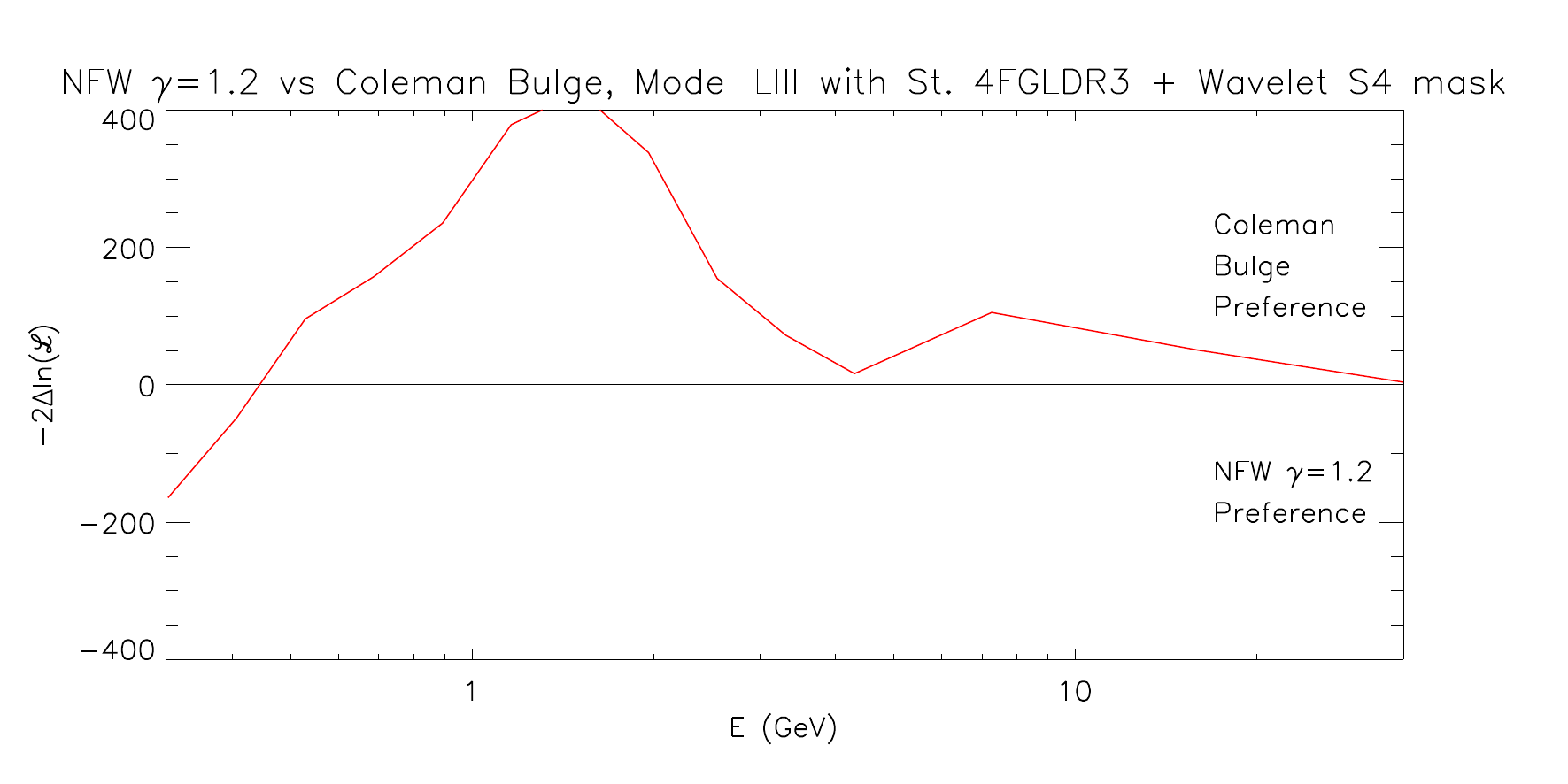}
\includegraphics[width=0.48\textwidth]{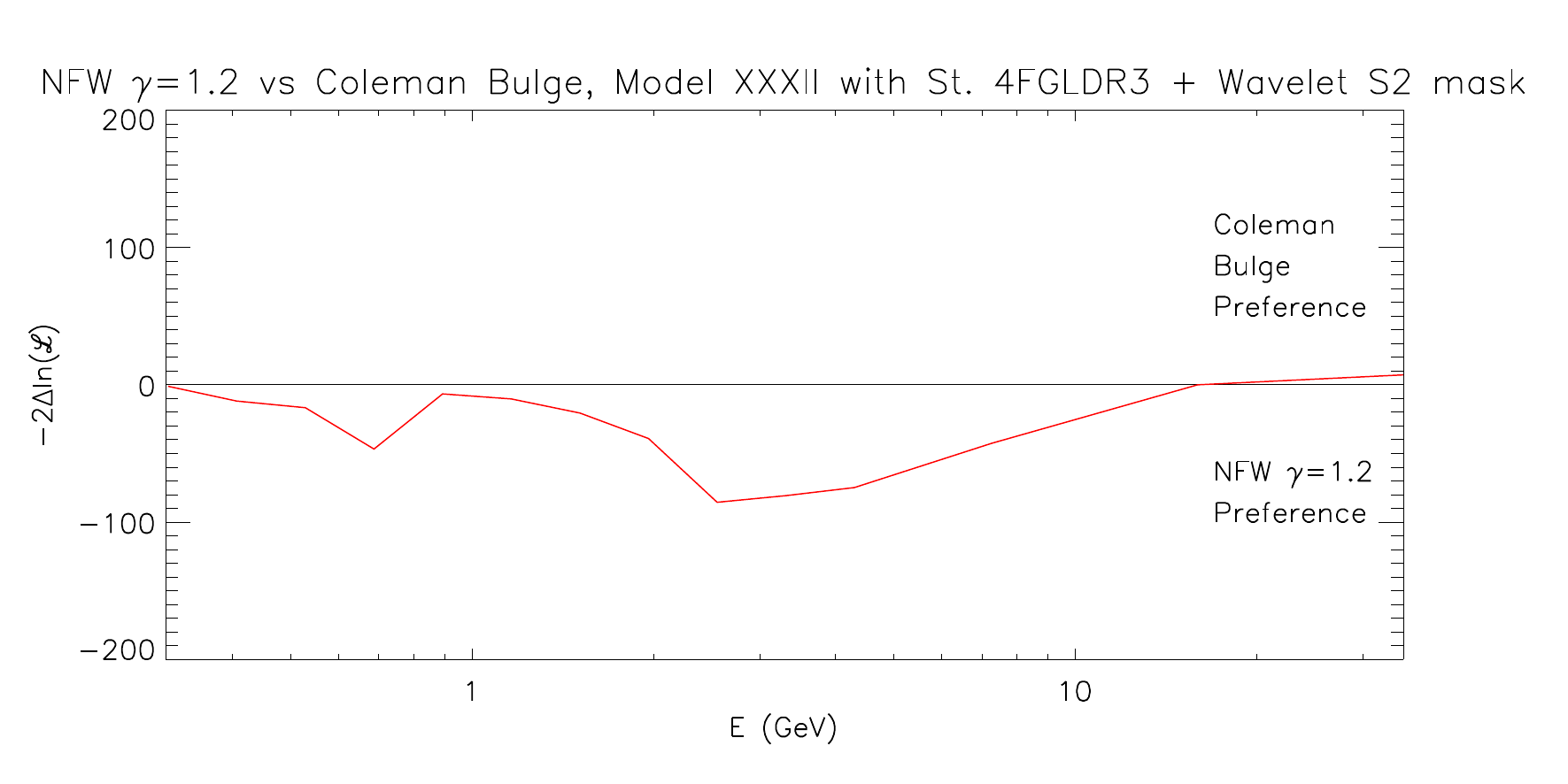}
\caption{For a sample of galactic diffuse models and masks, we show the energy bin-by-energy bin level of statistical preference between the dark matter annihilation ``NFW with $\gamma=1.2$'' profile vs. the ``Coleman Bulge'' profile.} 
\label{fig:StatPref_of_DM_and_Bulge}
\end{figure*}

In this appendix, we provide further results on the evolution with the energy of the statistical preference for the GCE to trace either the dark matter annihilation morphology or the ``Coleman Bulge'' one.
We test the cases where the GCE is purely tracing one or the other morphology. 
In Fig.~\ref{fig:StatPref_of_DM_and_Bulge}, we show for the same six masks as in section ~\ref{sec:morphism} and for specific 
galactic diffuse background models the energy bin-by-energy bin difference in negative log-likelihood (in $-2 \Delta \ln \mathcal{L}_j$) 
between the dark matter annihilation ``NFW with $\gamma=1.2$'' profile and the ``Coleman Bulge'' profile. The biggest differences show up at 
energies between 1 GeV and 5 GeV, with little statistical difference between the two profiles for $E > 10$ GeV where the number of detected photons is small. Also, at $E<0.7$ GeV, there is typically little statistical preference between the two options. At the lowest energies, the background emission is too dominant.    

\end{appendix}  
                  
\bibliography{GCE_bib}

\end{document}